\documentclass[11pt,tightenlines,eqsecnum,floats,aps,amsmath,amssymb,nofootinbib,prd,shownopacs,floatfix]{revtex4}
\usepackage[english]{babel}
\usepackage[utf8x]{inputenc}
\usepackage{amsmath,amsfonts,amssymb, extarrows,mathrsfs}
\usepackage{xcolor}
\usepackage{graphicx}
\usepackage{empheq}
\usepackage{hyperref}
\usepackage{bbm}
\usepackage{pdfpages}
\usepackage{makecell}
\newcommand{\be}{\begin{equation}}
\newcommand{\ee}{\end{equation}}
\newcommand{\ba}{\begin{eqnarray}}
\newcommand{\ea}{\end{eqnarray}}

\DeclareMathAlphabet{\mathpzc}{OT1}{pzc}{m}{it}
\newcommand*\bigcdot{\mathpalette\bigcdot@{.5}}
\begin{document}
\title{{Gauge invariant canonical cosmological perturbation theory with geometrical clocks in extended phase space - a review and applications}}

\author{Kristina Giesel}
\thanks{kristina.giesel@gravity.fau.de}
\author{Adrian Herzog}
\thanks{adri.a.herzog@studium.uni-erlangen.de}
\affiliation{Institute for Quantum Gravity, Department of Physics \\ FAU Erlangen -- N\"urnberg, 
Staudtstr. 7, 91058 Erlangen, Germany}

\begin{abstract}
The theory of cosmological perturbations is a well elaborated field and has been
successfully applied e.g. to model the structure formation in our universe and the prediction of the power spectrum of the cosmic microwave background. To deal with the 
diffeomorphism invariance of general relativity one generally introduces combinations of
the metric and matter perturbations which are gauge invariant up to the considered order in
the perturbations. For linear cosmological perturbations one works with the so-called
Bardeen potentials widely used in this context. However, there exists no common procedure to construct gauge invariant quantities also for higher order perturbations. Usually, one has to find new gauge
invariant quantities independently for each order in perturbation theory. With the relational formalism introduced by Rovelli and further developed by Dittrich and Thiemann, it is in principle possible to calculate manifestly gauge invariant quantities, that is quantities that are gauge invariant up to arbitrary order once 
one has chosen a set of so-called reference fields, often also called clock fields. This article contains a review of the relational formalism and its application to canonical general relativity following the work of Garcia, Pons, Sundermeyer and Salisbury. As the starting point for our application of this formalism to cosmological perturbation theory, we also review the Hamiltonian formulation of the linearized theory for perturbations around FLRW spacetimes. The main aim of our work will be to identify clock fields in the context of the relational formalism that can be used to reconstruct quantities like the Bardeen potential as well as the Mukhanov-Sasaki variable. This requires a careful analysis of the canonical formulation in the extended ADM-phase space where lapse and shift are treated as dynamical variables.
 The actual construction of such observables and further investigations thereof will be carried out in our companion paper.\end{abstract}
\maketitle

\newpage

\tableofcontents
\newpage

\section{Introduction}

The notion of observables in general relativity has been discussed and addressed since Einstein published his famous theory in the early 19'th century. Already Einstein realized that the general covariance of his theory leads to the problem that the dynamics of the spacetime metric are not uniquely determined by the field equations \cite{Einstein,Iftime}. The resolution to this problem is, that the value of the metric field in a specific coordinate system has no real physical meaning. The question, what specific value the field has at some point on the spacetime manifold can not be formulated in a coordinate independent way. Only quantities which are invariant under diffeomorphisms can be assigned a physical meaning. These quantities are denoted as observables. However, the question remained of how to actually construct such observables. Moreover, if physical quantities are independent of the choice of coordinates, and locally time is one of those coordinates, how can these observables possess a non-trivial time evolution? This problem of time is reflected in the canonical ADM-formulation of general relativity by the fact that the ADM-Hamiltonian consists of a linear combination of constraints only and hence general relativity is a fully constrained theory. This means that every Dirac observable in general relativity, that is a diffeomorphism invariant quantity on the ADM-phase space, by construction Poisson commutes with the ADM-Hamiltonian. Hence, the only possible contribution to a non-trivial time evolution of such an observable could come from an explicit time dependence. But if the observable would possess some explicit time dependence, it would not be invariant under time-reparametrizations and thus is not coordinate independent. 

A framework in which this problem can be resolved and a non-trivial evolution of Dirac observables can be formulated is the so-called relational formalism. These ideas date back to the seminal work of Bergmann, Komar and Kuchar \cite{Bergmann:1961zz,Bergmann:1961wa,1958PhRv..111.1182K,2011IJMPD..20....3K}. 
Its conceptual foundations were very much improved in the 1990s, see \cite{RovelliPartial,RovelliObservable} and references therein. The idea of this formalism is that in a generally covariant framework there are only fields and relations among those fields. It is not a gauge invariant question to ask, how some quantity evolves with respect to some specific foliation of spatial hypersurfaces and hence coordinate time, because diffeomorphisms might change this foliation. Instead, one should consider how fields `evolve' with respect to the values other fields take. This is a gauge invariant statement. Following this idea one wants to formulate general relativity in such a way that the evolution of fields is not described with respect to the four spacetime coordinates, denoted by $x^\mu$, but one considers the evolution with respect to values of four so-called reference or clock fields $T^\mu$, $\mu=0,1,2,3$. By means of these reference fields/clock fields one introduces physical spatial and temporal coordinates. These are the values the reference fields $T^\mu$ take and they will be denoted by $\tau^\mu$. The kinematics and dynamics of a given field $\phi$ are then formulated with respect to these physical coordinates.
The corresponding mathematical framework that allows an explicit construction of observables was also developed in the 1990's, see for instance \cite{Vytheeswaran:1994np} and references therein and rediscovered more recently by Dittrich \cite{Dittrich,Dittrich2} and Thiemann \cite{Thiemann2}. For a brief review see for instance \cite{Giesel:2008zz}. This allows to construct for each generic field $\phi$ an associated Dirac observable, that is a gauge invariant extension of $\phi$. Such a Dirac observable is a function of the values of the physical coordinate and encodes the value of $\phi$ at those values where the clock field takes the value $\tau^\mu$.  

An application of this formalism to general relativity can for instance be found in \cite{Thiemann3,Dust1,Dust2}, where one uses dust matter for the clock fields. These dust fields fill up the entire spacetime and therefore they supply a sufficient coordinatization for the other dynamical fields. An evolution of the observables  with respect to time and spatial coordinates is replaced by the evolution with respect to values of the dust fields, these provide the physical spatial and temporal coordinates. The introduction of the additional dust matter degrees of freedom can be understood as a dynamically coupled observer with respect to which the dynamics of the remaining fields is formulated. Since the observer is dynamically coupled, back-reaction effects will occur in general. However, as shown in \cite{Dust1}, for the dust model these can be tuned to be arbitrarily small. A further application to scalar-tensor-theories and LTB spacetimes can be found in \cite{Han:2015jsa} and \cite{Giesel:2009jp} respectively. Recently such matter reference systems were also used for models in loop quantum gravity \cite{Giesel:2007wn,Domagala:2010bm,Husain:2011tk}. For a review on scalar matter reference models see \cite{Giesel:2012rb}.

One can avoid introducing additional reference matter by using certain combinations of the metric components as reference fields. This is possible, because in vacuum general relativity the metric possesses only two physical degrees of freedom. However, as far as one later aims towards a quantization of the Poisson algebra of observables the reference matter systems are of great advantage because all of them yield to an  algebra of the observables with a standard form of the Poisson brackets. Therefore, finding representations of these algebras is not complicated, whereas in general this is exactly the point where reduced quantization could fail.

In this work we restrict to the classical theory and apply the relational formalism to linearized cosmological perturbation theory using clock fields chosen from the geometrical degrees of freedom. This is motivated from the framework of gauge invariant cosmological perturbation theory (see e.g. \cite{bardeen,Mukhanov}), where one uses combinations of the metric perturbation to construct gauge invariant quantities. These combinations are constructed by using the transformation behaviors of the linear perturbations under linearized diffeomorphisms. Common choices of these gauge invariant quantities are the so-called Bardeen potentials $\Psi$, $\Phi$ \cite{Mukhanov,bardeen2} and the Mukhanov-Sasaki variable $v$ \cite{Mukhanov2,Sasaki}. These linearized observables, however, are only gauge invariant up to first order in the perturbations. For higher order perturbation theory one has to find new gauge invariant quantities which are gauge invariant up to the considered order. As shown in \cite{Dust1,Dust2} the construction of the observables in the relational formalism provides a very natural setting to construct manifestly gauge invariant quantities, that means gauge invariant up to arbitrary order. 
The strategy one follows here is to choose a set of reference fields for full general relativity and then constructs manifestly gauge invariant quantities whose equations of motion can be understood as a gauge invariant version of Einstein's equations because each quantity involved in these equations is already manifestly gauge invariant. 
The latter equations are used as the starting point for cosmological perturbation theory. Hence, first one specializes the equations to FLRW spacetimes and secondly considers linear perturbations around the background. By construction one is only perturbing gauge invariant objects and hence any perturbation thereof will be again manifestly gauge invariant. As shown in \cite{Dust2} this framework reproduces the results obtained in linearized cosmological perturbation theory, when specialized to the case where the gauge invariant extensions are only considered up to linear order. The difference one can point out is that in the usual case one first perturbs the equations of motion and afterwards constructs gauge invariant extensions, whereas in \cite{Dust1,Dust2} one does exactly the opposite. One very convenient aspect of the latter approach is, that in this case one is able to choose reference fields for full general relativity gauge invariant extension up to any order are naturally given in the relational formalism once a set of reference fields has been chosen.

 In the case of linear cosmological perturbation theory the Bardeen potentials and the Mukhanov-Sasaki variable can be interpreted as linear gauge invariant extensions of cosmological perturbations whereas the extensions contain either purely metric perturbations and metric and scalar field perturbations respectively. In the relational formalism these gauge invariant extensions are constructed by means of the clock fields. Hence, this indicates that in the relational formalism the Bardeen potentials and the Mukhanov-Sasaki variable can be rediscovered choosing geometrical clocks. To find the explicit form of these geometrical clocks and thus re-derive the gauge invariant quantities used in cosmological perturbation theory will be the main topic of this review. As there is a natural interplay between the choice of clocks and a corresponding gauge fixing, this work might also help to understand the physical interpretation of the common gauges used in cosmological perturbation theory. 
 
To apply the observable framework to cosmological perturbation theory, we have to cast the latter into a Hamiltonian formulation. This has been done for instance by Langlois in \cite{Langlois} who also performed a scalar, vector and tensor decomposition of the perturbations in the reduced ADM-phase space, where lapse and shift are treated as Lagrange multipliers but not as dynamical degrees of freedom. A formulation of cosmological perturbation theory in the framework of the relational formalism based on the reduced phase space in terms of Ashtekar variables can be found in \cite{Dittrich:2007jx}. For other examples where cosmological perturbation theory has been analyzed in the canonical framework see for instance \cite{Pinho:2006ym,Falciano:2008nk}.

However, perturbations of lapse and shift are not considered in \cite{Langlois} which are needed in order to construct the Bardeen potentials. The Mukhanov-Sasaki variable does not involve lapse and shift perturbations and therefore Langlois was able to find the respective phase space function without considering lapse and shift perturbations. In our companion paper \cite{Giesel:2018opa} we will mainly focus on reproducing the Bardeen potentials and hence we need to take lapse and shift perturbations into account. For this purpose we apply the generalized formalism introduced by Pons et. al. \cite{Pons1,Pons2} for the full ADM phase space.
For simplicity we restrict to a flat FLRW background and a Klein-Gordon scalar field, minimally coupled to gravity, as the matter content in our work. 

This article is structured as follows: 

In the second section we review the relational formalism and the construction of observables discussed in \cite{Dittrich,Dittrich2,Thiemann2} and its application to canonical general relativity following the seminal work of Pons et al. in \cite{Pons1,Pons2,Pons3,Pons4}. This also provides the foundation for our companion paper \cite{Giesel:2018opa}, where the explicit construction of observables yielding the Bardeen potentials and the Mukhanov-Sasaki variable along with various other gauge invariant variables is presented. 

In the third section we discuss linear canonical cosmological perturbation theory and introduce the scalar-vector-tensor decomposition. We find expressions for the Bardeen potentials and the Mukhanov-Sasaki variable in phase space and derive their equations of motion. These equations of motion and the linearized constraints are compared to the linearized gauge invariant Einstein-equations of cosmological perturbation theory in \cite{Mukhanov} for consistency. This will be the starting point of our companion paper \cite{Giesel:2018opa} since the results involved in this paper give already a good hint what choice of geometrical clocks will be appropriate.

In the forth section we briefly sketch how to construct observables for cosmological perturbation theory. Actual applications of this formalism will be presented in a our companion paper \cite{Giesel:2018opa}. 

In section 5 we summarize the results, discuss open questions and give an outlook for possible future research directions of this topic. 

In Appendix A we derive the perturbed equations of motion for the full ADM-phase space of general relativity with an arbitrary background. 

For the benefit of the reader in the following we provide a list of our main notation and conventions that will be used throughout the article:
\begin{itemize}
\item We use the Einstein sum convention.
\item The signature $(-1,1,1,1)$ is used.
\item $c=1$ is used.
\item We use the $q-p$ Poisson bracket convention (e.g. $\{ q,p \}= 1$).
\item We suppress time and spatial dependencies  of tensor fields where possible for conciseness.
\item We assume that all quantities vanish at the boundary of the spatial manifold or at spatial infinity. This means that we can neglect boundary terms when performing integrations by parts.
\item We suppress $\Sigma$ in $\int_\Sigma\mathrm{d}^3x$ when it is clear that the integration is over the whole spatial hypersurface.
\item The Legendre map is denoted by $\mathcal{LM}$ and the corresponding inverse Legendre map by $\mathcal{LM}^*$.
\end{itemize}
\begin{center}
\textbf{Cosmology}
\begin{tabular}{r|l}
Notation & Meaning \\
\hline
$a$ & Cosmological scale factor \\
$A= a^2$ & The squared scale factor \\
$\mathcal{H} = \tfrac{\dot{a}}{a}$ & Hubble parameter \\
$\bar{N}$ & Background lapse, $\bar{N} = \sqrt{A}$: conformal time, $\bar{N} = 1$: proper time \\
$\tilde{\mathcal{H}} = \mathcal{LM}(\mathcal{H})$ & Hubble parameter in phase space \\
$\tilde{P} = - \frac{2\sqrt{A}\tilde{\mathcal{H}}}{\bar{N}}$ & $\propto$ momentum of $A$, $\bar{P}^{ab} = \tilde{P}\delta^{ab}$ \\
$\rho,p$ & Energy-density and pressure \\
$\bar{\varphi}, \bar{\pi}_\varphi$ & Background scalar field and its momentum
\end{tabular}
~\\
~\\
\textbf{Cosmological Perturbation Theory}
\begin{tabular}{r|l}
Notation & Meaning \\
\hline
$T^{(n)} \equiv \tfrac{1}{n!}\delta^nT$ & $n$'th perturbation of $T$ \\
$\phi,B,\psi,E,p_\phi,p_B,p_\psi,p_E$ & Scalar perturbations \\
$S^a,F_b,{p_S}_c,p_F^d$ & Transversal vector perturbations \\
$h^{TT}_{ab},p_{h^{TT}}^{cd}$ & Transversal traceless tensor perturbations \\
$\delta N, \delta N^a$ & Lapse and shift perturbations $\delta N = \bar{N}\phi$, $\delta N^a = B^{,a} + S^a$ \\
$\delta q_{ab}$ & Spatial metric perturbation $\delta q_{ab} = 2A( \psi\delta_{ab} + E_{,<ab>} + F_{(a,b)} + \tfrac{1}{2}h^{TT}_{ab} )$ \\
$\delta P^{ab}$ & Spatial momentum perturbation $\delta P^{ab} = 2\tilde{P}( p_\psi\delta^{ab} + p_E^{,<ab>} + p_F^{(a,b)} + \tfrac{1}{2}p_{h^{TT}}^{ab} )$ \\
$\delta \varphi,\delta \pi_\varphi$ & Scalar field perturbation and momentum thereof \\
$\mathcal{E}, \mathcal{P}$ & Matter part of energy/pressure perturbations \\
$\Phi,\Psi$ & Bardeen potentials \\
$\Upsilon$ & Gauge invariant perturbation related to the momentum of $\Psi$ \\
$v$ & Mukhanov-Sasaki variable \\
$\nu^a$ & Gauge invariant vector perturbation 
\end{tabular}
\end{center}

\begin{center}
\textbf{General notations, canonical general relativity}
\begin{tabular}{r|l}
Notation & Meaning \\
\hline
$a,b,c,... = 1,2,3$ & Spatial indices \\
$\mu,\nu,\rho,... = 0,...,3$ & Spacetime or temporal-spatial indices \\
$\partial_af, f_{,a}, \frac{\partial f}{\partial x^a}$ & Partial derivatives \\
$\nabla_\mu f, f_{;\mu}$ & Spacetime covariant derivatives \\
$D_a f, f_{|a}$ & Spatial covariant derivatives \\
$M$ & Spacetime manifold \\
$\Sigma$ & Spatial manifold \\
$X,Y,...$ & Spacetime coordinates \\
$x,y,...$ & Spatial coordinates \\
$g_{\mu\nu}$ & Spacetime metric \\
$\Gamma^\mu_{~\nu\rho}$ & Christoffel symbols \\
$R_{\mu\nu\rho}^{(4)~\sigma}$ & Spacetime Riemann tensor $[\nabla_\mu,\nabla_\nu]\omega_\rho = R_{\mu\nu\rho}^{~~~~\sigma}\omega_\sigma$ \\
$q_{ab}$ & Induced metric on $\Sigma$, spatial metric \\
$\Gamma^a_{bc}$ & Spatial Christoffel symbols \\
$P^{ab}$ & Conjugate momentum to $q_{ab}$ \\
$N,N^a$ & Lapse function and shift vector field \\
$\Pi,\Pi_a$ & Momenta of lapse and shift, primary constraints \\
$\lambda,\lambda^a$ & Lagrange multipliers associated to $\dot{N}$,$\dot{N}^a$ \\
$R^{(3)~d}_{abc}$ & Spatial Riemann tensor $[D_a,D_b]\omega_c = R^{(3)~d}_{abc}\omega_d$ \\
$K_{ab}$ & Extrinsic curvature \\
$T_{(ab...)}$ & Symmetrization \\
$T_{[ab...]}$ & Antisymmetrization \\
$\text{Tr}(T) = q^{ab}T_{ab}$ & Trace of the tensor $T$ \\
$T^T_{ab} = T_{ab} - \tfrac{1}{3}q_{ab}\text{Tr}(T)$ & Traceless part of $T$ \\
$T_{<ab>} = T_{(ab)}-\tfrac{1}{3}q_{ab}\text{Tr}(T)$ & Symmetric and traceless combination \\
$F[f]$ & Functional $F$ of a function $f$ \\
$f[f']$ & $ := \int_\Sigma\mathrm{d}^3x f(x)f'(x)$ \\
$\kappa$ & $:=16\pi G$, $G$: Newton's constant \\
$\{ q_{ab}(x),P^{cd}(y) \}$ & $ := \kappa\delta_{(a}^c\delta_{b)}^d\delta(x,y)$ \\
$\varphi,\pi_\varphi$ & Scalar field and its momentum \\
$\lambda_\varphi$ & Coupling constant of the scalar field action \\
$V(\varphi)$ & Scalar field potential
\end{tabular}
\end{center}

%\newpage

\begin{center}
\textbf{Observables in canonical general relativity} \\
\begin{tabular}{r|l}
Notation & Meaning \\
\hline
$T^\mu$ & Clock fields \\
$G^\mu = \tau^\mu-T^\mu$ & Gauge fixing constraints \\
$\mathcal{O}_{f,T}[\tau]$ & Observable of $f$ \\
$\mathcal{A}^\mu_\nu(x,y)$ & $:= \{ T^\mu(x),C_\nu(y) \}$, matrix for weak abelianization \\
$\mathcal{B}$ & $:= \mathcal{A}^{-1}$ \\
$\tilde{C}_\mu$ & weakly abelianized constraints for reduced ADM \\
$( \tilde{\Pi}_\mu,\tilde{\tilde{C}}_\nu )$ & weakly abelianized constraints for full ADM-phase space \\
$\{\cdot,\cdot\}^*$ & Dirac bracket with respect to constraint set $(G^\mu, \tilde{C}_\nu)$
\end{tabular}
\end{center}

\newpage

\section{Observables in canonical general relativity}\label{se:obsGR}
In this chapter we give a brief introduction to the relational formalism. In our presentation we mainly will follow the work by Rovelli in \cite{RovelliPartial,RovelliObservable}, Dittrich \cite{Dittrich,Dittrich2} and Thiemann \cite{Thiemann2}.
In particular we will discuss its application to general relativity. In \cite{RovelliPartial,RovelliObservable,Dittrich,Dittrich2,Thiemann2} only the reduced ADM-phase space was considered. A generalization to the full (extended) ADM-phase space where lapse and shift also treated as dynamical variables was introduced in \cite{Pons1,Pons3}. For the reason that later on we want to construct observables corresponding also to lapse and shift perturbations common in standard cosmological perturbation theory, we need the extended framework of Pons et al. here that will be reviewed in this chapter as well.

\subsection{Relational formalism and Dirac observables}\label{suse:obs}
The main idea is to take the background independence of general relativity seriously and define observables, these are gauge invariant quantities, not with respect to unphysical spacetime points in M but to use relations between dynamical fields instead. In the context of canonical general relativity such observables are called Dirac observables. The basic starting point for the relational formalism is a system that has a certain number of constraints. Let $n$ be the number of linear independent constraints generating the considered gauge transformations. Then the basic idea is to choose $n$ gauge variant phase space functions $T^\alpha$, $\alpha=1,...,n$, the so called \emph{clocks} or \emph{reference fields}. The clocks should be chosen in such a way that they can parametrize the gauge orbit, i.e. for each gauge the clocks take different values $\tau^\alpha$. We can now look at another phase space function $f$ that is gauge variant and does not depend on the clock degrees of freedom. Then following \cite{Dittrich,Dittrich2,Thiemann2} we can construct an observable $\mathcal{O}_{f,T}[\tau]$ associated to $f$ in the gauge, where the clock fields take the values $T^\alpha=\tau^\alpha$. 
As has been shown in \cite{Vytheeswaran:1994np,Dittrich} such observables $\mathcal{O}_{f,T}[\tau]$ are gauge invariant, as we will discuss later. 
The notion relational was chosen because the observable map returns the values of $f$ at those values where the reference fields $T^\alpha$ take the values $\tau^\alpha$. As a consequence $\mathcal{O}_{f,T}[\tau]$ can be understood as a function of $\tau^\alpha$, which are values of the fields $T^\alpha$ at different gauges and their evolution will be described with respect to the clock values.

In fact the picture behind this method is quite plausible. Note, that when doing measurements one does perform these measurements always with respect to a reference system (measurement apparatus) . As an example, if one would like to measure the gravitational acceleration on the surface of the earth, then one can let several objects fall from some initial height and measure the values $\tau_1$ and $\tau_2$ of some clock between the start of the falling and the collision with the ground. But the clock will in fact be some physical apparatus and not some abstract parameter $t \in \mathbb{R}$ running in the background. This is consistent with the basic foundation of general relativity that the notion of time is really an observer dependent quantity. Alternatively, we can also formulate this as follows: General relativity is a background independent theory, therefore we cannot assign a sensible physical meaning to field values at different points in the spacetime manifold. But a reasonable physical question is to ask, what value a field $\phi$ takes if another set of fields $T^\alpha$ takes values $\tau^\alpha$, because this can be expressed in background independent manner and hence can be formulated at the gauge invariant level and thus in terms of observables. 

In the next section we will discuss the application of the relational formalism to general relativity.

\subsection{Application to general relativity}

An excellent introduction to the general relational formalism is given in \cite{Dittrich}. In this work we will focus only on its application to general relativity. In particular our work will be based on generalization of Pons et. al. \cite{Pons1,Pons2,Pons3,Pons4}, where not only the reduced ADM-phase space is considered as in \cite{Dittrich2}, but also lapse and shift are taken into account as dynamical variables as in \cite{Pons3}. We will summarize the main aspects of this paper which are relevant for our work. 

Canonical general relativity coupled to matter in ADM-variables is characterized by the following phase space variables:
\begin{equation}
(q_{ab},P^{ab}) ~,~~ (N,\Pi) ~,~~ (N^a,\Pi_a) ~,~~ (\phi_I,\pi^I)~,
\end{equation}
which are fields on a 3d spatial manifold $\Sigma$ embedded in the 4d spacetime manifold $M$. $q_{ab}$ is the induced ADM-metric on $\Sigma$ and $P^{ab}$ the respective momentum. $N$ and $N^a$ are called lapse function and shift vector field respectively and $\Pi,\Pi_a$ represent their momenta. The entire matter degrees of freedom are encoded in $(\phi_I,\pi^I)$ denoting the matter fields and their associated momenta.
Note, that in our notation the momenta of the geometric quantities are defined in such a way that there appears a factor of $\kappa := 16\pi G$ in their Poisson brackets, that is
\begin{align}
\{ q_{ab}(x),P^{cd}(y)  \} &= \kappa \delta_{(a}^c\delta_{b)}^d\delta(x,y) & \{ N(x),\Pi(y) \} &= \kappa \delta(x,y) & \{ N^a(x),\Pi_b(y) \} = \kappa \delta^a_b\delta(x,y).
\end{align}
A Dirac constraint analysis shows that the system possesses four primary constraints $\Pi$, $\Pi_a$ which are exactly the momenta of lapse and shift and four secondary constraints called the Hamilton constraint $C$ and the spatial diffeomorphism constraint $C_a$. Their names stem from the fact that in the reduced ADM-phase space $C$ generates spatial diffeomorphisms within the spatial hypersurfaces, whereas $C$ is the generator for diffeomorphisms orthogonal to the spatial hypersurfaces. The secondary constraints consists of a geometric and a matter contribution, that is
$C=C_\text{geo}+C_\text{mat}$ and $C_a=C_{a,\text{geo}}+C_{a,\text{mat}}$, and are given by:
\begin{equation}
\label{eq:Cgeo}
C_\text{geo}= \frac{1}{\sqrt{\det q}} Q_{abcd}P^{ab}P^{cd}-\sqrt{\det q} R^{(3)}.
\end{equation}
\begin{equation}
\label{eq:Cageo}
C_{a,\text{geo}}= -2q_{ab}P^{bc}_{|c},
\end{equation}
with
\begin{equation}
Q_{abcd} := q_{ac}q_{bd} - \tfrac{1}{2}q_{ab}q_{cd}
\end{equation}
and the curvature scalar $R^{(3)}$ on $\Sigma$. The explicit form of $C_\text{mat}$ and $C_{a,\text{mat}}$ depends on the specific matter content used. The ADM-Hamiltonian has the following form:
\begin{equation}
\label{eq:HADM}
H=\frac{1}{\kappa} \int\limits_{\Sigma} \mathrm{d}^3 x \left[ NC+N^aC_a+\lambda \Pi+\lambda^a\Pi_a \right](x).
\end{equation}
$\lambda$ and $\lambda^a$ are Lagrange multipliers. The Hamiltonian equations of motion for the elementary phase space variables can be derived by calculating their Poisson brackets with the ADM-Hamiltonian. Together with the constraint equations these equations are equivalent to Einstein's equations in the Lagrangian framework. A special feature of general relativity is that the ADM-Hamiltonian consists purely of a linear combination of constraints showing that general relativity  is a fully constrained theory. This reflects the diffeomorphism invariance in the canonical formulation. 
A good introduction to the ADM formalism involving also the derivations of the results above can for instance be found in \cite{Thiemann}. We will review the canonical equations of motion in ADM-variables in chapter (\ref{se:cosmpert}). 

In what follows, we will use the notation $N^\mu$, $\mu=0,1,2,3$ corresponding to $N^0 \equiv N$ and analogously for the constraints and other functions to combine the respective temporal and spatial parts. Considering this, we can rewrite the set of first class constraints in the following form:
\begin{equation}
\Pi_\mu ~,~~ C_\mu ~,~~ \mu=0,1,2,3.
\end{equation}
The system may possess further constraints due to additional gauge symmetries of the matter fields. Nevertheless, these constraints will Poisson commute with the remaining constraints and can therefore be treated separately. Usually the phase space is reduced with respect to lapse and shift, because their equations of motion just involve the arbitrary Lagrange multipliers $\lambda$ and $\lambda^a$ and thus their dynamics is not uniquely determined unless particular values for $\lambda$ and $\lambda^a$ are chosen. This corresponds to a symplectic reduction with respect to the primary constraints $\Pi=0$ and $\Pi_a=0$ having the consequence that we can treat lapse and shift no longer as dynamical variables but as Lagrange multipliers. We will first discuss this case here for simplicity.

\subsubsection{Reduced ADM-phase space}
\label{sususe:obsGRred}
Since a symplectic reduction with respect to the primary constraints has been performed, we treat lapse and shift as Lagrange multipliers. The reduced phase space involves the following elementary variables:
\begin{equation}
(q_{ab},P^{ab}) ~,~~ (\phi_I,\pi^I),
\end{equation}
together with the four diffeormorphism constraints $C_\mu$. The associated gauge generator has the form:
\begin{equation}
\label{eq:GaugeGenGb}
G_{b,\vec{b}} = \frac{1}{\kappa} \int_\Sigma \mathrm{d}^3x \left[ b(t,x)C(x) +b^a(t,x)C_a(x)\right] =: \kappa^{-1} C_\mu[b^\mu] =: G_\textbf{b}.
\end{equation}
The factor of $\kappa^{-1}$ is just convention.

To construct observables in the reduced ADM-phase space, we have to choose a set of clock fields $T^\mu$ from the variables $q_{ab}$, $P^{ab}$, $\phi_I$ and $\pi^I$ such that these clocks parametrize the gauge orbits induced by the gauge generators $G_\textbf{b}$. This can be achieved by demanding:
\begin{equation}
\label{eq:detTC}
\det\left(\{ T^\mu(x),C_\nu(y) \}\right) \neq 0 \quad \text{f.a.}\,\, \mu,\nu = 0,\cdots,3,
\end{equation}
where $x,y$ denote local coordinates in $\Sigma$ and the requirements need to be satisfied at least locally in a region around $x,y$. How globally a set of chosen clocks can be used depends of course on the fact how global the condition above is satisfied. 
Because in the reduced case $G_\textbf{b}$ involves only the secondary constraints, the requirement applies only to them. 
Once we have chosen such a set clocks, the formalism developed in \cite{Vytheeswaran:1994np,Dittrich2} allows a construction of
observables in a formal power series for a generic phase space function $f$ being independent of the clock degrees of freedom. The observable map transforms $f$ to the gauge where the set of clocks $T^\mu(x)$ take the values $\tau^\mu(x)$. Formally,
\begin{equation}
\label{eq:obsformula1}
\mathcal{O}_{f,T}[\tau] := \left. \alpha_{G_\textbf{b}}(f) \right|_{\alpha_{G_\textbf{b}}(T)^\mu = \tau^\mu }.
\end{equation}
Note, that we suppressed the $\Sigma$-coordinate dependency for simplicity. Here, $\alpha_{G_\textbf{b}}(f)$ is the flow of $f$ under the gauge generator $G_\textbf{b}$, that is
\begin{equation}
\alpha_{G_\textbf{b}}(f) := \exp\left( \{ \cdot,G_\textbf{b} \} \right)f.
\end{equation}
In case one chooses as $f$ clock fields, then the clock fields $T^\mu$ are mapped onto the values $\tau^\mu$. The observable $\mathcal{O}_{f,T}[\tau]$ is a functional of the functions $\tau^\mu$. It is not hard to show that the above constructed observable is indeed gauge invariant using that $\alpha_{G_{\textbf{b}}}$ is a family of gauge transformation (see e.g. \cite{Dittrich}). An explicit construction of observables requires to solve
\begin{equation}
\alpha_{G_\textbf{b}}(T)^\mu = \tau^\mu,
\end{equation}
for $b^\mu$. This becomes particularly simple if one chooses a set of clocks and constraints $\tilde{C}_\mu$, which defines the same constraints surface than the original set of constraints $C_\mu$, that build, at least weakly, a canonically conjugate pair, that is 
\begin{equation}
\{ T^\mu(x),\tilde{C}_\nu(y) \} \approx \delta^\mu_\nu \delta(x,y),
\end{equation}
where the weak equivalence corresponds to the constraint hypersurface of the $C_\mu$'s. Due to the assumption in (\ref{eq:detTC}), this is possible, at least locally. Let us define the following distributional matrix:
\begin{equation}
\mathcal{A}^\mu_\nu(x,y) := \{ T^\mu(x),C_\nu(y) \}.
\end{equation}
By assumption this matrix is invertible. Hence, we can use its inverse to define a new but equivalent set of first class constraints:
\begin{equation}
\tilde{C}_\mu(x) := \int\mathrm{d}^3y~\mathcal{B}^\nu_\mu(y,x)C_\nu(y) ~~~\text{with  } \mathcal{B} := \mathcal{A}^{-1}.
\end{equation}
The advantage of this equivalent set $\{\tilde{C}_\mu(x)\}$ is that this set of constraints is indeed weakly canonically conjugate to the clock fields:
\begin{eqnarray*}
\{T^\mu(x)\,\tilde{C}_\nu(y)\}&=&\int d^3z \{T^\mu(x),\mathcal{B}^{\rho}_\nu(z,y)C_\rho(z)\}
\approx \int d^3z  \{T^\mu(x),C_\rho(z)\}\mathcal{B}^{\rho}_\nu(z,y) \\
&=&\int d^3z \mathcal{A}^\mu_\rho(x,z)\mathcal{B}^{\rho}_\nu(z,y)
=\delta^\mu_\nu \delta(x,y).
\end{eqnarray*}
Moreover, one can show that the constraint set $\tilde{C}_\mu$ Poisson commutes up to second order in the secondary constraints:
\begin{equation}
\{ \tilde{C}_\mu(x),\tilde{C}_\nu(y) \} = \mathcal{O}(C^2).
\end{equation}
This has the effect that associated Hamiltonian vector fields $\chi_\mu := \{ \cdot,\tilde{C}_\mu \}$ are weakly abelian. Hence, this method of using the above defined constraint set $\tilde{C}_\mu$ is referred to as \emph{weak abelianization}. Now we can rewrite the gauge generator using $\tilde{C}_\mu$ as:
\begin{equation}
\tilde{G}_\xi := \frac{1}{\kappa} \int_\Sigma \mathrm{d}^3x~ \xi^\mu(t,x)\tilde{C}_\mu(x)
\end{equation}
and in order to relate $\tilde{G}_\xi$ to generator to the former gauge generator $G_\mathbf{b}$ in (\ref{eq:GaugeGenGb}) we have to set:
\begin{equation}
\xi^\mu(x,t) = \int\mathrm{d}^3y \mathcal{A}^\mu_\nu(x,y)b^\nu(y).
\end{equation}
Now, as before we can solve for the $\xi^\mu$ in terms of the clock fields $T^\mu$ and their values $\tau^\mu$ and obtain:
\begin{equation}
\alpha_{\tilde{G}_\xi}(T)^\mu \approx T^\mu +\xi^\mu \overset{!}{=} \tau^\mu \quad\quad
 \Rightarrow \xi^\mu \approx \tau^\mu -T^\mu =: G^\mu.
\end{equation}
Given, this we can in principle construct observables in terms of a formal power series with the observable formula in (\ref{eq:obsformula1}). This yields
\begin{equation}
\label{eq:obsformula2}
\mathcal{O}_{f,T}[\tau] \approx \left. \alpha_{\tilde{G}_\xi}(f) \right|_{\xi^\mu=G^\mu }.
\end{equation}
We call $G^\mu$ the \emph{gauge fixing constraints} as for the reason that in the gauge $G^\mu = 0$ we get $\mathcal{O}_{f,T}[\tau] = f$.
The explicit form of the formal power series in terms of the gauge fixing constraints $G^\mu$ is given by:
\begin{eqnarray}
\label{eq:obsformula3}
\mathcal{O}_{f,T}[\tau] &=& f + \sum\limits_{n=1}^{\infty} \frac{1}{n!} \int\mathrm{d}^3x_1 ... \int\mathrm{d}^3x_n~ G^{\mu_1}(x_1) ... G^{\mu_n}(x_n) \{...\{ f,\tilde{C}_{\mu_1}(x_1) \}, ...\tilde{C}_{\mu_n}(x_n) \}.\nonumber\\
\end{eqnarray}
We will now see that in the reduced ADM phase space the so far arbitrary Lagrange multipliers for lapse and shift $N^\mu$ are fixed by requiring that these gauge fixing conditions are stable under the evolution. The Hamiltonian of the reduced ADM-phase space simplifies to:
\begin{equation}
H_\text{red} = \frac{1}{\kappa} \int_\Sigma \mathrm{d}^3x~ N^\mu(x) C_\mu (x).
\end{equation}
We have to ensure stability of the gauge conditions $G^\mu$, that is
\begin{eqnarray}
\dot{G}^\mu(x) &=& \partial_tG^\mu(x) + \{ G^\mu(x), H_\text{red} \}= \partial_tG^\mu(x) - \frac{1}{\kappa} \int_\Sigma \mathrm{d}^3y~ N^\nu(y) \mathcal{A}^\mu_\nu(x,y) \overset{!}{=} 0.
\end{eqnarray}
The equation above can be satisfied if we choose lapse and shift as
\begin{equation}
\label{eq:Nmustability}
N^\mu(x) \overset{!}{=} \kappa \int_\Sigma \mathrm{d}^3y~ \mathcal{B}^\mu_\nu(x,y) \frac{\partial G^\nu(y)}{\partial t}
\end{equation}
and hence the arbitrary Lagrange multiplier $N^\mu$ are determined now.
~\\
~\\
\paragraph*{\centerline{1.a Evolution of Dirac observables}}
~\\

Next to the construction of observables, which would rather describe the kinematics of the theory, of course we are also interested in their dynamics. By construction all Dirac observables commute with the ADM-Hamiltonian, so it is apparently clear that it cannot be the generator of their evolution. This is related to the problem of time in general relativity where the Hamiltonian consists of a linear combination of the first class constraints only. However, in order to define also the dynamics relational, we take into account that the observables constructed with (\ref{eq:obsformula1}) depend on the parameters $\tau^\mu$, which are associated with the values of the clocks $T^\mu$ at a specific gauge. Since one of the clock fields is associated with physical time and without loss of generality we choose for this $T^0$ with parameter $\tau^0$, we can use formulate equations of motion for $\mathcal{O}_{f,T}$ as an evolution with respect to $\tau^0$, that is
\begin{equation}
\dot{\mathcal{O}}_{f,T}[\tau]:=\frac{\partial \mathcal{O}_{f,T}}{\partial \tau^0(x)}[\tau].
\end{equation}
As shown in \cite{Dittrich},\cite{Thiemann2} the above expression can again be expressed as an observable:
\begin{equation}
\label{eq:obstauevo}
\frac{\partial \mathcal{O}_{f,T}}{\partial \tau^0(x)}[\tau] \approx \mathcal{O}_{ \{f,\tilde{C}_0(x) \},T }[\tau].
\end{equation}
An interesting question is whether (\ref{eq:obstauevo}) can be written in the form
\begin{equation}
\frac{\partial \mathcal{O}_{f,T}}{\partial \tau^0(x)}[\tau]\simeq
\{\mathcal{O}_{f,T}[\tau],{\bf H}_{\rm phys}\},
\end{equation}
for an appropriate choice of a so called physical Hamiltonian ${\bf H}_{\rm phys}$ that is itself again an observable and is the generator of the physical evolution, that is with respect to physical time $\tau^0$. As shown in \cite{Thiemann2} this is always possible if the reference fields are chosen in such a way that the total Hamiltonian constraint can be written linearly in the momenta of the clock field associated with time. This important observation has for instance been used for the matter reference models in \cite{Dust1,Dust2,Domagala:2010bm,Husain:2011tk,Giesel:2016gxq,Giesel:2017mfc}.
~\\
~\\
\paragraph*{\centerline{1.b How to choose clock fields?}}
~\\

There are in principle two strategies for choosing appropriate clocks. Either one can choose clocks from the matter fields $\phi_I$. This has been done in several more recent models \cite{Dust1,Dust2,Domagala:2010bm,Husain:2011tk} to derive the associated (partially) reduced phase space of general relativity. In \cite{Dust1,Dust2} four dust fields were considered as clock fields and a reduction could be performed with respect to the spatial diffeomorphism and Hamiltonian constraint. In contrast in \cite{Domagala:2010bm,Husain:2011tk} the authors introduce one clock field only and therefore obtain the partially reduced phase space for general relativity and cosmology respectively since only a reduction with respect to the Hamiltonian constraint is taken into account. 

The other possibility is to take combinations from the geometric degrees of freedom, i.e. from $(q_{ab},P^{ab})$ as clocks. This has the advantage, that even in the absence of matter one can apply the relational formalism and define a physical time evolution. However, for both cases the construction of the observables will in general be complicated as the constraints are not necessarily linear in the elementary phase space variables. As a consequence, it may be difficult to find appropriate clocks such that the power series (\ref{eq:obsformula3}) can be calculated in explicit form up to arbitrary high orders. However, often one can at least compute the observables up to a specific order in $G^\mu$, which is justified for small deviations from the gauge constraint surface $G^\mu=0$. Hence, in such situations a perturbative treatment might still be possible and exactly this will also be considered for our work here. We will present more details on the perturbative approach in section (\ref{se:obsinpert}).

\subsubsection{Extended (or full) ADM-phase space}
\label{sususe:obsGRfull}
As mentioned before in the extended ADM-phase space we treat lapse $N$ and shift $N^a$ as dynamical variables. As far as the reduced ADM-phase space is considered the secondary constraints $C$ and $C_a$ represent the generators of diffeomorphisms for all elementary phase space variables on-shell, that is if the equations of motion are satisfied. More precisely, the group they generate is the so called Bergmann-Komar group \cite{bergmannkomar} and the action of the corresponding generators agrees only on-shell with that of the diffeomorphism group.  
However, these secondary constraints commute with lapse and shift. This implicates, that $C$ and $C_a$ do not implement the full diffeomorphism group
as in the Lagrangian picture diffeomorphisms can change $g_{00}$ and $g_{0i}$ and correspondingly on phase space lapse and shift should also transform non-trivially. Lapse and shift have the special property that their momenta coincide with the primary constraints $\Pi=\Pi_a=0$ and thus if we reduce with respect to them, as done for reduced ADM-phase space, the just become Lagrange multipliers. However, in the extended ADM-phase space the question arises how a generator for diffeomorphisms on the extended phases space looks like that generates also diffeomorphisms for lapse and shift variables. This question was answered in full detail in an article by by Pons and Garcia \cite{Pons1} where they discussed the problem of projectability\footnote{In fact we also assumed that the descriptors $b$ and $b^a$ do only depend on the coordinates $x^\mu$ and not on the fields. In particular it is crucial that the descriptors do not depend on lapse and shift in order for the diffeomorphism to be projectable to phase space (see e.g. \cite{Pons4} or \cite{bergmannkomar})} 
 of Noether gauge symmetries from the tangent bundle to phase space in the case of constrained systems involving first and second class constraints. As in the ADM-phase space formulation of general relativity there are no second-class constraints, we will briefly review their work for systems with first class constraints only.
~\\
~\\
\paragraph*{\centerline{2.a Diffeomorphism on extended phase space: Finitely many degrees of freedom}}
~\\

For simplicity, first we discuss their formalism in the context of systems with finitely many degrees of freedom relevant for classical mechanics. Assume we start with a Lagrangian theory with finite configuration-space $Q = \{q^1,q^2,...,q^N\}$, the corresponding tangent bundle $TQ$ and a given Lagrangian $L(q,\dot{q};t)$. Recall that, when the Hessian
\begin{equation}
W := \frac{\partial^2L}{\partial \dot{q}^I \partial \dot{q}^J}
\end{equation}
is singular, then the Legendre map $\mathcal{LM}:TQ\to T^*Q$ can only be applied partially and primary constraints $\phi_\alpha$ appear. It can be shown that the kernel of $W$ is spanned by the following vector fields \cite{Pons1}:
\begin{equation}
\Gamma_\alpha = \gamma^I_\alpha \frac{\partial}{\partial \dot{q}^I}~~,~~~\gamma^I_\alpha := \mathcal{LM}^*\frac{\partial \phi_\alpha}{\partial p_I} = \mathcal{LM}^*\{q^I,\phi_\alpha\},
\end{equation}
where $ \mathcal{LM}^*:T^*Q\to TQ$ denotes the inverse Legendre map and $p_I$ the conjugate momenta to $q^I$. Following \cite{Pons1} we introduce the notion of projectability, that is a function $f$ on the tangent bundle is projectable onto phase space if and only if
\begin{equation}
\Gamma_\alpha f = 0.
\end{equation}
Let us denote the generator of an infinitesimal Noether-symmetry in the Lagrangian picture by $G_L$ and Pons et al show in \cite{Pons1} that   $G_L$ is projectable to a function $G$ in phase space. Moreover, it is shown that the infinitesimal change of the configuration variables under this symmetry can be written as:
\begin{equation}
\delta_{G_L} q^I = \mathcal{LM}^*\{ q^I,G \} - r^\alpha\gamma^I_\alpha.
\end{equation}
The coefficients $r^\alpha$ are up to now arbitrary and will be determined below. The second term on the righthand side involving $\gamma^I_\alpha$ shows that the phase space function $G$ does not generate the complete gauge orbit of its associated Noether symmetry. In order to fix the $r^\alpha$ following \cite{Pons1} we consider the time derivative of the gauge generator $G$. In case that there are no second class constraints this reads
\begin{equation}
\label{eq:timeDerG}
\frac{dG}{dt} = \frac{\partial G}{\partial t}+\{ G,H \}.
\end{equation}
Note, that $G_L$ is the generator of a Noether-symmetry, which is a conserved quantity. Thus, its time derivative vanishes when the equations of motion are satisfied \cite{Pons1}. As we can relate $G_L$ to $G$ by $G_L = \mathcal{LM}^* G$ we can conclude that its time derivative shown in (\ref{eq:timeDerG}) is just a linear combination of constraints. Let us denote the primary constraints as $\{ \phi_\alpha \}$ and all secondary and higher order ones as $\{ \phi^1_\alpha \}$, which are all assumed to be first class, i.e. their algebra is closed. Hence, we write:
\begin{equation}
\label{eq:dotGgeneral}
\frac{dG}{dt} = \frac{\partial G}{\partial t} + \{ G,H \} = A^\alpha\phi^1_\alpha + ~\mathrm{primary}~\mathrm{constraints}.
\end{equation}
The relation between $r^\alpha$ and $A^\alpha$ is derived in \cite{Pons1} and given by:
\begin{equation}
r^\alpha = \mathcal{LM}^*A^\alpha.
\end{equation}
Let us assume that $\{ A^\alpha,q^I \}=0$, i.e. $A^\alpha$ does not depend on the momenta $p_I$. Since this will be the case for the diffeomorphism generator in general relativity we will restrict to this case here. A straight forward calculation shows that the action of $G_L$ on the configuration variables can be expressed as:
\begin{align}
\delta_{G_L}q^I &= \mathcal{LM}^* \{ q^I,G \} - r^\alpha \gamma^I_\alpha \nonumber \\
~ &= \mathcal{LM}^* \{ q^I,G \} - (\mathcal{LM}^* A^\alpha)(\mathcal{LM}^*\{ q^I,\phi_\alpha \}) \nonumber \\
~ &= \mathcal{LM}^* \{ q^I,G - A^\alpha\phi_\alpha \}.
\end{align}
Given this, it is natural to define a modified gauge generator $G'$ of the form:
\begin{equation}
\label{eq:G'general}
G' = G - A^\alpha \phi_\alpha .
\end{equation}
This generator  generates exactly the gauge transformations $\delta_{G_L}q^I = \mathcal{LM}^* \delta_{G'}q^I := \{ q^I,G' \}$. Furthermore, it can be shown that the same $G'$ also generates gauge transformations for $p_I$ which correspond to $\delta_{G_L}\left( \frac{\partial L}{\partial \dot{q}^I} \right)$ in the Lagrangian formalism if the equations of motion are fulfilled. 
This is however no drawback as we demand equivalence of the two formalisms only on the level of the equations of motion likewise the case of the reduced ADM-phase space. 
~\\
~\\
\paragraph*{\centerline{2.b Diffeomorphism on extended phase space: Application to general relativity}}
~\\

Let us now apply this technique to the case of the diffeomorphism generator in general relativity. The secondary constraints $C$ and $C_a$ generate diffeomorphisms in the reduced ADM-phase space. The ADM-Hamiltonian vanishes on the constraint hyersurface since it is linear in the primary and secondary constraints. We have:
\begin{align}
\label{eq:GundH}
G_{b,\vec{b}}&= \frac{1}{\kappa} \left( C[b] + C_a[b^a] \right) & H = \frac{1}{\kappa} \left( C[N] + C_a[N^a] + \lambda\Pi + \lambda^a\Pi_a \right).
\end{align}
We use the notation $F[f] = \int\mathrm{d}^3x F(x)f(x)$, where $F(x)$ is the integral kernel of a distribution $F$ and $f(x)$ a smooth test function. Next, we want to determine the modified gauge generator $G'_{b,\vec{b}}$, that is the analoge of (\ref{eq:G'general}) in the case of general relativity. For this purpose we need to compute the time derivative of $G_{b,\vec{b}}$ explicitly and we first
consider the partial time derivative of the gauge generator, we obtain \cite{Pons1}:
\begin{equation}
\frac{\partial G_{b,\vec{b}}}{\partial t} = \frac{1}{\kappa} \left( C[\dot{b}] + C_a[\dot{b}^a] \right) ~,
\end{equation}
where $b$ and $b^a$ are 1-parameter families (the $t$-dependencies are suppressed) of smooth test functions on $\Sigma$.
Note, that these test functions are assumed be independent on phase space and thus $\dot{b} = b_{,t}$, $\dot{b}^a = b^a_{,t}$. 
In order to compute the Poisson bracket between $G_{b,\vec{b}}$ and the ADM-Hamiltonian $H$ that is also involved in (\ref{eq:dotGgeneral}), we use the hypersurface deformation algebra of the secondary constraints (see for instance in \cite{Thiemann}). This yields further: 
\begin{equation}
\{ G_{b,\vec{b}},H \} = \frac{1}{\kappa}\left( C[b^aN_{,a}-N^ab_{,a}] + C_a[q^{ab}(bN_{,b}-Nb_{,b})-N^ab^b_{,a}+b^aN^b_{,a}] \right).
\end{equation}
Using (\ref{eq:dotGgeneral}) and (\ref{eq:G'general}) we find for the modified gauge generator (compare e.g. with \cite{Pons4}):
\begin{equation}
\label{eq:G'}
G'_{b,\vec{b}} = \frac{1}{\kappa}\left( C[b] + C_a[b^a] + \Pi[\dot{b}+b^aN_{,a}-N^ab_{,a}] + \Pi_a[\dot{b}^a+q^{ab}(bN_{,b}-Nb_{,b})-N^ab^b_{,a}+b^aN^b_{,a}] \right).
\end{equation}
In some situations we want to discuss diffeomorphisms in $\Sigma$ ($b=0$) and perpendicular to it ($b^a=0$) separately. Therefore it is useful to define the modified constraints $C'$ and $C'_a$:
\begin{align}
\label{eq:C'C'_a}
C'[b] &:= C[b] + \Pi[\dot{b}-N^ab_{,a}] + \Pi_a[q^{ab}(bN_{,b}-Nb_{,b})] \nonumber \\
C'_a[b^a] &:= C_a[b^a] +\Pi[b^aN_{,a}] + \Pi_a[\dot{b}^a-N^ab^b_{,a}+b^aN^b_{,a}].
\end{align}
Now we can analyze, how the elementary phase space variables transform under $G'_{b,\vec{b}}$. Introducing the notation $\delta_{G'}f:=\{f,G'\}$ a simple and straight forward computations lead to the following transformation behavior (compare e.g. with \cite{Pons4}):
\begin{align}
\label{eq:dG'metric}
\delta_{G'_{b,\vec{b}}}N &= b_{,t}-N^ab_{,a}+b^aN_{,a} \nonumber \\
\delta_{G'_{b,\vec{b}}}N^a &= q^{ab}(bN_{,b}-Nb_{,b})+b^a_{,t}+b^bN^a_{,b}-N^bb^a_{,b} \nonumber \\
\delta_{G'_{b,\vec{b}}}q_{ab} &= \left. \dot{q}_{ab} \right|_{N=b,N^a=b^a}
\end{align}
and
\begin{align}
\label{eq:dG'momenta}
\delta_{G'_{b,\vec{b}}}\Pi &= (\Pi_aq^{ab}b)_{,b}+\Pi_aq^{ab}b_{,b}+(\Pi b^a)_{,a} \nonumber \\
\delta_{G'_{b,\vec{b}}}\Pi_a &= \Pi b_{,a}+(\Pi_ab^b)_{,b}+\Pi_bb^b_{,a} \nonumber \\
\delta_{G'_{b,\vec{b}}}P^{ab} &= \left. \dot{P}^{ab} \right|_{N=b,N^a=b^a} + q^{c(a}q^{b)d}\Pi_c(bN_{,d}-Nb_{,d}).
\end{align}
It should be noted, that the constraint hypersurface $\Pi=\Pi_a=0$ is left invariant under $G'$, which is necessary, because otherwise $G'_{b,\vec{b}}$ would transform to a non-physical sector. Moreover, $\delta_{G'_{b,\vec{b}}}P^{ab}$ contains an additional term proportional to $\Pi_a$. But on the physical sector this term vanishes. When performing the inverse Legendre map $\mathcal{LM}^*$ this term vanishes anyway. Straight forward but quite lengthy calculations show that $\delta_{G'_{b,\vec{b}}}N$ and $\delta_{G'_{b,\vec{b}}}N^a$ do in fact coincide with $\delta_{G_L}N$, $\delta_{G_L}N^i$ on the tangent bundle in the adapted frame $X^\mu_a =\delta^\mu_a$ (see e.g. \cite{Dust2} section 4.4). 

Note, that the full diffeomorphism generator $G'_{b,\vec{b}}$ (\ref{eq:G'}) is not a general linear combination of the 8 constraints. The specific form comes from the requirement of matching diffeomorphisms in the phase space with diffeomorphisms in the Lagrangian formulation. One therefore does not choose 8 independent clocks, but only 4, as one only has 4 descriptor functions $b,\vec{b}$ in $G'_{b,\vec{b}}$. The extra terms in $G'_{b,\vec{b}}$ are proportional to the primary constraints $\Pi_\mu$. Hence, a gauge transformation acting on fields other than lapse and shift reduces to the gauge transformation generated by $G_{b,\vec{b}}$ in (\ref{eq:GundH}) and if the clocks $T^\mu$ also do not depend on lapse and shift, also $\mathcal{A}^\mu_\nu$ does not change. Therefore, as long as these two conditions are fulfilled, the formalism discussed in the previous section is still valid and can thus be applied. Nevertheless, one may want to construct observables with respect to lapse and shift or may use lapse and shift as clock fields.

As lapse and shift are dynamical in the extended ADM-phase space the equations in (\ref{eq:Nmustability}) become secondary gauge fixing constraints ensuring the stability of this gauge. We have
\begin{align}
G^{(2)\mu}(x) &:= \dot{G}^{\mu}(x)=[\dot{\tau}^\nu - \dot{T}^\mu](x) \nonumber \\
~ &= [\partial_t \tau^\nu - \partial_tT^\mu](x) - \kappa^{-1}\int_\Sigma \mathrm{d}^3y~ \mathcal{A}^\mu_\nu(x,y) N^\nu(y).
\end{align}
In the extended phase space the weak abelianization has to be performed for the secondary and the primary constraints. As a consequence the formula for the observables needs to be modified accordingly. In particular the descriptors and their time derivatives appearing in the diffeomorphism generator $G'_{b,\vec{b}}$ are replaced by $G^\mu$ and $G^{(2)\mu}$ respectively. In what follows we will assume that the clocks $T^\mu$ do \textbf{not} depend on lapse and shift as well as $\partial_tT^\mu$. If this is not the case a weak abelianization will in general be much more difficult. Let us define the following set of constraints:
\begin{align}
\mathfrak{G}^I &:= (G^\mu,G^{(2)\mu}) & \mathfrak{C}_I &:= (C_\mu,\Pi_\mu) ~~~I=1,\cdots,8.
\end{align}
As in the reduced case, we are interested in the matrix build from the Poisson brackets among the individual constraints:
\begin{equation}
\mathfrak{A}^I_J := -\{ \mathfrak{G}^I,\mathfrak{C}_J \} = - \left[ \begin{matrix}
\mathcal{A}^\mu_\nu & 0 \\
\{ \dot{T}^\mu,C_\nu \} & \mathcal{A}^\mu_\nu
\end{matrix} \right].
\end{equation}
Note, that we have used the identity $\dot{T}^\mu=\partial_tT^\mu + \int\mathrm{d}^3x~ \mathcal{A}^\mu_\nu(\cdot,x) N^\nu(x)$ for the last entry of the above matrix and as before we still use the definition  $\mathcal{A}^\mu_\nu = \{ T^\mu,C_\nu \}$. Let us denote the inverse of 
$\mathfrak{A}$ by $\mathfrak{B}$. As shown in \cite{Pons3} $\mathfrak{B}$ can be easily computed and has the following form:
\begin{equation}
\mathfrak{B}^I_J = (\mathfrak{A}^{-1})^I_J = \left[ \begin{matrix}
\mathcal{B}^\mu_\nu & 0 \\
S^\mu_\nu & \mathcal{B}^\mu_\nu
\end{matrix} \right],
\end{equation}
with
\begin{equation}
S^\mu_\nu(x,y) = -\int\mathrm{d}^3z\int\mathrm{d}^3v~ \mathcal{B}^\mu_\rho(x,z)\mathcal{B}^\sigma_\nu(v,y)\{ \dot{T}^\rho(z),C_\sigma(v) \}.
\end{equation}
The abelianized constraints can be constructed by using $\mathfrak{B}$, leading to:
\begin{equation}
\tilde{\mathfrak{C}}_I(x) = \int\mathrm{d}^3y~\mathfrak{B}^J_I(y,x)\mathfrak{C}_J(y)
\end{equation}
We define the equivalent abelian set of constraints by $\tilde{\mathfrak{C}}_I(x) =: (\tilde{\tilde{C}}_\mu,\tilde{\Pi}_\mu)$, which expressed in terms of the original primary and secondary constraints reads:
\begin{align}
\label{eq:AbelConstrExt}
\tilde{\Pi}_\mu(x) &= \int\mathrm{d}^3y~\mathcal{B}^\nu_\mu(y,x)\Pi_\nu(y) \nonumber \\
\tilde{\tilde{C}}_\mu(x) &= \int\mathrm{d}^3y~\mathcal{B}^\nu_\mu(y,x) \left[ C_\nu(y) - \int\mathrm{d}^3z\int\mathrm{d}^3v~ \mathcal{B}^\sigma_\rho(v,z)\{ \dot{T}^\rho(z),C_\nu(y) \}\Pi_\sigma(v) \right].
\end{align}
To present the final observable formula in the extended ADM-phase space we take advantage of a result proven in \cite{Pons3} that shows that the complicated form of $G'_{b,\vec{b}}$ in (\ref{eq:G'})
is equivalent to the simpler gauge generator in terms of $\tilde{\tilde{C}}_\mu$, $\tilde{\Pi}_\mu$ up to second order in the constraints, that is
\begin{equation}
G'_{b,\vec{b}} +\mathcal{O}(C^2) = \tilde{G}'_{\xi,\vec{\xi}} =: \kappa^{-1}\left( \tilde{\Pi}_\mu[\dot{\xi}^\mu] + \tilde{\tilde{C}}_\mu[\xi^\mu] \right), 
\end{equation}
with $\xi^\mu = \int\mathrm{d}^3x \mathcal{A}^\mu_\nu(\cdot,x)b^\nu(x)$. Consequently, the construction of observables can be based on the simpler gauge generator $\tilde{G}'_{\xi,\vec{\xi}}$ yielding:
\begin{equation}
\label{eq:obsGR1}
\mathcal{O}_{f,T}[\tau] \approx \left. \alpha_{ \tilde{G}'_{\xi,\vec{\xi}} }(f) \right|_{\xi^\mu=G^\mu,\dot{\xi}^\mu=G^{(2)\mu}}.
\end{equation}
Using the definitions of $\mathfrak{G}^I = (G^\mu,G^{(2)\mu})$ and $\tilde{\mathfrak{C}}_I(x) = (\tilde{\tilde{C}}_\mu,\tilde{\Pi}_\mu)$, we can use formula (\ref{eq:obsformula3}) also for the extended ADM-phase space and as before obtain a formal power series for the observables:
\begin{equation}
\label{eq:obsGR2}
\mathcal{O}_{f,T}[\tau] =f+ \sum\limits_{n=1}^{\infty} \frac{1}{n!} \int\mathrm{d}^3y_1...\int\mathrm{d}^3y_n~ \mathfrak{G}^{I_1}(y_1)...\mathfrak{G}^{I_n}(y_n) \{...\{ f,\tilde{\mathfrak{C}}_{I_1}(y_1) \},...\tilde{\mathfrak{C}}_{I_n}(y_n) \}.
\end{equation}
A difference to the former reduced case is that now the equations of motion additionally contain functional derivatives of the observables with respect to $\dot{\tau}^\mu$. However, such terms can again be written as an observable associated to a Poisson bracket and hence also be expressed at the gauge invariant level. We find:
\begin{align}
\label{eq:ObsExt}
\frac{\mathrm{d}}{\mathrm{d}t}\mathcal{O}_{f,T}[\tau] &\approx \mathcal{O}_{\partial_tf,T}[\tau] + \int \mathrm{d}^3y~ \eta'^\mu(y) \mathcal{O}_{ \{f,\tilde{\tilde{C}}_\mu(y)\} ,T}[\tau] + \int \mathrm{d}^3y~ \dot{\eta}'^\mu(y) \mathcal{O}_{ \{f,\tilde{\Pi}_\mu(y)\} ,T}[\tau],
\end{align}
with $\eta^\mu := \dot{\tau}^\mu$ and $\eta'^\mu = \eta^\mu - \partial_t T^\mu$ evaluated at $T^\mu = \tau^\mu$.
This concludes the section about Dirac observables in general relativity. It should be mentioned that, despite the fact that one can in principle construct observables for all ADM-variables at least locally with the formula in (\ref{eq:obsGR2}), it still remains to find appropriate clock fields $T^\mu$. Moreover, as the constraints in general relativity are non-linear functions of the phase space variables it will be difficult to find clocks such that the power series in (\ref{eq:obsGR2}) will be fully computable in practice. If this turns out to be too complicated we still have the possibility to work in a perturbative framework, i.e. one calculates observables at a 'region' close to the gauge constraint surface $\mathfrak{G}^I=0$. In this case one can neglect higher order terms of $\mathfrak{G}^I$ in the power series and truncate the series accordingly. As an example the first order approximation of an observable would be given by:
\begin{equation}
\mathcal{O}_{f,T}[\tau] \approx f + \int\mathrm{d}^3y~ \mathfrak{G}^I(y)\{ f,\tilde{\mathfrak{C}}_I(y) \}.
\end{equation}
In the next chapter we will discuss linear cosmological perturbation theory in the Hamiltonian formulation. This will be the starting point for our application of the relational formalism to cosmological perturbation theory. Likewise for the constraints, in this context we will assume that the gauge fixing constraints satisfied at zeroth order and thus $G^\mu$ counts as a first order perturbation.

Finally, let us remark that even if the observables constructed from formula (\ref{eq:ObsExt}) cannot be written down in closed form, as long as we are able to compute their associated Poisson algebra, we can derive the corresponding reduced phase. Whether a quantization of the reduced theory is possible crucially depends on how complicated the observable algebra actually is, see e.g. \cite{Thiemann2} for a discussion. For an application to general relativity using dust clock fields see e.g. \cite{Dust1} and \cite{Dust2} and for a reduced phase space quantization using dust fields see \cite{Giesel:2007wn}.

\newpage

\section{Linear cosmological perturbation theory in Hamiltonian formulation}
\label{se:cosmpert}

In this section we understand the cosmological perturbation theory in the framework of Dirac observables and clocks. 
The conventional approach to study cosmological perturbations is as follows. One starts with perturbing the geometric and matter degrees of 
freedom around a homogeneous and isotropic background.  The perturbations which are generally assumed to be small fluctuations over the 
homogeneous background are decomposed in to scalar, vector and tensor perturbations. The latter are 
associated with gravitational waves, and the vector perturbations tend to die out in an expanding universe \cite{bardeen}. It is the dynamics of scalar perturbations, which are associated with 
energy density and curvature perturbations, that 
capture the structure formation in the universe. Though perturbing general relativity to linear order seems to be a straight forward task, care must be exercised in interpreting the physical content of 
the perturbations. Since perturbations are a priori not gauge invariant quantities, it is common in 
cosmological perturbation theory to construct perturbations which are invariant under diffeomorphisms, at least up to the corresponding order in 
perturbation theory. Another possibility is to fix a specific gauge and work with the gauge fixed quantities. Several common choices of gauges are
discussed in the literature, such as the synchronous gauge, the spatially flat gauge, the longitudinal gauge and the comoving gauge. A detailed discussion of these gauges,  along with some other common gauges, with respect to the clock fields is 
performed in a companion work \cite{Giesel:2018opa}.

Let us recall that conventionally cosmological perturbation theory has been mostly discussed in the Lagrangian picture of general relativity. This means one either perturbs 
the Einstein field-equations up to first order, or the Einstein-Hilbert action up to second order, and derives the equations of motion of the perturbations thereof. 
Alternatively one may discuss cosmological perturbations in a Hamiltonian framework. This has been done e.g. by Halliwell and Hawking \cite{Hawking}
and  by Langlois \cite{Langlois}. We will mostly refer to the latter approach as it is more closely related to the usual formalism of linear 
cosmological perturbation theory, including a scalar-vector-tensor decomposition. However, in Langlois' seminal work lapse and shift perturbations are not taken into account 
 due to which his analysis can not be compared in full generality with the Lagrangian picture. Only for the gauge invariant variables which do not involve lapse and shift perturbations, such a correspondence can be studied. An example is the 
 'Mukhanov-Sasaki-variable' $v$ in \cite{Mukhanov2} which is derived independently in the Hamiltonian framework as the variable $Q$ in \cite{Langlois}. It is important to note that the 
 correspondence no longer holds in `Langlois' treatment if gauge invariant variables involving lapse and shift perturbations are studied. In particular, if one wants to compare the results between the Lagrangian and Hamiltonian treatments 
 for the perturbed gauge invariant equations of 
motion in terms of the Bardeen potentials $\Psi$ and $\Phi$ \cite{Mukhanov} one must take lapse and shift perturbations into account by generalizing to the extended phase space discussed earlier. 

In this section we will build up cosmological perturbation theory around a flat ($k=0$) FLRW cosmology background by perturbing canonical general relativity in ADM 
variables with a scalar field $\varphi$ minimally coupled to gravity. We will start this section with reviewing the equations of motion for general relativity and a minimally coupled scalar field. Given this we 
are going to consider linearized perturbations around a generic relativistic background and derive the equations of motion of the perturbations on the extended ADM phase space. As an application of this we choose the FLRW k=0 solutions as the background and specialize the equations of motion to this case. A scalar-vector-tensor decomposition of the perturbations is performed. Notations are kept as
closely to standard cosmological perturbation theory (see \cite{Mukhanov}) as possible. The transformation behavior of these perturbations, including lapse and shift 
perturbations, can be derived using the full generator of diffeomorphisms $G'_{b,\vec{b}}$ presented in (\ref{eq:G'}). With this knowledge gauge invariant quantities can be constructed, which 
correspond to the Bardeen potentials. Finally, the gauge invariant equations of motion are derived and compared to the results of Mukhanov, Feldman and Brandenberger \cite{Mukhanov}, see appendix \ref{append:GIEOM} for more details.
 
\subsection{General relativity with a minimally coupled scalar field}

This section aims to derive the equations of motion for gravity plus a minimally coupled scalar field. Such a matter content is suitable for instance for simple (single field) inflation models (see e.g. \cite{bardeen} Ch. 3). In this case the total action consists of the Einstein-Hilbert action plus the scalar field action that has the following form
\begin{equation}
\label{eq:Sphi}
S_\varphi=\frac{1}{2\lambda_\varphi}\int\limits_{M}\mathrm{d}^4X \sqrt{|\det g|} \left( -g^{\mu\nu} \varphi_{,\mu}\varphi_{,\nu}-V(\varphi), \right).
\end{equation}
where $\lambda_\varphi$ is the coupling constant and $V(\varphi)$ denotes the potential associated with the scalar field $\varphi$. As usual in order to derive at the corresponding Hamiltonian formulation of the theory, one performs a 3+1 split of the action and expresses it in terms of ADM variables, see for instance \cite{Thiemann} for a pedagogical introduction to this formalism. The canonical conjugate momenta of the scalar field configuration variables $\varphi$ are given by:
\begin{equation}
\mathcal{LM}^*\pi_\varphi := \lambda_\varphi\frac{\delta S}{\delta \dot{\varphi}}=\sqrt{{\rm det}(q)}\varphi_n,
\end{equation}
where $\varphi_n$ denotes the partial derivative projected along the timelike vector field $n$ orthogonal to $\Sigma$, that is $\varphi_n=n^\mu\varphi_{,\mu}$ and $\varphi,\pi_\varphi$ satisfy the usual Poisson bracket algebra, that is $\{\varphi(x),\pi_\varphi(y)\} = \delta(x,y)$, whereas all remaining Poisson brackets vanish. Similarly, for the ADM momenta $P^{ab}$ we obtain:
\begin{equation}
\label{eq:PabinKab}
\mathcal{LM}^* P^{ab} := \kappa\frac{\delta S}{\delta \dot{q}_{ab}} = \sqrt{\det q} (K^{ab}-q^{ab}K),
\end{equation}
where we wrote $P^{ab}$ in terms of the extrinsic curvature $K_{ab}$, that involves the velocity of the ADM metric,  (see e.g. \cite{Thiemann}):
\begin{equation}
\label{eq:Kab}
K_{ab} = \frac{1}{2N}\left( \dot{q}_{ab} - \left(\mathcal{L}_{\vec{N}}q\right)_{ab} \right).
\end{equation}
While the spatial curvature $R^{(3)}$ describes the intrinsic curvature on $\Sigma$ the \emph{extrinsic curvature} tensor $K_{ab}$ describes the local bending of the stacked spatial slices amongst each other at different spacetime points. This is precisely the information of how the normal vector $n^\mu$ of the spatial slices changes in spacetime. Note, that for the reason that the the total action does not involve velocities associated with lapse and shift their momenta $\Pi,\Pi_a$ become the already mentioned primary constraints.

Now we perform a Legendre transformation for all variables but lapse and shift and express the total action on phase space. For this purpose we need to solve the equation involving the momenta for the associated velocities. In the case of the scalar field this is straight forward. For the ADM-momentum $P^{ab}$ this gets simplified if we introduce the so-called \emph{supermetric} and its inverse:
\begin{align}
Q_{abcd} &:= q_{ac}q_{bd}-\tfrac{1}{2}q_{ab}q_{cd} & (Q^{-1})^{abcd} &:= q^{ac}q^{bd}-q^{ab}q^{cd}.
\end{align}
The second tensor is the inverse of the first one in the following sense:
\begin{equation}
Q_{abcd} (Q^{-1})^{cdef} = \delta_a^e\delta_b^f = (Q^{-1})^{efcd}Q_{cdab}.
\end{equation}
Furthermore, it possesses the following symmetries: $Q_{abcd} = Q_{cdab} = Q_{badc}$ and similarly for $Q^{-1}$. Using the supermetric and its inverse one can easily solve (\ref{eq:PabinKab}) for the extrinsic curvature:
\begin{equation}
\mathcal{LM}~K_{ab} = \frac{1}{\sqrt{\det q}} Q_{abcd}P^{cd} = \frac{1}{\sqrt{\det q}}\left( P_{ab} - \tfrac{1}{2}q_{ab}P \right)
\end{equation}
and given (\ref{eq:Kab}) also for the velocity $\dot{q}_{ab}$.

The standard constraint stability analysis shows that the total Hamiltonian and spatial diffeomorphism constraint, now including the standard geometry as well as a scalar field contribution, are secondary constraints. These ensure the stability of the primary constraints. The total constraints are of the form  $C=C_{\rm geo}+C_\varphi$ and  $C_a=C_{a,\rm{geo}}+C_{a,\varphi}$, where we set $C_{\rm{mat}}=C_{\varphi}$ and likewise for $C_a$ because the scalar field is the only matter content. The explicit forms of the $C_\varphi$ and $C_{a,\varphi}$ are given below:
\begin{equation}
\label{eq:Cphi}
C_\varphi= \frac{\kappa}{2} \left( \frac{\lambda_\varphi}{\sqrt{\det q}} \pi_\varphi^2+\frac{\sqrt{\det q}}{\lambda_\varphi} \left( q^{ab}\varphi_{,a}\varphi_{,b}+V(\varphi) \right) \right),
\end{equation}
\begin{equation}
\label{eq:Caphi}
C_{a,\varphi}=\kappa\pi_\varphi \varphi_{,a}.
\end{equation}
As mentioned before the ADM-Hamiltonian consists of a linear combination of constraints only and it can be directly read off from the ADM action on phase space. The Hamiltonian equations of motion can be derived by calculating the Poisson brackets of the elementary phase space variables with the ADM-Hamiltonian. We introduce the following notation for the time derivative:
\begin{equation}
\frac{\mathrm{d}}{\mathrm{d}t}f=\dot{f}:=\frac{\partial f}{\partial t}+\{f,H\}.
\end{equation}
The time derivative of the ADM-metric $\dot{q}_{ab}$ is relatively simple to calculate, however for $\dot{P}^{ab}$ one needs to calculate the variation $\frac{\delta R^{(3)}}{\delta q_{ab}}$. For this purpose one rewrites the explicit expression of $R^{(3)}$ in terms of Christoffel symbols and considers the variation of the latter. A very similar calculation has to be performed when we discuss linear perturbations of the equations of motion. Therefore, for more details on perturbations of the curvature scalar we refer the reader to chapter (\ref{sususe:curpert}). 
Finally, the following Hamiltonian equations of motion can be derived:
\begin{equation}
\label{eq:dotq}
\dot{q}_{ab}=\frac{2N}{\sqrt{\det q}} Q_{abcd}P^{cd}+(\mathcal{L}_{\vec{N}}q)_{ab}.
\end{equation}
\begin{align}
\label{eq:dotP}
\dot{P}^{ab} &= N\left[ \frac{1}{2}Cq^{ab}-\sqrt{\det q} (Q^{-1})^{abcd}R^{(3)}_{cd}-\frac{2}{\sqrt{\det q}} (P^{ac}P^{bd}-\frac{1}{2}P^{ab}P^{cd})q_{cd} \right] \nonumber \\
~ &~ + \sqrt{\det q} (D^aD^b N-q^{ab}D_cD^c N)+(\mathcal{L}_{\vec{N}}P)^{ab} \nonumber \\
~ &~ +\frac{\kappa N}{2} \frac{\sqrt{\det q}}{\lambda_\varphi} \left[ (Q^{-1})^{abcd}\varphi_{,c}\varphi_{,d} - V(\varphi)q^{ab} \right].
\end{align}
\begin{align}
\dot{N}=\lambda ~&,~~ \dot{N}^a=\lambda^a \nonumber \\
\dot{\Pi}=-C ~&,~~ \dot{\Pi}_a= -C_a
\end{align}
and for the scalar field:
\begin{align}
\label{eq:dotphidotpiphi}
\dot{\varphi} &= \frac{\lambda_\varphi}{\sqrt{\det q}}N\pi_\varphi + N^a\varphi_{,a} \nonumber \\
\dot{\pi}_\varphi &= \left( \frac{\sqrt{\det q}}{\lambda_\varphi}Nq^{ab}\varphi_{,a}\right)_{,b} - \frac{1}{2}\frac{\sqrt{\det q}}{\lambda_\varphi}N\frac{\mathrm{d}V}{\mathrm{d}\varphi}(\varphi) + (N^a\pi_\varphi)_{,a}.
\end{align}
These equations together with the constraints $C=0,C_a=0,\Pi=0,\Pi_a=0$ are equivalent to the Einstein equations in the Lagrangian framework.  
As in the Lagrangian picture of general relativity, the equations of motion are highly nonlinear. Therefore, for most cases approximations and assumptions have to be made. Later on in this work we will perturb the equations of motion to first order around a flat FLRW cosmology background. As a preparation for this we present perturbation theory in general relativistic context in the next section.

\subsection{General relativistic perturbation theory}

In this section we review the formulation of general relativistic perturbation theory, that is we consider perturbations around some specific solution of Einstein's equations, but keep this solution generic so far. This known solution is referred to as \emph{background solution}.
We want to derive the equations of motion for linear perturbations for an arbitrary background and specialize those to a FLRW background in the next section. We denote the solutions of the ADM-phase space variables for a generic background with a bar, i.e.
\begin{align}
(\bar{q}_{ab},\bar{P}^{ab}) &&& (\bar{N},\bar{\Pi}) &&& (\bar{N}^a,\bar{\Pi_a}) &&& (\bar{\varphi},\bar{\pi}_\varphi)
\end{align}
are the \emph{background quantities}. Their dynamics is already fixed by the fact that they satisfy Einstein's equations. The Lagrange multipliers $\bar{\lambda}^\mu$ are assumed to be fixed as well and the background constraints vanish identically. For perturbation theory, we consider small deviations from this background solution. In this case all phase space variables considered will consist of the respective fixed background contribution plus a yet undetermined perturbation part which is denoted with a leading `$\delta$' symbol. In the case of the extended ADM-phase space we have:
\begin{align}
q_{ab} &= \bar{q}_{ab} + \delta q_{ab} & P^{ab} &= \bar{P}^{ab} + \delta P^{ab} &
N &= \bar{N} + \delta N & N^a &= \bar{N}^a + \delta N^a \nonumber \\
\varphi &= \bar{\varphi} + \delta \varphi & \pi_\varphi &= \bar{\pi}_\varphi+\delta \pi_\varphi & \Pi &= \bar{\Pi} + \delta \Pi & \Pi_a &= \bar{\Pi}_a+\delta \Pi_a.
\end{align}
Each quantity with a $\delta$ is assumed to be small compared to the background quantities. We can expand phase space functions with respect to these small perturbations as:
\begin{equation}
F(\Phi) = \sum\limits_{n=0}^{\infty} F^{(n)}(\bar{\Phi},\delta \Phi),
\end{equation}
with $\Phi=(q_{ab},P^{ab},N,N^a,\Pi,\Pi_a,\varphi,\pi_\varphi)$ and
\begin{equation}
\label{eq:F(n)}
F^{(n)}(\bar{\Phi},\delta \Phi)(x) = \frac{1}{n!} \int_\Sigma\mathrm{d}^3y_1...\int_\Sigma\mathrm{d}^3y_n \frac{\delta^nF}{\delta \Phi_{I_1}(y_1)...\delta \Phi_{I_n}(y_n)}(\bar{\Phi})(x)\delta \Phi_{I_1}(y_1)...\delta \Phi_{I_n}(y_n).
\end{equation}
Here we perform a summation over all indices $I_1$,..., $I_n$, which are representatives for the different elementary phase space variables $q_{ab}$, $P^{ab}$ etc. We will often use the simplified notation:
\begin{equation}
F^{(n)} \equiv \frac{1}{n!}\delta^nF,
\end{equation}
with the appropriate definition of $\delta^nF$.

\subsubsection{Equations of motion for linear perturbations}
\label{sususe:curpert}
This section involves a brief review of the equations of motion of linear perturbations around a general background solution of general relativity. Two main ingredients that enter these equations are the linearized perturbations of the scalar curvature $\delta R$ and the constraints $\delta C, \delta C_a$ respectively, that will be treated separately and first in our discussion.  

We start with the linearized perturbations of the scalar curvature $\delta R^{(3)}$. The superscript $^{(3)}$ will be suppressed in this section for simplicity. As the curvature scalar is the Ricci tensor contracted with the inverse metric its perturbation is given by:
\begin{equation}
\delta R = -\bar{q}^{ac}\bar{q}^{bd}\delta q_{cd}\bar{R}_{ab} + \bar{q}^{ab}\delta R_{ab}.
\end{equation}
Considering that we can write the Ricci tensor in terms of Christoffel symbols we obtain:
\begin{align}
R_{abc}^{~~~d} &= \Gamma^d_{~ac,b} - \Gamma^d_{~bc,a} + \Gamma^e_{~ac}\Gamma^d_{~be} - \Gamma^e_{~bc}\Gamma^d_{~ae} \nonumber \\
R_{ab} = R_{acb}^{~~~c} &= \Gamma^c_{~ab,c} - \Gamma^c_{~cb,a} + \Gamma^d_{~ab}\Gamma^c_{~cd} - \Gamma^d_{~bc}\Gamma^c_{~ad}.
\end{align}
Since the `$\delta$'-operation satisfies the Leibniz rule, the perturbation of the Ricci tensor can be written as:
\begin{align}
\label{deltaRab}
\delta R_{ab} &= \delta \Gamma^c_{~ab,c} - \delta \Gamma^c_{~cb,a} + \delta \Gamma^d_{~ab}\bar{\Gamma}^c_{~cd} + \bar{\Gamma}^d_{~ab}\delta \Gamma^c_{~cd} - \delta \Gamma^d_{~bc}\bar{\Gamma}^c_{~ad} - \bar{\Gamma}^d_{~bc}\delta \Gamma^c_{~ad}.
\end{align}
Next, we will show that $\delta \Gamma^c_{~ab}$ is a tensor on the background manifold $\bar{\Sigma}$. We denote the covariant derivative of a generic tensor $T$ on $\bar{\Sigma}$  by $T_{|a} := \bar{D}_aT$. The expression for $\delta R_{ab}$ can be simplified by rewriting the covariant derivative in terms of the Christoffel symbols with respect to the background metric. This yields:
\begin{equation}
\delta R_{ab} = \delta \Gamma^c_{~ab|c} - \delta \Gamma^c_{~cb|a}.
\end{equation}
The perturbation of the Christoffel symbols $\delta \Gamma^c_{~ab}$ can be calculated explicitly and one arrives at the following result:
\begin{equation}
\delta \Gamma^c_{~ab} = \frac{1}{2}\bar{q}^{cd}( \delta q_{ad|b} + \delta q_{bd|a} - \delta q_{ab|d} ).
\end{equation}
This shows that $\delta \Gamma$ is indeed a tensor on $\bar{\Sigma}$. This allows to express $\delta R_{ab}$ in terms of $\delta q_{ab}$:
\begin{equation}
\label{eq:delR}
\delta R_{ab} = \frac{1}{2}\bar{q}^{cd}( \delta q_{ad|bc} + \delta q_{bd|ac} - \delta q_{ab|dc} - \delta q_{cd|ba} ).
\end{equation}
Now, the linear perturbations of the scalar curvature $\delta R$ can be obtained from contracting $\delta R_{ab}$ with the inverse background metric $\bar{q}^{ab}$ leading to
\begin{equation}
\delta R = \bar{q}^{ab}\delta R_{ab}=( \bar{D}^a\bar{D}^b - \bar{q}^{ab}\bar{\boldsymbol{\Delta}} -\bar{R}^{ab} )\delta q_{ab},
\end{equation}
with $\bar{\boldsymbol{\Delta}} := \bar{q}^{ab}\bar{D}_a\bar{D}_b$. Note, that we raise and lower all indices with respect to $\bar{q}_{ab}$. 
Note, that one treats the perturbations as tensors on the background manifold $(\bar{\Sigma},\bar{q}_{ab})$. Covariant derivatives thereof are with respect to $\bar{q}_{ab}$ as well as other operations like the trace of a tensor. Care has to be taken, when comparing this to the full theory, because the perturbation of a tensor $T_{|a}$on $\Sigma$ is not equal to $\delta T_{|a}$ on the background space, as $D_a \neq \bar{D}_a$ and analogously for the trace operation $\text{Tr}$.

Next, we discuss the linear perturbations of the constraints.  The dynamics of the background theory are assumed to be already fixed. Hence, the specific combinations of background quantities in $\bar{C}$ and $\bar{C}_a$ vanish identically and we have:
\begin{equation}
\bar{C}=0,\quad\bar{C}_a=0,\quad \overline{\Pi}=0,\quad \overline{\Pi}_a=0
\end{equation}
The perturbations $\delta\Pi,\delta\Pi_a$ are independent, whereas $\delta C=\delta C_{\rm geo}+\delta C_{\varphi}$ and $\delta C_a=\delta C_{a\rm{geo}}+\delta C_{a,\varphi}$ are functions of the elementary phase space variables and can be expressed in terms of their perturbations. At the linear order we get:
\begin{align}
\label{eq:deltaCgeogeneral}
\delta C_\text{geo} &= \left[-\frac{1}{2} \bar{C}_\text{geo}\bar{q}^{ab}+\sqrt{\det\bar{q}} (\bar{Q}^{-1})^{abcd} \bar{R}^{(3)}_{cd}+\frac{2}{\sqrt{\det\bar{q}}}(\bar{P}^{ac}\bar{P}^{bd}-\frac{1}{2}\bar{P}^{ab}\bar{P}^{cd})\bar{q}_{cd} \right]\delta q_{ab} \nonumber \\
~ &+ \frac{2}{\sqrt{\det\bar{q}}}\bar{Q}_{abcd}\bar{P}^{cd}\delta P^{ab} - \sqrt{\det\bar{q}} \bar{q}^{ab}\delta R^{(3)}_{ab}.
\end{align}
\begin{align}
\label{eq:deltaCphigeneral}
\delta C_\varphi &= \left[-\frac{1}{2} \bar{C}_\varphi\bar{q}^{ab}-\frac{\kappa}{2}\frac{\sqrt{\det\bar{q}}}{\lambda_\varphi}\left( (\bar{Q}^{-1})^{abcd}\bar{\varphi}_{,c}\bar{\varphi}_{,d}-V(\bar{\varphi})\bar{q}^{ab} \right) \right]\delta q_{ab} \nonumber \\
~ &+ \frac{\lambda_\varphi \kappa}{\sqrt{\det\bar{q}}}\bar{\pi}_\varphi \delta \pi_\varphi + \frac{\kappa\sqrt{\det\bar{q}}}{\lambda_\varphi} \left(\bar{q}^{ab}\bar{\varphi}_{,a}\delta \varphi_{,b}+\frac{1}{2}\frac{\mathrm{d}V}{\mathrm{d}\varphi}(\bar{\varphi})\delta \varphi \right).
\end{align}
\begin{equation}
\label{eq:deltaCageogeneral}
\delta C_{a,\text{geo}}= -2 (\bar{P}^{bc}_{|c}) \delta q_{ab} -2\bar{q}_{ab}(\delta P^{bc}_{|c} + \bar{P}^{cd}\delta \Gamma^b_{cd}).
\end{equation}
\begin{equation}
\label{eq:deltaCaphigeneral}
\delta C_{a,\varphi}=\kappa(\bar{\varphi}_{,a}\delta \pi_\varphi + \bar{\pi}_\varphi \delta \varphi_{,a}).
\end{equation}
Similar to the full theory, the stability requirement for the primary constraints $\delta\Pi$ and $\delta\Pi_a$ yields the linearized secondary constraints $\delta C$ and $\delta C_a$. Note, that for simplicity, we displayed only the linear order here. Taking higher orders of $\delta$ into account the actual constraint stabilization yields the secondary constraints:
\begin{equation}
C = \sum\limits_{n=1}^{\infty} C^{(n)} = \sum\limits_{n=1}^{\infty} \frac{1}{n!}\delta^nC
\end{equation}
and analogously for $C_a$.

As explained in appendix \ref{appendPoissonPert} the perturbed equations of motion are generated by the second order perturbation of the Hamiltonian, that has in our case the following form: 
\begin{equation}
H^{(2)} = \frac{1}{\kappa} \int_\Sigma \mathrm{d}^3x~\left[ \bar{N} \tfrac{1}{2}\delta^2C + \bar{N}^a\tfrac{1}{2}\delta^2C_a + \delta N \delta C + \delta N^a \delta C_a + \delta \lambda \delta \Pi + \delta \lambda^a\delta \Pi_a \right](x),
\end{equation}
where more details can be found in appendix \ref{appendA}. Note, that we allowed a perturbation of the Lagrange multipliers $\lambda,\lambda^a$. In order to compute linearized equations of motion two equivalent option exist. Either these equations can be derived by directly perturbing the full equations of motion in (\ref{eq:dotq}) and (\ref{eq:dotP}) or by calculating $H^{(2)}$ and derive the EOMs from the associated Hamiltonian equations shown in the appendix in (\ref{eq:deltadotPhi}). The reader is referred to appendix \ref{appendA} for more detailed calculations using the latter approach. At this place we will just display the final results for the Hamiltonian form of the equations of motion. In case of the ADM-metric we obtain:
\begin{align}
\label{eq:deltadotq}
\delta \dot{q}_{ab} &= \frac{2}{\sqrt{\det\bar{q}}}(\bar{P}_{ab}-\frac{1}{2}\bar{q}_{ab}\bar{P}) \delta N \nonumber \\
~ &+ \bar{N}\left[ -\frac{1}{\sqrt{\det\bar{q}}}\bar{Q}_{abef} \bar{P}^{ef}\bar{q}^{cd}+\frac{1}{\sqrt{\det\bar{q}}}(4\delta^c_a\bar{P}^d_b-\delta^c_a\delta^d_b\bar{P}-\bar{q}_{ab}\bar{P}^{cd}) \right] \delta q_{cd} \nonumber \\
~ &+ \bar{N} \frac{2}{\sqrt{\det\bar{q}}}\bar{Q}_{abcd}\delta P^{cd} + 2 \delta N_{(a|b)} +(\mathcal{L}_{\vec{N}}\delta q)_{ab}
\end{align}
For the ADM-momentum we split the resulting equation into two parts in order to present it in a convenient form:
\begin{align}
\label{eq:deltadotPgeo}
\left. \delta \dot{P}^{ab} \right|_\text{geo} &=  \left[ \frac{1}{2} \bar{C}_\text{geo} \bar{q}^{ab} - \sqrt{\det\bar{q}}(\bar{Q}^{-1})^{abcd}\bar{R}^{(3)}_{cd}-\frac{2}{\sqrt{\det\bar{q}}}(\bar{P}^{ac}\bar{P}^b_c-\frac{1}{2}\bar{P}^{ab}\bar{P}) \right]\delta N \nonumber \\
~ &+\sqrt{\det\bar{q}}(\bar{Q}^{-1})^{abcd}\delta N_{|cd} +\bar{N}\left[-\frac{1}{4} \bar{C}_\text{geo} (\bar{q}^{ab}\bar{q}^{cd}+2\bar{q}^{ac}\bar{q}^{bd}) \right. \nonumber \\
~ &- \left. \sqrt{\det\bar{q}} \left( \frac{1}{2}\bar{q}^{ab}\bar{q}^{ce}\bar{q}^{df}+\frac{1}{2}\bar{q}^{ae}\bar{q}^{bf}\bar{q}^{cd}+\bar{q}^{ac}\bar{q}^{bd}\bar{q}^{ef}-2\bar{q}^{c(a}\bar{q}^{b)e}\bar{q}^{df} \right) \bar{R}^{(3)}_{ef} \right. \nonumber \\
~ &+ \left. \frac{2}{\sqrt{\det\bar{q}}} \left( -\bar{P}^{ac}\bar{P}^{bd}+\frac{1}{2}\bar{P}^{ab}\bar{P}^{cd} + \frac{1}{2}( \bar{P}^a_e\bar{P}^{be} - \tfrac{1}{2}\bar{P}\bar{P}^{ab} )\bar{q}^{cd} + \frac{1}{2}( \bar{P}^c_e\bar{P}^{de} - \tfrac{1}{2}\bar{P}\bar{P}^{cd} )\bar{q}^{ab} \right) \right] \delta q_{cd} \nonumber \\
~ &+\frac{\bar{N}}{\sqrt{\det\bar{q}}}\left[ \bar{Q}_{cdef}\bar{P}^{ef}\bar{q}^{ab}-\left( 4\delta^{(a}_c\bar{P}^{b)}_d-\delta^a_c\delta^b_d\bar{P}-\bar{q}_{cd}\bar{P}^{ab} \right) \right] \delta P^{cd} \nonumber \\
~ &-\sqrt{\det\bar{q}} \bar{N} (\bar{q}^{ac}\bar{q}^{bd}-\frac{1}{2}\bar{q}^{ab}\bar{q}^{cd}) \delta R^{(3)}_{cd} \nonumber \\
~ &+ \sqrt{\det \bar{q}} \left( \bar{q}^{ab}\bar{q}^{ce}\bar{q}^{df} + \bar{q}^{ac}\bar{q}^{bd}\bar{q}^{ef} - 2\bar{q}^{c(a}\bar{q}^{b)e}\bar{q}^{df} + \frac{1}{2}(\bar{Q}^{-1})^{abef}\bar{q}^{cd} \right) \bar{N}_{|ef}\delta q_{cd} \nonumber \\
~ &- \sqrt{\det \bar{q}} (\bar{Q}^{-1})^{abcd}\delta \Gamma^e_{~cd}\bar{N}_{|e} +\left( \mathcal{L}_{\delta \vec{N}}\bar{P} \right)^{ab} + (\mathcal{L}_{\vec{N}}\delta P)^{ab}.
\end{align}
\begin{align}
\label{eq:deltadotPphi}
\left. \delta \dot{P}^{ab} \right|_\varphi &= \left[\frac{1}{2} \bar{C}_\varphi\bar{q}^{ab}+\frac{\kappa}{2}\frac{\sqrt{\det\bar{q}}}{\lambda_\varphi}\left( (\bar{Q}^{-1})^{abcd}\bar{\varphi}_{,c}\bar{\varphi}_{,d}-V(\bar{\varphi})\bar{q}^{ab} \right) \right]\delta N \nonumber \\ ~ &+ \bar{N}\left[-\frac{1}{4}\bar{C}_\varphi(\bar{q}^{ab}\bar{q}^{cd}+2\bar{q}^{ac}\bar{q}^{bd}) + \frac{\kappa}{2}\frac{\sqrt{\det\bar{q}}}{\lambda_\varphi} \left( \bar{q}^{ac}\bar{q}^{bd}V(\bar{\varphi}) \right. \right. \nonumber \\
~ &+ \left. \left. \left( \frac{1}{2}\bar{q}^{ab}\bar{q}^{ce}\bar{q}^{df}+\frac{1}{2}\bar{q}^{ae}\bar{q}^{bf}\bar{q}^{cd}+\bar{q}^{ac}\bar{q}^{bd}\bar{q}^{ef}-2\bar{q}^{c(a}\bar{q}^{b)e}\bar{q}^{df} \right) \bar{\varphi}_{,e}\bar{\varphi}_{,f} \right) \right] \delta q_{cd} \nonumber \\
~ &+\bar{N}\frac{\kappa}{2}\left[ \left( \frac{\lambda_\varphi}{\sqrt{\det \bar{q}}} \bar{\pi}_\varphi \delta \pi_\varphi - \frac{1}{2}\frac{\sqrt{\det \bar{q}}}{\lambda_\varphi}\frac{\mathrm{d}V}{\mathrm{d}\varphi}(\bar{\varphi})\delta \varphi \right)\bar{q}^{ab} + \frac{\sqrt{\det \bar{q}}}{\lambda_\varphi}\left( 2\bar{q}^{c(a}\bar{q}^{b)d} - \bar{q}^{ab}\bar{q}^{cd} \right)\bar{\varphi}_{,c}\delta \varphi_{,d} \right],
\end{align}
with $\delta \dot{P}^{ab} = \left. \delta \dot{P}^{ab} \right|_\text{geo} + \left. \delta \dot{P}^{ab} \right|_\varphi$. \\

As can be seen, especially the expression for $\delta \dot{P}^{ab}$ is quite complicated. We will see, however, that things simplify considerably for the case of a FLRW $k=0$ cosmology background. The equations of motion for lapse and shift perturbations are simply:
\begin{equation}
\delta \dot{N} = \delta \lambda~,~~\delta \dot{N}^a=\delta \lambda^a.
\end{equation}
\begin{equation}
\delta \dot{\Pi} = -\delta C~,~~\delta \dot{\Pi}_a = -\delta C_a.
\end{equation}
For the scalar field perturbations the following equations of motion can be derived:
\begin{align}
\label{eq:deltadotphi}
\delta \dot{\varphi} &= \delta N \frac{\lambda_\varphi}{\sqrt{\det \bar{q}}}\bar{\pi}_\varphi + \bar{N}\frac{\lambda_\varphi}{\sqrt{\det \bar{q}}}\left( \delta \pi_\varphi - \frac{1}{2}\bar{\pi}_\varphi \bar{q}^{ab}\delta q_{ab} \right) + \delta N^a\bar{\varphi}_{,a} + \bar{N}^a\delta \varphi_{,a}.
\end{align}
\begin{align}
\label{eq:deltadotpiphi}
\delta \dot{\pi}_\varphi &= \left[ \bar{N}\frac{\sqrt{\det \bar{q}}}{\lambda_\varphi} \left( \frac{\delta N}{\bar{N}} \bar{q}^{ab}\bar{\varphi}_{,b} - (\bar{q}^{ac}\bar{q}^{bd}-\tfrac{1}{2}\bar{q}^{ab}\bar{q}^{cd})\bar{\varphi}_{,b}\delta q_{cd} + \bar{q}^{ab}\delta \varphi_{,b} \right) \right]_{,a} \nonumber \\
~ &- \bar{N} \frac{\sqrt{\det \bar{q}}}{\lambda_\varphi} \left( \frac{\delta N}{\bar{N}} \frac{1}{2}\frac{\mathrm{d}V}{\mathrm{d}\varphi}(\bar{\varphi})  + \frac{1}{4}\frac{\mathrm{d}V}{\mathrm{d}\varphi}(\bar{\varphi})\bar{q}^{ab}\delta q_{ab} + \frac{1}{2}\frac{\mathrm{d}^2V}{\mathrm{d}\varphi^2}(\bar{\varphi})\delta \varphi \right)  \nonumber \\
~ & + \left( \delta N^a\bar{\pi}_\varphi + \bar{N}^a\delta \pi_\varphi \right)_{,a}.
\end{align}

\subsection{Cosmological perturbation theory in extended ADM-phase space}
In this section we apply the results of the previous section to a specific solution of Einstein's equations, namely we will consider a Friedman-Lemaître-Robertson-Walker (FLRW) spacetime with $k=0$. Note, that the $k=0$ case is in agreement with current observations of the CMB anisotropies (see e.g. \cite{Planck}), so that we do not consider $k=\pm 1$ here. For $k=0$ models all background quantities are independent of the spatial coordinates and hence do only depend on temporal coordinates. 

\subsubsection{FLRW Background solution}

Before specializing the dynamical equations for the linearized perturbations to a FLRW background, we briefly discuss the form and properties of the FLRW solution. In the flat ($k=0$) case the spacetime metric has the following form:
\begin{equation}
\mathrm{d}s^2 = -N^2(t)\mathrm{d}t^2 + a^2(t)\delta_{ab}\mathrm{d}x^a\mathrm{d}x^b~.
\end{equation}
Note, that we have chosen an adaptive frame $X^\mu_a = \delta^\mu_a$, where $X^\mu_{,a}$ are tangential vector fields constructed from the embedding $X^\mu$ of $\Sigma$ into $M$. The choice of the lapse function $N(t)$ is arbitrary and a change thereof corresponds to time-reparametrizations. There are two commonly used particular choices: \emph{Proper time} corresponds to $N=1$ and \emph{conformal time} to $N=a$. Both conventions are broadly used in literature. We therefore don't specify $N$ further in this work, to be able to compare our results with both conventions. In our analysis we choose for the configuration variables associated with the FLRW metric:
\begin{equation}
A := a^2.
\end{equation}
Rewriting this again in terms of the ADM-variables we obtain:
\begin{align}
\bar{q}_{ab} &= A(t)\delta_{ab} & \bar{N} &= \bar{N}(t) & \bar{N}^a &= 0.
\end{align}
The background curvature vanishes and hence for Cartesian coordinates the covariant derivative on $\bar{\Sigma}$ reduces to a partial derivative, that is:
\begin{align}
\bar{R}^{(3)}_{ab} &= 0 & \bar{D}_a &= \partial_a.
\end{align}
Using the relation between the extrinsic curvature and the momenta $P^{ab}$ in (\ref{eq:PabinKab}) we find:
\begin{equation}
\mathcal{LM}^*\bar{P}^{ab} = -\frac{1}{\sqrt{A}N}\dot{A}\delta^{ab}~.
\end{equation}
We have used that $\sqrt{\det\bar{q}} = A^{3/2}$. Hence, the background momenta of the spatial metric can be described by a single, only time dependent parameter which will be denoted as $\tilde{P}$ in what follows:
\begin{equation}
\bar{P}^{ab} =: \tilde{P}(t) \delta^{ab}.
\end{equation}

The matter content of a homogeneous and isotropic universe can be described by two parameters called \emph{energy-density} $\rho$ and \emph{pressure} $p$. These quantities are encoded in the energy-momentum tensor $T^\mu_\nu$:
\begin{align}
\rho &:= -T^0_0 & p &:= T^1_1 = T^2_2 = T^3_3.
\end{align}
As for the geometric part the configuration variables and conjugate momenta do also only depend on time:
\begin{equation}
\bar{\varphi} = \bar{\varphi}(t),\quad \bar{\pi}_\varphi=\bar{\pi}_\varphi(t).
\end{equation}
On phase space the energy-momentum tensor for a homogeneous isotropic scalar field can be expressed in terms of the variables $\bar{\varphi},\bar{\pi}_\varphi$ and $A$, as we will see below.

The standard Dirac constraint algorithm yields the primary constraints:
\begin{equation}
\bar{\Pi} = \bar{\Pi}_a = 0
\end{equation}
and demanding their stability leads to the following secondary constraints:
\begin{align}
\label{eq:barCbarCa}
\bar{C} &= -\frac{3}{2}\sqrt{A}\tilde{P}^2 + \kappa A^{3/2}\rho = 0 \nonumber \\
\bar{C}_a &= 0.
\end{align}
Given, this we can easily construct the corresponding ADM-Hamiltonian using (\ref{eq:HADM}) and compute the Hamiltonian equations of motion. For the geometric sector we get:
\begin{align}
\label{eq:dotAdotPcosm}
\dot{A} &= -\bar{N}\sqrt{A}\tilde{P} & \dot{\tilde{P}} &= \bar{N}\left( \frac{1}{4}\frac{\tilde{P}^2}{\sqrt{A}} + \frac{\kappa}{2}\sqrt{A}p \right).
\end{align}
The equations of motion for lapse and shift yield the following expression for the Lagrange-multipliers:
\begin{equation}
\dot{\bar{N}}=\bar{\lambda},\quad \dot{\bar{N}}^a=\bar{\lambda}^a=0.
\end{equation}
The background scalar field degrees of freedom evolve according to:
\begin{align}
\label{eq:dotbarvarphidotbarpivarphi}
\dot{\bar{\varphi}} &= \bar{N}\frac{\lambda_\varphi}{A^{3/2}}\bar{\pi}_\varphi & \dot{\bar{\pi}}_\varphi = -\bar{N}\frac{A^{3/2}}{\lambda_\varphi}\frac{1}{2}\frac{\mathrm{d}V}{\mathrm{d}\varphi}(\bar{\varphi}).
\end{align}
Note, that the $\lambda_\varphi$ from the scalar field action may not be confused with the Lagrange-multiplier $\lambda$ in the full Hamiltonian.

If one is rather used to work in the Lagrangian formalism, then these equations might not look too familiar. For this reason we will rewrite them in a form involving the Hubble parameter that also enters the conventional form of the Friedmann equations. In the Lagrangian picture 
 the \emph{Hubble-parameter} is defined as the relative velocity of the scale factor.
\begin{equation}
\mathcal{H} = \frac{\dot{a}}{a}.
\end{equation}
Note, that the dot is either with respect to proper time if $N=1$, to conformal time, when $N=a$ or to some different time parametrization. We want to define a corresponding quantity $\tilde{\mathcal{H}}$ in phase space, such that $\mathcal{LM}^*\tilde{\mathcal{H}} = \mathcal{H}$. Using $\tfrac{\dot{a}}{a}=\tfrac{1}{2}\tfrac{\dot{A}}{A}$ and the equation of motion for $A$ in (\ref{eq:dotAdotPcosm}) we find:
\begin{equation}
\tilde{\mathcal{H}} := - \frac{\bar{N}\tilde{P}}{2\sqrt{A}}.
\end{equation}
We will use $\tilde{\mathcal{H}}$ instead of $\tilde{P}$ at some places, because some equations look more concise in terms of this parameter. Moreover this notation is beneficial for comparison with literature. In particular we can formulate the background equations of motion as:
\begin{align}
\dot{A} &= 2\tilde{\mathcal{H}}A & \dot{\tilde{P}} &= -\frac{1}{2}\tilde{\mathcal{H}}\tilde{P} + \frac{\kappa}{2}\bar{N}\sqrt{A}p.
\end{align}
If we consider the expressions for the energy and momentum density involved in the energy momentum tensor on phase space 
\begin{align}
\label{eq:rhop}
\rho &= \frac{1}{2}\left( \frac{\lambda_\varphi}{A^3}\bar{\pi}_\varphi^2 + \frac{1}{\lambda_\varphi}V(\bar{\varphi}) \right) & p &=\frac{1}{2}\left( \frac{\lambda_\varphi}{A^3}\bar{\pi}_\varphi^2 - \frac{1}{\lambda_\varphi}V(\bar{\varphi}) \right).
\end{align}
we can derive dynamical equations for $\rho$ and $p$ by using the definition of $\rho$, $p$ from above and the equations of motion for the background scalar field in (\ref{eq:dotbarvarphidotbarpivarphi}) we finally obtain:
\begin{align}
\label{eq:dotrhodotp}
\dot{\rho} &= -3\tilde{\mathcal{H}}(\rho+p) & \dot{p} &= -3\tilde{\mathcal{H}}(\rho + p) - \frac{\bar{N}}{A^{3/2}}\bar{\pi}_\varphi \frac{\mathrm{d}V}{\mathrm{d}\varphi}(\bar{\varphi}).
\end{align}

\subsubsection{Equations of motion for linear perturbations around a FLRW background}

To discuss the Hamiltonian equations for linear perturbations around a FLRW $k=0$ background, we specialize the perturbed equations of motion, derived in section \ref{sususe:curpert} and displayed in (\ref{eq:deltadotq}), (\ref{eq:deltadotPgeo}) and (\ref{eq:deltadotPphi}) to the FLRW $k=0$ background. For the elementary variables of geometry we obtain:
\begin{align}
\label{eq:deltadotqflat}
\delta \dot{q}_{ab} &= 2\tilde{\mathcal{H}}A\delta_{ab}\frac{\delta N}{\bar{N}}-2\tilde{\mathcal{H}} \left( \delta^c_a\delta^d_b-\frac{1}{2}\delta_{ab}\delta^{cd} \right) \delta q_{cd} - 4\tilde{\mathcal{H}} \frac{A}{\tilde{P}} \left( \delta_{ac}\delta_{bd}-\frac{1}{2}\delta_{ab}\delta_{cd} \right) \delta P^{cd} + 2 \delta N_{(a,b)}.
\end{align}
\begin{align}
\label{eq:deltadotPflat}
\delta \dot{P}^{ab} &= \frac{1}{4} \frac{1}{\sqrt{A}} \tilde{P}^2\delta^{ab} \delta N +\frac{1}{\sqrt{A}} \left( \partial^a \partial^b-\delta^{ab}\Delta \right)\delta N \nonumber \\
~ &- \bar{N}\left[ \frac{1}{A^{3/2}}\tilde{P}^2\left( \frac{5}{4}\delta^{ac}\delta^{bd}-\frac{3}{8}\delta^{ab}\delta^{cd} \right)\delta q_{cd} +\frac{1}{\sqrt{A}}\left( \delta^{ac}\delta^{bd}-\frac{1}{2}\delta^{ab}\delta^{cd} \right)\delta R^{(3)}_{cd} \right. \nonumber \\
~ &+ \left. \frac{1}{\sqrt{A}}\tilde{P}\left( \delta^a_c\delta^b_d-\frac{1}{2}\delta^{ab}\delta_{cd} \right)\delta P^{cd} \right] +\tilde{P}\left( \delta N^c_{,c} \delta^{ab}-2\delta N^{(a,b)} \right) \nonumber \\
~ &+ \frac{\kappa}{2}\bar{N}\sqrt{A} \left[ p \delta^{ab} \frac{\delta N}{\bar{N}}-\left( p \delta^{ac}\delta^{bd}+\frac{1}{2}\rho\delta^{ab}\delta^{cd} \right) \frac{\delta q_{cd}}{A} + \mathcal{P}\delta^{ab} \right]~,
\end{align}
where we have defined:
\begin{equation}
\mathcal{P} := \frac{\lambda_\varphi}{A^3}\bar{\pi}_\varphi \delta \pi_\varphi - \frac{1}{2\lambda_\varphi}\frac{\mathrm{d}V}{\mathrm{d}\varphi}(\bar{\varphi})\delta \varphi .
\end{equation}
This expression contains all scalar field perturbation parts. Furthermore the following similar definition will be useful later:
\begin{equation}
\mathcal{E} := \frac{\lambda_\varphi}{A^3}\bar{\pi}_\varphi \delta \pi_\varphi + \frac{1}{2\lambda_\varphi}\frac{\mathrm{d}V}{\mathrm{d}\varphi}(\bar{\varphi})\delta \varphi .
\end{equation}
One can interpret the above defined quantities as the parts of the pressure and energy-density perturbations which contain the scalar field perturbations.
In the derivation of the equations of motion above, the following identities were used:
\begin{equation}
\bar{q}^{ab}\delta R^{(3)}_{ab} = \frac{1}{2A^2}\delta^{ab}\delta^{cd}( \delta q_{ad,bc} + \delta q_{bd,ac} - \delta q_{ab,cd} - \delta q_{cd,ab} ) = \frac{1}{A^2}\left( \partial^a\partial^b - \delta^{ab}\Delta \right)\delta q_{ab}
\end{equation}
and
\begin{equation}
\left( \bar{q}^{ac}\bar{q}^{bd} - \frac{1}{2}\bar{q}^{ab}\bar{q}^{cd} \right) \delta R^{(3)}_{cd} = \frac{1}{2A^3}\left( 2\delta^{c(a}\partial^{b)}\partial^d - \delta^{ab}\partial^c\partial^d - \delta^{cd}\partial^a\partial^b - (\delta^{ac}\delta^{bd}-\delta^{ab}\delta^{cd})\Delta \right) \delta q_{cd}.
\end{equation}
The perturbed secondary constraints can be derived from the general form of the linearized secondary constraints in (\ref{eq:deltaCgeogeneral}), (\ref{eq:deltaCphigeneral}), (\ref{eq:deltaCageogeneral}) and (\ref{eq:deltaCaphigeneral}), leading to:
\begin{align}
\label{eq:deltaCcosmo}
\delta C &= -\frac{1}{4}\frac{\tilde{P}^2}{\sqrt{A}}\delta^{ab}\delta q_{ab} - \sqrt{A}\tilde{P}\delta_{ab}\delta P^{ab} - \frac{1}{\sqrt{A}}\left( \partial^a\partial^b - \delta^{ab}\Delta \right) \delta q_{ab} \nonumber \\
~ & + \kappa\left( -\frac{\sqrt{A}}{2}p \delta^{ab} \delta q_{ab} + \mathcal{E} \right).
\end{align}
\begin{align}
\label{eq:deltaCacosmo}
\delta C_a &= -2A \delta_{ab}\delta P^{bc}_{,c} - 2\tilde{P} \left( \delta^b_a\partial^c - \frac{1}{2}\delta^{bc}\partial_a \right) \delta q_{bc} + \kappa \bar{\pi}_\varphi \delta \varphi_{,a}.
\end{align}
The form of the linearized constraints agrees with the results in \cite{Langlois} equations (19) and (20), albeit Langlois' notation differs slightly from our. In particular: $\kappa_\text{Langlois} = \tfrac{1}{2}\kappa$, $V_\text{Langlois} = 2V$, $\varphi$ is named $\phi$, $A = \mathrm{e}^{2\alpha}$, $\bar{q}_{ab}$, $\bar{P}^{ab}$ are named $\gamma_{ij}$ and $\pi^{ij}$ respectively, $\pi_\alpha = 6A\tilde{P}$  and the Poisson bracket of $\gamma_{ij}$ with $\pi^{ij}$ is defined without $\kappa$. The exact form of the linearized constraints are also involved in the dynamical equations for the lapse and shift perturbations that have the following form
\begin{equation}
\delta \dot{N} = \delta \lambda,\quad 
\delta \dot{\Pi} = -\delta C, \quad \delta \dot{N}^a=\delta \lambda^a,\quad\delta \dot{\Pi}_a = -\delta C_a.
\end{equation}

Lastly the perturbed scalar field equations of motion can be derived using (\ref{eq:deltadotphi}) and (\ref{eq:deltadotpiphi}). They are given by:
\begin{align}
\delta \dot{\varphi} &= \delta N \frac{\lambda_\varphi}{A^{3/2}}\bar{\pi}_\varphi + \bar{N}\frac{\lambda_\varphi}{A^{3/2}}\left( \delta \pi_\varphi - \frac{1}{2A} \bar{\pi}_\varphi \delta^{ab}\delta q_{ab} \right) \nonumber \\
\delta \dot{\pi}_\varphi &=  - \delta N \frac{A^{3/2}}{\lambda_\varphi}\frac{1}{2}\frac{\mathrm{d}V}{\mathrm{d}\varphi}(\bar{\varphi}) + \bar{N}\frac{A^{3/2}}{\lambda_\varphi}\left( \frac{1}{A}\Delta \delta \varphi - \frac{1}{4A}\frac{\mathrm{d}V}{\mathrm{d}\varphi}(\bar{\varphi})\delta^{ab}\delta q_{ab} - \frac{1}{2}\frac{\mathrm{d}^2V}{\mathrm{d}\varphi^2}(\bar{\varphi})\delta \varphi \right) \nonumber \\
~ & \hspace{1em} + \bar{\pi}_\varphi \delta N^a_{,a}.
\end{align}

\subsubsection{Scalar-vector-tensor decomposition of the perturbations}
\label{sususe:scavete}

We have seen that the equations of motion for canonical cosmological perturbation theory can be derived straight forwardly from the ADM-formulation of general relativity. However, the equations of motion are still complicated coupled tensorial  linear partial differential equations. It is common in linear cosmological perturbation theory to decompose the metric perturbation $\delta q_{ab}$ into scalar, vector and tensor degrees of freedom because they decouple at linear order. As a consequence the equations of motion for scalar, vector and tensor perturbations can be solved separately. A covariant treatment of this scalar-vector-tensor decomposition in cosmological perturbation theory has been discussed for instance by Steward in \cite{Steward}. Furthermore, York shows in \cite{York} that this decomposition is in fact an orthogonal decomposition. We will briefly review the scalar-vector-tensor decomposition for a generic background closely following the presentation in \cite{York}. Afterwards we apply such a decomposition to the case of interest, namely the choice of a flat FLRW background.

We will start by reviewing the Helmholtz decomposition of a vector field and covector field respectively. 
Let $\Sigma$ be a spatial manifold of dimension 3 with metric $q$ and $V$ a vector field, that is a section on  the tangent bundle $T\Sigma$. Later on we will replace this general spatial manifold by some background manifold $(\bar{\Sigma},\bar{q})$. We can decompose this vector field into a scalar and a transversal part as follows:
\begin{equation}
V^a = D^a\hat{V}+V_\perp^a
\end{equation}
where $D$ is the (unique torsion free and metric-compatible) covariant derivative on $\Sigma$ and $D_aV_\perp^a=0$. We find
\begin{equation}
\hat{V} = \boldsymbol{\Delta}^{-1} D_aV^a~~,~~V_\perp^a = V^a - D^a\hat{V},
\end{equation}
with $\boldsymbol{\Delta}=q^{ab}D_aD_b$ and $\boldsymbol{\Delta}^{-1}$ is the Green's function of the Laplace-Beltrami equation $\boldsymbol{\Delta}u=f$. Assuming the existence of the Green's function in $\Sigma$ it satisfies $\boldsymbol{\Delta}\boldsymbol{\Delta}^{-1}(x,y) = \delta(x,y)$. We use the following abbreviation for the Green's function:
\begin{equation}
\boldsymbol{\Delta}^{-1}f(x) = \int_{\Sigma} \mathrm{d}^3y~\boldsymbol{\Delta}^{-1}(x,y)f(y).
\end{equation}
Using the transversality of $V^a_\perp$ one finds that the following projectors project onto the scalar and transversal part respectively:
\begin{align}
(\hat{P}_SV) &:= \boldsymbol{\Delta}^{-1} D_aV^a = \hat{V} \nonumber \\
(\hat{P}_\perp V)^a &:= (\delta^a_b-D^a\boldsymbol{\Delta}^{-1} D_b)V^b = V_\perp^a.
\end{align}
Covector fields can be decomposed analogously. For this purpose let $\omega$ be a covector field, that is a section on the cotangent bundle $T^*\Sigma$. In this case the corresponding scalar and transversal projectors can then be defined as follows:
\begin{align}
(\hat{P}_S\omega) &:= \boldsymbol{\Delta}^{-1} D^a\omega_a = \hat{\omega} \nonumber \\
(\hat{P}_\perp \omega)_a &:= (\delta^b_a-D_a\boldsymbol{\Delta}^{-1} D^b)\omega_b = \omega^\perp_a.
\end{align}

The tensor field that we deal with is a symmetric (0,2)-tensor field, that we denote by $T$. We want to generalize the Helmholtz decomposition for covector fields such that we can  decompose $T$ in a similar way. As a first step a decomposition of $T$ into its trace and a traceless part is performed yielding:
\begin{equation}
T_{ab} = \frac{1}{3}q_{ab}\text{Tr}(T) + T^T_{ab},
\end{equation}
with $\text{Tr}(T):= q^{ab}T_{ab}$ and $T^T_{ab}$ the traceless part of $T$. Subsequently we can decompose $T^T$ into a transversal part $T^{TT}$ and some residue denoted as $T^{T'}$. $T^{TT}$ has to fulfill the following 3 conditions:
\begin{equation}
D^aT^{TT}_{ab}=0.
\end{equation}
Considering the traceless condition together with the symmetry of $T$ we can conclude that $T^{TT}$ has two independent degrees of freedom. 
Thus there are three more degrees of freedom left sitting in $T^{T'}$. We parametrize these by introducing a covector field $A_a$ and rewriting $T^{T'}_{ab}$ in terms of it:
\begin{equation}
T^{T'}_{ab}= 2D_{<a}A_{b>} := 2\left(D_{(a}A_{b)} - \frac{1}{3}q_{ab}D^cA_c \right).
\end{equation}
As one can easily see, $T^{T'}_{ab}$ is traceless and has three independent degrees of freedom encoded in $A_a$.

Likewise to the Helmholtz decomposition we want to introduce the associated projectors for this decomposition. For this purpose we need to use the following identity:
\begin{equation}
D^bD_aA_b = D_aD^bA_b + R^b_aA_b.
\end{equation}
Using this we find:
\begin{equation}
D^bT^T_{ab} = D^bT^{T'}_{ab} = \left( \boldsymbol{\Delta}\delta_a^b + \frac{1}{3}D_aD^b + R_a^b \right)A_b.
\end{equation}
This motivates to introduce the following projectors:
\begin{align}
\label{eq:projectorsgeneral}
(\hat{P}_{Tr}T) &:= \frac{1}{3}q^{ab}T_{ab} = \frac{1}{3}\text{Tr}(T) \nonumber \\
(\hat{P}_{L}T)_a &:= \left( M^{-1} \right)_a^b \left[ D^cT_{bc}-\frac{1}{3}D_b\text{Tr}(T) \right] = A_a \nonumber \\
(\hat{P}_{TT}T)_{ab} &:= T_{ab} - (\hat{P}_{Tr}T)q_{ab} - 2D_{<a}(\hat{P}_{L}T)_{b>} = T^{TT}_{ab},
\end{align}
with
\begin{equation}
M_a^b := \delta_a^b\boldsymbol{\Delta}+ \frac{1}{3}D_aD^b +R_a^b.
\end{equation}
Now the question arises whether the inverse of $M_a^b$ exists. As shown in \cite{York} the inverse to the operator $\mathbf{D}$
\begin{equation}
\label{eq:AtoLpart}
(\mathbf{D}A)_a := D^b\left(D_{(a}A_{b)}-\frac{1}{3}q_{ab}D^cA_c\right)
\end{equation}
exists and we easily compute that $2(\mathbf{D}A)_a = M_a^b A_b$. Hence, $\left( M^{-1} \right)_a^b$ does indeed exist.

Next, we can perform a Helmholtz decomposition to further decompose $A_a$ into its longitudinal and transversal part:
\begin{equation}
A_a = D_a\hat{A}+A^\perp_a~~,~~D^aA^\perp_a=0.
\end{equation}
Given this, it is convenient to introduce the longitudinal scalar projector denoted by $\hat{P}_{LS} = \hat{P}_S \circ \hat{P}_L$ and the longitudinal transversal projector denoted by $\hat{P}_{LT} = \hat{P}_\perp \circ \hat{P}_L$. Their explicit form reads:
\begin{align}
\label{eq:PLSPLTgeneral}
(\hat{P}_{LS}T) &:= \boldsymbol{\Delta}^{-1}D^a (\hat{P}_{L}T)_a = \hat{A} \nonumber \\
(\hat{P}_{LT}T)_a &:= (\hat{P}_{L}T)_a - (\hat{P}_{LS}T)_{,a} = A^\perp_a.
\end{align}

What we have finally arrived at is a decomposition of the tensor field $T$ into two scalars, one transversal vector and one transversal and traceless tensor encoding exactly the six degrees of freedom a symmetric (0,2) tensor fields has in three dimensions. This scalar-vector-tensor decomposition can expressed in terms of the projectors introduced above and reads:
\begin{equation}
T_{ab} = (\hat{P}_{Tr}T)q_{ab} + 2D_{<a}D_{b>}(\hat{P}_{LS}T) + 2D_{(a}(\hat{P}_{LT}T)_{b)} + (\hat{P}_{TT}T)_{ab}.
\end{equation}
One can perform this decomposition analogously for symmetric (2,0)-tensor fields (or tensor field densities). We will also use the notation $\hat{P}_I$, $I=Tr,LS,LT,TT$ for those tensors, i.e. for $S^{ab}$ a symmetric (2,0) tensor (density) field we have $(\hat{P}_{Tr}S) = \frac{1}{3}q_{ab}S^{ab}$, etc. In the next section we consider such decomposition for a FLRW background.
~\\
~\\
\paragraph*{\centerline{3.a Scalar-vector-tensor decomposition for a flat FLRW background:}} 
~\\
For the FLRW $k=0$ case we apply the scalar-vector-tensor decomposition to $q_{ab} = A\delta_{ab}$. Considering that the Ricci tensor vanishes, we obtain for the covariant derivative, the Laplacian and the inverse of $M_a^b$:
\begin{align}
D^a &= \frac{1}{A}\partial^a & \boldsymbol{\Delta} &= \frac{1}{A}\Delta & (M^{-1})^a_b &= A\Delta^{-1}\left( \delta^a_b - \frac{1}{4}\Delta^{-1}\partial_a\partial^b \right).
\end{align}
The projectors therefore take the following form:
\begin{align}
\label{eq:projectorscosmology}
(\hat{P}_SV) &= A\Delta^{-1} \partial_aV^a \nonumber \\
(\hat{P}_\text{Tr}T) &= \frac{1}{3A}\delta^{ab}T_{ab} \nonumber \\
(\hat{P}_{LS}T) &= \frac{3}{4}\Delta^{-2}\partial^{<a}\partial^{b>}T_{ab} \nonumber \\
(\hat{P}_{LT}T)_a &= \Delta^{-1}\left( \delta_a^b\partial^c -\tfrac{1}{3}\partial_a\delta^{bc} \right) T_{bc} - \frac{4}{3}\partial_a (\hat{P}_{LS}T).
\end{align}
$\hat{P}_\perp$ and $\hat{P}_{TT}$ can be derived from the other projectors. $\Delta^{-1}$ stands for the Green's function of the Poisson equation and we denote the associated integral kernel of $\Delta^{-1}$ by $G(x,y)$, that is:
\begin{equation}
\Delta^{-1}f(x) = \int_{\bar{\Sigma}}\mathrm{d}^3y~ G(x,y)f(y).
\end{equation}

In conventional cosmological perturbation theory (see e.g. \cite{Mukhanov} or \cite{bardeen}) it is common to work with the following perturbed quantities:
\begin{align}
\label{eq:decomposedmetricdef}
\phi &:= \frac{\delta N}{\bar{N}} & B &:= \frac{1}{A}(\hat{P}_S\delta \vec{N}) & S^a &:= (\hat{P}_\perp\delta \vec{N})^a \nonumber \\
\psi &:= \frac{1}{2}(\hat{P}_\text{Tr}\delta q) & E &:= \frac{1}{A}(\hat{P}_{LS}\delta q) & F_a &:= \frac{1}{A}(\hat{P}_{LT}\delta q) & h^{TT}_{ab} &:= \frac{1}{A}(\hat{P}_{TT}\delta q)_{ab}.
\end{align}
Within the literature these perturbations are sometimes defined with different sign conventions. In \cite{Mukhanov} a different sign for $\psi$ is used and in \cite{Dust2} a different sign for $\phi$ is considered. Often $E$ is defined such that $\Delta E = 0$ or that it contains a trace part \cite{Dust2}. In this case one works with $E_{,ab}$ instead of $E_{,<ab>}$ in the expression for $\delta q_{ab}$. Overall one has to be careful as also different signatures are used and the shift vector field may be defined with a different sign as for instance in \cite{bardeen}.
Up to only background dependent factors these quantities are precisely the quantities which one obtains by applying the Helmholtz decomposition to the shift vector field perturbation $\delta \vec{N}$ and the scalar-vector-tensor decomposition to the spatial metric perturbation $\delta q_{ab}$. The perturbed quantities can thus be written in terms of four scalars, two transversal vector fields and one traceless-transversal tensor field:
\begin{align}
\delta N &= \bar{N}\phi \nonumber \\
\delta N^a &= B^{,a} + S^a \nonumber \\
\delta q_{ab} &= 2A\left( \psi \delta_{ab} + E_{,<ab>} + F_{(a,b)} + \tfrac{1}{2}h^{TT}_{ab} \right).
\end{align}

Often cosmological perturbation theory is done in the Lagrangian picture, where one works on the tangent bundle and the relevant quantities are the configuration variables ($\psi,E,\cdots$) and their associated velocities ($\dot{\psi},\dot{E},\cdots$). For our approach that is formulated on phase space and thus in the cotangent bundle we work with the perturbations of the configuration variables and their conjugate momenta. Analogously to the choice of variables for the decomposition of $\delta q_{ab}$ we also consider a scalar-vector-tensor decomposition of the momenta and, using the projectors, define the following variables:
\begin{align}
\label{eq:decomposedmomentadef}
p_\psi &:= \frac{1}{2}\frac{1}{A\tilde{P}}(\hat{P}_\text{Tr}\delta P) & p_E &:= \frac{1}{A^2\tilde{P}} (\hat{P}_{LS}\delta P) & p_F^a &:= \frac{1}{A\tilde{P}} (\hat{P}_{LT}\delta P)^a & p_{h^{TT}}^{ab} &:= \frac{1}{\tilde{P}}(\hat{P}_{TT}\delta P)^{ab}
\end{align}
This allows to rewrite $\delta P^{ab}$ in the following form:
\begin{equation}
\delta P^{ab} = 2\tilde{P}\left( p_\psi\delta^{ab} + p_E^{,<ab>} + p_F^{(a,b)} + \tfrac{1}{2}p_{h^{TT}}^{ab} \right).
\end{equation}
For the temporal-temporal and temporal-spatial part of the perturbed metric we introduce the following decomposition of the associated momenta:
\begin{eqnarray}
p_\phi &:=&\frac{1}{\overline{N}}\delta\Pi\nonumber\\
p_B &:=&\frac{1}{A}(\hat{P}_S\delta\vec{\Pi}),\quad p^S_{a}=(\hat{P}_\perp\delta\vec{\Pi})_a
\end{eqnarray}
where as before $\delta\Pi,\delta\vec{\Pi}$ denote the conjugate momenta of the perturbed lapse function $\delta N$ and the perturbed shift vector $\delta\vec{N}$.

The aim of the next section is to discuss the corresponding equations of motion for these decomposed quantities.

\subsubsection{Equations of motion}

Each of the perturbed quantities introduced above can be written as some projector $\hat{P}$ acting on a perturbation of the ADM-phase space quantities $\delta Q$. The projectors may however be explicitly time dependent and hence for the equations of motion of the decomposed perturbations one makes use of the following relation:
\begin{equation}
\frac{\mathrm{d}}{\mathrm{d}t}(\hat{P}\delta f) = (\dot{\hat{P}}\delta f) + ( \hat{P}\delta \dot{f}).
\end{equation}
For the scalar metric perturbations we use $\psi = \frac{1}{6A}\delta^{ab}\delta q_{ab}$ and $E = \frac{1}{A}(\hat{P}_{LS}\delta q)$ as well as the equation for $\delta \dot{q}_{ab}$ (\ref{eq:deltadotqflat}) to derive the following two equations of motion:
\begin{align}
\label{eq:dotpsiE}
\dot{\psi} &= 2\tilde{\mathcal{H}}\left( p_\psi - \frac{1}{2}\psi \right) + \tilde{\mathcal{H}}\phi + \frac{1}{3}\Delta B \nonumber \\
\dot{E} &= -4\tilde{\mathcal{H}}(E+p_E) + B.
\end{align}
To derive the equations of motion for $p_\psi$ and $p_E$ one first has to express $\delta R^{(3)}_{ab}$ in terms of the decomposed perturbations.
\begin{equation}
\label{eq:delRdecomposed}
\delta R^{(3)}_{ab} = - \frac{4}{3}\Delta \left( \psi -\frac{1}{3}\Delta E \right)\delta_{ab} - \left( \psi -\frac{1}{3}\Delta E \right)_{,<ab>} - \frac{1}{2}\Delta h^{TT}_{ab}.
\end{equation}
Using the equation of motion for $\delta P^{ab}$ in (\ref{eq:deltadotPflat}) and the projectors projecting to $p_\psi$ and $p_E$ the following equations of motion can be derived:
\begin{align}
\label{eq:dotGammaSigma}
\dot{p_\psi} &= \frac{1}{6}\frac{\bar{N}^2}{A\tilde{\mathcal{H}}}\Delta\left( \phi+\psi-\frac{1}{3}\Delta E \right) +\left( -\frac{1}{2}\tilde{\mathcal{H}} + \frac{\kappa}{4}\frac{\bar{N}^2}{\tilde{\mathcal{H}}}p \right)\left( p_\psi-\frac{1}{2}\psi \right) - \frac{\kappa}{8}\frac{\bar{N}^2}{\tilde{\mathcal{H}}}\delta\tilde{T} \nonumber \\
~ &-\frac{1}{2}\left( \frac{1}{2}\tilde{\mathcal{H}} + \frac{\kappa}{4}\frac{\bar{N}^2}{\tilde{\mathcal{H}}}p \right)\phi + \frac{1}{6}\Delta B \nonumber \\
\dot{p_E} &= -\frac{1}{4}\frac{\bar{N}^2}{A\tilde{\mathcal{H}}} \left( \phi+\psi-\frac{1}{3}\Delta E \right)+\left(\frac{5}{2}\tilde{\mathcal{H}} + \frac{\kappa}{4}\frac{\bar{N}^2}{\tilde{\mathcal{H}}}p \right) (E+p_E) -B~,
\end{align}
where we introduced the spatial energy momentum perturbation:
\begin{equation}
\delta \tilde{T} := \frac{1}{A^{3/2}}\left( \mathcal{P} - 3(\rho+p)\psi \right)
\end{equation}
One can derive the equations of motion for the vector and tensor analogously by applying the corresponding projectors to $\delta q_{ab}$ and $\delta P^{ab}$ but we will not need them in this work.
The equations of motion of the scalar matter field perturbation expressed in terms of the decomposed perturbations are:
\begin{align}
\delta \dot{\varphi} &= \bar{N}\frac{\lambda_\varphi}{A^{3/2}}\bar{\pi}_\varphi\left( \phi -3\psi + \frac{\delta \pi_\varphi}{\bar{\pi}_\varphi} \right) \nonumber \\
\delta \dot{\pi}_\varphi &= \bar{N}\frac{A^{3/2}}{\lambda_\varphi}\left[  -\frac{1}{2}\frac{\mathrm{d}V}{\mathrm{d}\varphi}(\bar{\varphi})(\phi+3\psi) +\frac{1}{A}\Delta\delta\varphi -\frac{1}{2}\frac{\mathrm{d}^2V}{\mathrm{d}\varphi^2}(\bar{\varphi})\delta\varphi \right].
\end{align}
The perturbations of the primary constraints $\delta \Pi$, $\delta \Pi_a$ are stabilized by the first order perturbations of the secondary constraints:
\begin{align}
\delta \dot{\Pi} &= - \delta C & \delta \dot{\Pi}_a &= -\delta C_a
\end{align}
The linearized secondary constraints can also be expressed in terms of the projected quantities and they have the following form:
\begin{align}
\label{eq:deltaCdeltaCa}
\delta C &=  -\frac{3}{2}\sqrt{A}\tilde{P}^2(\psi+4p_\psi) + 4\sqrt{A}\Delta\left( \psi -\frac{1}{3}\Delta E \right) + \kappa A^{3/2} \left( -3p\psi + \mathcal{E} \right) \nonumber \\
\delta C_a &= -4A\tilde{P}\left( p_\psi - \frac{1}{2}\psi + \frac{2}{3}\Delta (E+p_E) \right)_{,a} - 2A\tilde{P}\Delta(F_a+p_F^b\delta_{ab}) + \kappa \bar{\pi}_\varphi\delta \varphi_{,a}.
\end{align}
Here we used the form of the linearized constraints in (\ref{eq:deltaCcosmo}) and (\ref{eq:deltaCacosmo}) respectively. $\delta C$ can be cross-checked with e.g. \cite{Dust2} equation (19). Note, that in \cite{Dust2} one uses $\bar{N}=\sqrt{A}$, $\phi = 0$, the $A$ in \cite{Dust2} actually corresponds to the small $a$ as well as $\Xi$ denotes the scalar field. Moreover $\psi$ corresponds to $\psi + \tfrac{1}{3}\Delta E$ and all perturbed quantities in \cite{Dust2} are actually observables constructed with the dust clocks.
The perturbed spatial diffeomorphism constraint can be decomposed into a scalar part $\delta \hat{C}$ and a transversal part $\delta C^\perp_a$:
\begin{align}
\delta \hat{C} &= -4A\tilde{P}\left( p_\psi - \frac{1}{2}\psi + \frac{2}{3}\Delta (E+p_E) \right) + \kappa \bar{\pi}_\varphi\delta \varphi \nonumber \\
\delta C^\perp_a &= 2A\tilde{P}(F_a+p_F^b\delta_{ab}).
\end{align}
The dynamics of lapse and shift perturbations are related to yet undetermined functions which are the perturbations of the Lagrange-multipliers $\lambda = \bar{\lambda}+\delta \lambda$, $\lambda^a = \bar{\lambda}^a+\delta \lambda^a$:
\begin{align}
\label{eq:dotphidotB}
\dot{\phi} &= -\frac{\dot{\bar{N}}}{\bar{N}}\phi + \frac{\delta \lambda}{\bar{N}} & \dot{B} &= \delta \hat{\lambda} & \dot{S}^a &= \delta \lambda^a_\perp ,
\end{align}
with $\delta \lambda^a = \delta \hat{\lambda}_{,a} + \delta \lambda^a_\perp$. For the associated conjugate momenta we obtain:
\begin{align}
\dot{p}_\phi &= -\frac{\dot{\bar{N}}}{\bar{N}^2}\delta\Pi-\frac{1}{\bar{N}}\delta C & \dot{p}_B &= -\delta\hat{C} & \dot{p}_{S^a} &= -\delta C^a_\perp
\end{align}

\subsubsection{Gauge invariant perturbations on the extended ADM-phase space}

In the last section we derived the equations of motion for linear scalar, vector and tensor perturbations around a FLRW $k=0$ background in the Hamiltonian formulation. However, these individual perturbations are not gauge-invariant quantities. There exist two possible ways to deal with the gauge freedom in this context. One option is that we choose a particular gauge, but this may lead to unphysical gauge modes, when solving the equations of motion (see eg.: \cite{bardeen,Mukhanov,Durrer}). Moreover, for the choice of a specific gauge to find the physical meaning of such gauge dependent perturbations is difficult as for instance discussed in \cite{Durrer}. The other possibility is to construct quantities, which are gauge invariant. In the context of cosmological perturbation theory, usually one does not construct manifestly gauge invariant quantities, that is quantities that are invariant under gauge transformation up to any order. Instead one choses a fixed order in the perturbative expansion and requires that gauge invariant quantities are gauge invariant up to corrections that are of higher order than the chosen order in the perturbation theory. Applied to our case, we need to construct quantities that are invariant under gauge transformation at least up to first order and thus invariant under infinitesimal diffeomorphisms.
 
In the literature several common choices of gauge invariant variables exist. Bardeen \cite{bardeen2} introduced the so-called \emph{Bardeen potentials} $\Psi$ and $\Phi$, which are also used for the gauge invariant equations of motion in \cite{Mukhanov}. Another important gauge-invariant quantity is the \emph{Mukhanov-Sasaki-variable} $v$ introduced by Sasaki in \cite{Sasaki} and by Mukhanov in \cite{Mukhanov2}, which can be understood as a gauge invariant extension of the scalar field perturbation by means of adding appropriate geometric degrees of freedom. 

In this section we will discuss how the linear perturbations transform under infinitesimal gauge transformations and how this can be used to construct gauge invariant quantities on the extended ADM phase space. 
~\\
~\\
\paragraph*{\centerline{5.a Diffeomorphisms on extended ADM-phase space}}
~\\

We assume that our choice of coordinates is already fixed for the background quantities. Therefore for linear perturbation theory we restrict to infinitesimal diffeomorphisms, i.e. the change of the tensor fields induced by those diffeomorphisms is of order of the perturbations $\epsilon$.
In chapter \ref{sususe:obsGRfull} it was discussed, that the generator of 4-diffeomorphisms in the full ADM-phase space is given by $G'_{b,\vec{b}}$ in (\ref{eq:G'}). In our case the descriptors $b$ and $\vec{b}$ will be of order $\epsilon$. Let $Q$ be a tensor field on the phase space. Using that linear perturbations are of order $\epsilon$ we find up to linear order the following transformation behavior under diffeomorphisms generated by $G'_{b,\vec{b}}$:
\begin{equation}
Q \to Q'= Q + \{ Q,G'_{b,\vec{b}} \} = \bar{Q} + \delta Q + \overline{\{ Q,G'_{b,\vec{b}} \}} + \mathcal{O}(\epsilon^2)
\end{equation}
where the bar at the Poisson bracket indicates that it is evaluated on the FLRW background. Introducing $\delta_{G'}Q := Q'-Q$ we obtain:
\begin{equation}
\delta_{G'_{b,\vec{b}}}Q = \overline{\{ Q,G'_{b,\vec{b}} \}} + \mathcal{O}(\epsilon^2).
\end{equation}
As this term does not contain any terms of order $\epsilon^0$, we see that the background quantity $\bar{Q}$ does not change at all under such a diffeomorphism.
For first order perturbations we take only terms of order $\epsilon$ into account. Hence the change of the linear perturbation $\delta Q$ of $Q$ is simply given by:
\begin{equation}
(\delta_{G'_{b,\vec{b}}}\delta Q)^{(1)} = \overline{\{ Q,G'_{b,\vec{b}} \}}.
\end{equation}
Note, that even if $Q$ contains first order perturbations only, the action of the diffeomorphism generator in general yields additional terms of order $\epsilon^2$ and higher. Let us decompose $\vec{b}$ into its transversal and scalar part, that is $\vec{b}^a = \hat{b}^{,a}+b^a_\perp$. Using equations (\ref{eq:dG'metric}) and (\ref{eq:dG'momenta}) we can show that for a flat FLRW background the elementary ADM-phase space variables transform under infinitesimal diffeomorphisms according to:

\begin{align}
\label{eq:dG'metricFRW}
\delta_{G'_{b,\vec{b}}}N &= b_{,t} \nonumber \\
\delta_{G'_{b,\vec{b}}}N^a &= -\frac{\bar{N}}{A}b^{,a}+\hat{b}^{,a}_{,t}+b^a_{\perp,t} \nonumber \\
\delta_{G'_{b,\vec{b}}}q_{ab} &= 2A \left[ \left( \frac{\tilde{\mathcal{H}}}{\bar{N}}b+\frac{1}{3}\Delta \hat{b} \right)\delta_{ab} + \hat{b}_{,<ab>}+b^c_{\perp,(a}\delta_{b)c} \right].
\end{align}
and
\begin{align}
\label{eq:dG'momentaFRW}
\delta_{G'_{b,\vec{b}}}\Pi &= \bar{\Pi}\Delta \hat{b} = 0 \nonumber \\
\delta_{G'_{b,\vec{b}}}\Pi_a &= \bar{\Pi}b_{,a} = 0 \nonumber \\
\delta_{G'_{b,\vec{b}}}P^{ab} &= 2\tilde{P} \left[ \left( \left( -\frac{1}{4}\frac{\tilde{\mathcal{H}}}{\bar{N}}-\frac{\kappa}{8}\frac{\bar{N}}{\tilde{\mathcal{H}}}p \right)b +\frac{1}{6}\Delta \left( \frac{\bar{N}}{A\tilde{\mathcal{H}}}b+\hat{b}\right) \right)\delta^{ab} - \left( \frac{\bar{N}}{4A\tilde{\mathcal{H}}}b+\hat{b}\right)^{,<ab>}-b_{\perp}^{(a,b)} \right].
\end{align}
\begin{align}
\label{eq:trafoscalarfield}
\delta_{G'_{b,\vec{b}}}\varphi &= \frac{\lambda_\varphi}{A^{3/2}}\bar{\pi}_\varphi b & \delta_{G'_{b,\vec{b}}}\pi_\varphi &= \bar{\pi}_\varphi\Delta\hat{b}-\frac{1}{2}\frac{A^{3/2}}{\lambda_\varphi}\frac{\mathrm{d}V}{\mathrm{d}\varphi}(\bar{\varphi})b.
\end{align}
All terms of order higher than linear oder in $\epsilon$ have been neglected. 
~\\
~\\
\paragraph*{\centerline{5.b Scalar-vector-tensor decomposition}}
~\\

The equations encoding the transformation behavior of the linear perturbations have been brought into a form suitable to apply the scalar-vector-tensor decomposition. As all projectors $\hat{P}_\text{Tr}$, $\hat{P}_{LS}$, etc. depend on background quantities only and background quantities do not change under the diffeomorphisms considered, the following relation holds:
\begin{equation}
\hat{P}\delta_{G'_{b,\vec{b}}}Q = \delta_{G'_{b,\vec{b}}}\hat{P}Q
\end{equation}
Hence, acting with the scalar, vector and tensor projectors on (\ref{eq:dG'metricFRW}) and (\ref{eq:dG'momentaFRW}) yields the transformation behaviors of the decomposed perturbations:
\begin{align}
\label{eq:trafoscalarsmetric}
\delta_{G'_{b,\vec{b}}}\phi &= \frac{1}{\bar{N}}b_{,t} ~&~ \delta_{G'_{b,\vec{b}}}B &=-\frac{\bar{N}}{A}b+\hat{b}_{,t} \nonumber \\
\delta_{G'_{b,\vec{b}}}\psi &= \frac{\tilde{\mathcal{H}}}{\bar{N}}b+\frac{1}{3}\Delta\hat{b} ~&~ \delta_{G'_{b,\vec{b}}}E &= \hat{b}
\end{align}
and
\begin{align}
\label{eq:traforscalarmomenta}
\delta_{G'_{b,\vec{b}}}p_\psi &=-\left( \frac{1}{4}\frac{\tilde{\mathcal{H}}}{\bar{N}}+\frac{\kappa}{8}\frac{\bar{N}}{\tilde{\mathcal{H}}}p \right)b+\frac{1}{6}\Delta \left( \frac{\bar{N}}{A\tilde{\mathcal{H}}}b+\hat{b}\right) ~&~ \delta_{G'_{b,\vec{b}}}p_E &= -\frac{\bar{N}}{4A\tilde{\mathcal{H}}}b-\hat{b} \nonumber\\
\delta_{G'_{b,\vec{b}}}p_\phi &=0 & \delta_{G'_{b,\vec{b}}}p_B &= 0
\end{align}
as well as 
\begin{eqnarray}
\delta_{G'_{b,\vec{b}}}S^a&=&b^a_{\perp,t}~,~~
\delta_{G'_{b,\vec{b}}}{p_{S}}_{a}=0 \nonumber\\
\delta_{G'_{b,\vec{b}}}F_a&=&b_{\perp a},~~\delta_{G'_{b,\vec{b}}}p_F^a=-b^a_{\perp},\nonumber\\
\delta_{G'_{b,\vec{b}}}h^{TT}_{ab}&=&0 ~~\text{and}~~ \delta_{G'_{b,\vec{b}}}p_{h^{TT}}^{ab}=0,
\end{eqnarray}
where we used that $h^{TT}_{ab}$ and $p_{h^{TT}}^{ab}$ are already gauge invariant. Note that the momenta associated with $\phi$ and $S^a$ are linear in the primary constraints and thus do not change under these transformations. In the Lagrangian framework usually an infinitesimal diffeomorphism is parametrized by $\epsilon^\mu$ with $x^\mu\to x^\mu+\epsilon^\mu$ where $x^\mu$ denotes spacetime coordinates. 
To compare these results with the literature (e.g. \cite{Mukhanov}) we have to express $\epsilon^\mu$ in terms of the descriptors $b$ and $b^a$. This relation is given by $\epsilon^\mu = b n^\mu + X^\mu_a b^a$ and for an adaptive frame we have $X^\mu_a = \delta^\mu_a$ and $n^\mu = N^{-1}(1,-N^a)$.
~\\
~\\
\paragraph*{\centerline{5.c Gauge invariant perturbations}}
~\\

In order to construct quantities that are gauge invariant up to first order in $\epsilon$, one can take specific combinations of the perturbations such that they don't change under an infinitesimal diffeomorphism. This is the usual approach in cosmological perturbation theory, where gauge invariant quantities are constructed order by order and hence these quantities are only gauge invariant up to the respective order. Let us discuss this for the tensor, vector and scalar part separately. The tensor perturbations are already gauge invariant. For the vector perturbations one can choose the following gauge invariant combination:
\begin{equation}
\nu^a := F_b\delta^{ab}+p_F^a.
\end{equation}
In general there are many possibilities to choose gauge invariant combinations of the scalar perturbations. In order to be able to compare to the results in the Lagrangian framework, we will use the phase space analoges of the Bardeen potentials $\Psi$, $\Phi$ and the Mukhanov-Sasaki variable $v$. 
~\\
~\\
\paragraph*{\centerline{5.c.1 Bardeen potentials and Mukhanov-Sasaki variable on extended ADM phase space}}
~\\

We adapt the form of the Bardeen potentials given in e.g. \cite{Mukhanov}, \cite{bardeen2} to our notations and define:
\begin{equation}
\mathcal{LM}^*\Psi := \psi - \frac{1}{3}\Delta E + \frac{\mathcal{H}A}{\bar{N}^2}(B-\dot{E}).
\end{equation}
To find the corresponding function on the phase space, we have to use the equation of motion for $E$ in (\ref{eq:dotpsiE}) in order to replace the velocity $\dot{E}$ by its associated momentum. This yields:
\begin{equation}
B-\dot{E} \overset{\mathcal{LM}}{\longrightarrow} 4\tilde{\mathcal{H}}(E+p_E).
\end{equation}
Inserting this into the above equation, we end up with the following form of the Bardeen potential $\Psi$ on the phase space:
\begin{align}
\label{eq:Psi}
\Psi &= \psi - \frac{1}{3}\Delta E + \frac{4\tilde{\mathcal{H}}^2A}{\bar{N}^2}(E+p_E) \nonumber \\
~ &= \psi - \frac{1}{3}\Delta E + \tilde{P}^2(E+p_E).
\end{align}
As before $\Psi$ can be understood as a first order gauge invariant extension of $\psi$. For the second Bardeen potential $\Phi$ we define:
\begin{equation}
\mathcal{LM}^*\Phi := \phi + \frac{1}{\bar{N}}\frac{\mathrm{d}}{\mathrm{d}t}\left( \frac{A}{\bar{N}}(B-\dot{E}) \right).
\end{equation}
To find the corresponding phase space function we use $\tilde{\mathcal{H}} = -\frac{\bar{N}\tilde{P}}{2\sqrt{A}}$ and compute:
\begin{align}
\dot{\tilde{\mathcal{H}}} &= \tilde{\mathcal{H}}\frac{\dot{\bar{N}}}{\bar{N}}-\frac{3}{2}\tilde{\mathcal{H}}^2-\frac{\kappa}{4}\bar{N}^2p \nonumber \\
\dot{E}+\dot{p_E} &= -\frac{1}{4}\frac{\bar{N}^2}{A\tilde{\mathcal{H}}}\left( \phi+\psi-\frac{1}{3}\Delta E \right) -\frac{3}{2}\tilde{\mathcal{H}}(E+p_E)+\frac{\kappa}{4}\frac{\bar{N}^2}{\tilde{\mathcal{H}}}p(E+p_E)~,
\end{align}
where we considered the equations of motion for $\tilde{P}$, $A$, $E$ and $p_E$. Inserting these results into $\Phi$ yields
\begin{equation}
\Phi = -\Psi.
\end{equation}
Hence, on phase space the Bardeen potentials coincide on the level of the equations of motion up to an overall minus sign. The same result is also obtained in the Lagrangian picture, as there one gets from the perturbed gauge invariant Einstein-equations that $\Phi = - \Psi$ if the perturbed gauge invariant energy momentum tensor  $\delta {T^{(gi)}}^\mu_\nu$ satisfies
\begin{equation}
\delta T^{(gi)i}_j \propto \delta^i_j, 
\end{equation}
see e.g. \cite{Mukhanov}. It can easily be shown that this is the case for a minimally coupled scalar field. 

As we are working on phase space, we need another gauge invariant potential which will be related to the momentum of $\Psi$. For this purpose we define:
\begin{equation}
\label{eq:Upsilon}
\Upsilon := p_\psi - \frac{1}{6}\Delta E + \frac{2}{3}\Delta(E+p_E) -\left( \frac{1}{4}\tilde{P}^2 + \frac{\kappa}{2}Ap \right)(E+p_E).
\end{equation}
This quantity can be understood as a gauge invariant extension of $p_\psi$. As we will discuss later this quantity appears in the equation of motion for $\Psi$.

In order to define the Mukhanov-Sasaki variable on phase space we need to define gauge invariant variables for the scalar field degrees of freedom. We use the transformation behavior of the scalar field and its momentum in (\ref{eq:trafoscalarfield}) as well as those of the geometrical perturbations in (\ref{eq:trafoscalarsmetric}) and (\ref{eq:traforscalarmomenta}). This allows us to find gauge invariant extensions of $\delta \varphi$ and $\delta \pi_\varphi$ by adding appropriate combinations of the scalar perturbations $\psi$, $E$, $p_\psi$ and $p_E$ to them. This yields:
\begin{align}
\label{eq:deltavarphigi}
\delta \varphi^{(gi)} &:= \delta \varphi + \bar{\pi}_\varphi \frac{4\tilde{\mathcal{H}}\lambda_\varphi}{\sqrt{A}\bar{N}}(E+p_E) \\
\delta \pi_\varphi^{(gi)} &:= \delta \pi_\varphi -\bar{\pi}_\varphi\Delta E -2\frac{\tilde{\mathcal{H}}A^{5/2}}{\lambda_\varphi \bar{N}}\frac{\mathrm{d}V}{\mathrm{d}\varphi}(\bar{\varphi})(E+p_E),
\end{align}
where again the label $(gi)$ denotes the gauge invariant variable. In principle we could also choose different combinations of the scalar perturbations in order to construct other gauge invariant matter perturbations. However, this specific combination will appear in the equation of motion for $\Upsilon$. A common choice for the gauge invariant scalar field perturbation is the Mukhanov-Sasaki variable (compare e.g. \cite{Langlois} appendix A). It is constructed from the following combination of matter and geometry perturbations, which on the extended ADM-phase space reads:
\begin{align}
\label{eq:mukhsasak}
v &:= \delta \varphi - \frac{\mathcal{LM}(\dot{\bar{\varphi}})}{\tilde{\mathcal{H}}}\left( \psi -\frac{1}{3}\Delta E\right) \nonumber \\
~ &= \delta \varphi -\frac{\lambda_\varphi}{A^{3/2}\tilde{\mathcal{H}}}\bar{N}\bar{\pi}_\varphi \left( \psi -\frac{1}{3}\Delta E\right) \nonumber \\
~ &= \delta \varphi^{(gi)} -\frac{\lambda_\varphi}{A^{3/2}\tilde{\mathcal{H}}}\bar{N}\bar{\pi}_\varphi \left( \psi -\frac{1}{3}\Delta E\right) - \bar{\pi}_\varphi \frac{4\tilde{\mathcal{H}}\lambda_\varphi}{\sqrt{A}\bar{N}}(E+p_E) \nonumber \\
~ &= \delta \varphi^{(gi)} - \frac{\lambda_\varphi}{A^{3/2}\tilde{\mathcal{H}}}\bar{N}\bar{\pi}_\varphi \Psi .
\end{align}
Hence, we have found the analoges of the Bardeen potentials and the Mukhanov-Sasaki variable on phase space and we will discuss their equations of motion in the next section.

\subsubsection{Gauge invariant equations of motion}

Using the equations of motion for the scalar and matter perturbations as well as the background quantities one can derive the gauge invariant equations of motion, i.e. equations of motion for $\Psi$ and $\Upsilon$. Their derivation is discussed in detail  in appendix (\ref{append:GIEOM}). For conciseness we will just state the results here. The equations of motion for $\Psi$ and $\Upsilon$ are of the form:
\begin{align}
\label{eq:dotPsi&dotUpsilon}
\dot{\Psi} &= -2\tilde{\mathcal{H}}(\Psi-\Upsilon) \nonumber \\
\dot{\Upsilon} &= -\frac{1}{2}\left( \tilde{\mathcal{H}}-\frac{\kappa}{2}\frac{\bar{N}^2}{\tilde{\mathcal{H}}}p \right) \Upsilon
+ \frac{1}{2}\tilde{\mathcal{H}} \Psi -\frac{\kappa}{8}\frac{\bar{N}^2}{\tilde{\mathcal{H}}}\delta \tilde{T}^{(gi)}.
\end{align}
From the linearized secondary constraints $\delta C$, $\delta C_a$ one can derive the following equations:
\begin{align}
3\tilde{\mathcal{H}}^2(\Psi-2\Upsilon) + \frac{\bar{N}^2}{A}\Delta \Psi &\approx \frac{\kappa}{4}\bar{N}^2\delta \tilde{T}_0^{0(gi)} \nonumber \\
2\tilde{\mathcal{H}}\left( \Upsilon-\frac{1}{2}\Psi \right)_{,a} &\approx \frac{\kappa}{4}\bar{N}^2\delta \tilde{T}_a^{0(gi)}~,
\end{align}
where on the right-hand side of the equations $\delta \tilde{T}^{(gi)}$, $\delta \tilde{T}_0^{0(gi)}$ and $\delta \tilde{T}_a^{0(gi)}$ are related to the gauge invariant perturbed energy-momentum tensor $\delta T_\nu^{\mu(gi)}$ as follows:
\begin{align}
\delta \tilde{T}^{(gi)} &:= \mathcal{LM}\left( \tfrac{1}{3}\delta^i_j \delta T_i^{j(gi)} \right) & \delta \tilde{T}_0^{0(gi)} &:= \mathcal{LM} \delta T_0^{0(gi)} & \delta \tilde{T}_a^{0(gi)} &:= \mathcal{LM} \delta T_a^{0(gi)}.
\end{align}
Their explicit expressions can be found in appendix \ref{append:GIEOM}. The equation of motion together with the perturbed secondary constraints should be equivalent to the gauge invariant perturbed Einstein equations involving a minimally coupled scalar field in the matter sector. That the above gauge invariant equations of motion together with the linearized secondary constraints are equivalent to the gauge invariant linearized Einstein equations as given in e.g. \cite{Mukhanov} is also shown in appendix \ref{append:GIEOM}. Hence, our phase space formulation of cosmological perturbation theory is consistent with previous results in the Lagrangian approach. 

This concludes the chapter about linear cosmological perturbation theory in the Hamiltonian formulation. In the next chapter we will discuss a perturbative version of the relational formalism discussed in chapter \ref{se:obsGR}. In our subsequent paper \cite{Giesel:2018opa} we will then apply this perturbative relational formalism to cosmological perturbation theory which might then be a powerful framework for discussing higher order cosmological perturbation theory.

\newpage

\section{Perturbation theory in the context of the relational formalism}
\label{se:obsinpert}

In section \ref{suse:obs} we reviewed the relational formalism introduced by Rovelli \cite{RovelliPartial,RovelliObservable} following the works of Dittrich \cite{Dittrich,Dittrich2} and Thiemann \cite{Thiemann2} and its application to general relativity according to Pons et al. \cite{Pons2,Pons3}.
In this section we quickly summarize this formalism and then discuss a perturbative treatment thereof. Applications to cosmological perturbation theory will then be subject to our companion paper \cite{Giesel:2018opa}.

\subsection{Scalar-vector-tensor decomposition for observables}
Before we discuss the actual scalar-vector-tensor decomposition, we briefly summarize the important points relevant for the construction of observables. It was shown in section (\ref{suse:obs}) equation (\ref{eq:obsformula3}) that $\mathcal{O}_{f,T}[\tau]$ can be represented as a power series in the $G^\mu$ if $f$ does not depend on lapse and shift.
\begin{equation}
\label{eq:obsGR0}
\mathcal{O}_{f,T}[\tau] = f + \sum\limits_{n=1}^{\infty}\frac{1}{n!}\int\mathrm{d}^3y_1 ...\int\mathrm{d}^3y_n G^{\mu_1}(y_1) ... G^{\mu_n}(y_n)\{ ... \{ f,\tilde{C}_{\mu_1}(y_1) \}, ..., \tilde{C}_{\mu_n}(y_n) \},
\end{equation}
with the weakly abelianized secondary constraints:
\begin{equation}
\tilde{C}_\mu(x) = \int\mathrm{d}^3y \mathcal{B}^\nu_\mu(y,x)C_\nu(y).
\end{equation}
The distributional matrix $\mathcal{B}$ is defined as the inverse of the matrix obtained by computing the Poisson brackets of the clocks with the secondary constraints:
\begin{align}
\mathcal{A}^\mu_\nu(x,y) &:= \{ T^\mu(x),C_\nu(y) \} & \mathcal{B} &:= \mathcal{A}^{-1}.
\end{align}
Note, that all quantities may have explicit time dependencies, which is suppressed in our notation but should be kept in mind.

When $f$ does depend on lapse $N$ and shift $N^a$, one additionally has to take the primary constraints $\Pi_\mu$ and the time derivatives of the gauge fixing constraints
\begin{equation}
G^{(2)\mu} := \dot{G}^\mu = \dot{\tau}^\mu - \partial_tT^\mu - \{ T^\mu,H \}
\end{equation}
into account. Observables can be constructed using the primary, secondary and gauge fixing constraints:
\begin{align}
\mathfrak{G}^I &:= (G^\mu,G^{(2)\mu}) & \mathfrak{C}_I &:= (C_\mu,\Pi_\mu) ~~~I=1,...,8.
\end{align}
Let us recall the set of constraints introduced in (\ref{eq:AbelConstrExt}) where we defined $\tilde{\mathfrak{C}}_I(x) =: (\tilde{\tilde{C}}_\mu,\tilde{\Pi}_\mu)$ with:
\begin{align}
\label{eq:tildePitildetildeC}
\tilde{\Pi}_\mu(x) &= \int\mathrm{d}^3y~\mathcal{B}^\nu_\mu(y,x)\Pi_\nu(y) \nonumber \\
\tilde{\tilde{C}}_\mu(x) &= \int\mathrm{d}^3y~\mathcal{B}^\nu_\mu(y,x) \left[ C_\nu(y) - \int\mathrm{d}^3z\int\mathrm{d}^3v~ \mathcal{B}^\sigma_\rho(v,z)\{ \dot{T}^\rho(z),C_\nu(y) \}\Pi_\sigma(v) \right].
\end{align}
Using these definitions the observable $\mathcal{O}_{f,T}[\tau]$ can again be written as a power series which has already been stated in (\ref{eq:obsGR2}):
\begin{equation}
\label{eq:ObsGen}
\mathcal{O}_{f,T}[\tau] = f + \sum\limits_{n=1}^{\infty} \frac{1}{n!} \int\mathrm{d}^3y_1...\int\mathrm{d}^3y_n~ \mathfrak{G}^{I_1}(t,y_1)...\mathfrak{G}^{I_n}(t,y_n) \{...\{ f,\tilde{\mathfrak{C}}_{I_1}(y_1) \},...\tilde{\mathfrak{C}}_{I_n}(y_n) \}.
\end{equation}
Note, that all formulas are evaluated on-shell, i.e. we neglect terms proportional to the primary and secondary constraints. Using this, one can see that the above formula does reduce to equation (\ref{eq:obsGR0}) if $f$ does not depend on lapse and shift.

In the context of cosmological perturbation theory we introduced projectors $\hat{P}_\text{Tr}$, $\hat{P}_{LS}$, etc. projecting onto the scalar, vector and tensor parts of the perturbations. For the reason that later on we want to construct observables associated with these scalar, vector and tensor perturbations, the question arises how the observables shown in the equation above need to be modified if we consider instead of a generic phase space function $f$ its corresponding projection $\hat{P}f$. We will discuss the application of this  for the case that our $f$ will be chosen among the linear perturbations of the geometry or matter sector. Note that the action of the projectors $\hat{P}$ only affects the function $f$ in the observable formula. This can be seen from the observable formula in (\ref{eq:ObsGen}). The projectors $\hat{P}$ involve background quantities as well as derivative operators with respect to the variables the function $f$ depends on. However, the constraints and the gauge fixing constraints are evaluated at a different, independent variable and thus the action of $\hat{P}$ on them becomes trivial. Furthermore, the iterated Poisson bracket involved in the observable formula is evaluated with respect to the phase space of the linear perturbations. As discussed earlier for perturbations around a k=0 FLRW background the relevant projectors depend only on differential operators and background quantities. In other words as far as the phase space of the perturbations is considered those projectors are phase space independent. Given this, we have for all projectors $\hat{P}$ considered in our further computations that the following relation holds:
\begin{align}
\mathcal{O}_{\hat{P}f,T}[\tau] &= \hat{P} \mathcal{O}_{f,T}[\tau].
\end{align}
In the following we will use this fact only for the case of linear perturbations theory which is sufficient for the computations in this article. To apply the observable framework to cosmological perturbation theory we have to treat the observable framework in a perturbative way. This will be discussed in the next section.

\subsection{Linear perturbations of observables}

We consider linear perturbations of the observable formula for functions $f$ independent of lapse and shift. For this case we can use the observable formula (\ref{eq:obsGR0}). The first order perturbation yields:
\begin{align}
\delta \mathcal{O}_{f,T}[\tau] &= \delta f + \int\mathrm{d}^3y~ \delta G^\mu(y)\overline{\mathcal{O}_{ \{ f,\tilde{C}_\mu(y) \},T }[\tau] } \nonumber \\
~ &+ \sum\limits_{n=1}^{\infty} \frac{1}{n!} \int\mathrm{d}^3y_1...\int\mathrm{d}^3y_n~ \bar{G}^{\mu}(y_1) ... \bar{G}^{\mu_n}(y_n) \{ ... \{ f,\tilde{C}_{\mu_1}(y_1) \}, ...,\tilde{C}_{\mu_n}(y_n) \}^{(1)}.
\end{align}
Let us look at the gauge fixing constraints first. The functions $\tau^\mu$ involved therein should be in the range of the clocks $T^\mu$, that is there exists a gauge such that $T^\mu=\tau^\mu$. Now, for a generic background we assume that on the background the gauge fixing conditions are satisfied and thus
\begin{align}
\bar{G}^\mu &= \bar{\tau}^\mu-\bar{T}^\mu=0,\quad\Leftrightarrow\quad \bar{T}^\mu=\bar{\tau}^\mu .
\end{align}
Considering the linear perturbations of the gauge fixing constraint $\delta G^\mu$ we obtain
\begin{equation*}
\delta G^\mu=G^\mu-\bar{G}^\mu=\tau^\mu - T^\mu-\bar{\tau}^\mu+\bar{T}^\mu=\delta \tau^\mu-\delta T^\mu 
\end{equation*}
Given this the formula for the linear perturbations of the observables simplifies to
\begin{align}
\label{eq:pertobsGen}
\delta \mathcal{O}_{f,T}[\tau] &= \delta f +\int\mathrm{d}^3y\delta G^\mu(y)\overline{\{ f,\tilde{C}_\mu(y) \}} \nonumber \\
~ &\approx \delta f + \int\mathrm{d}^3y\int\mathrm{d}^3z~ \delta G^\mu(y)\bar{\mathcal{B}}^\nu_\mu(z,y)\overline{\{ f,C_\nu(z) \}}.
\end{align}
As we will discuss in our companion paper in more detail, the range of values allowed for $\bar{\tau}^\mu$ of course depends on the specific choice of clocks $\bar{T}^\mu$ and their linear perturbations respectively. A particular choice of $\delta T^\mu$ can therefore be related to a specific gauge. However, the constructed observables are not restricted to this gauge, but can be understood as gauge invariant extensions away from the gauge fixing constraint surface.

So far we assumed that $f$ does not depend on lapse and shift. Otherwise one has to perturb the observable formula (\ref{eq:obsGR2}) instead. Using again the assumption that the gauge fixing constraints and its time derivatives on the background are satisfied, that is $\bar{G}^\mu=0$ and $\bar{G}^{(2)\mu}=0$ the linearized observable formula takes the following form:
\begin{align}
\label{eq:pertobs2}
\delta \mathcal{O}_{f,T}[\tau] &= \delta f + \int\mathrm{d}^3y \left[ \delta \dot{G}^\mu(y)\overline{\{ f,\tilde{\Pi}_\mu(y) \}} + \delta T^\mu(y)\overline{\{ f,\tilde{\tilde{C}}_\mu(y) \}} \right] \nonumber \\
~ &\approx \delta f + \int\mathrm{d}^3y \int\mathrm{d}^3z \bar{\mathcal{B}}^\nu_\mu(z,y) \left[ \delta \dot{G}^\mu(y)\overline{\{ f,\Pi_\nu(z) \}} - \delta G^\mu(y) \left( \overline{\{ f,C_\nu(z) \}} \right. \right. \nonumber \\
~ &\hspace{8em} + \left. \left. \int\mathrm{d}^3w \int\mathrm{d}^3v~ \bar{\mathcal{B}}^\rho_\sigma(w,v) \overline{\{ \dot{T}^\sigma(v),C_\nu(z) \}}~ \overline{\{ f, \Pi_\rho(w) \}} \right) \right].
\end{align}

In a subsequent paper we will apply this framework to cosmological perturbation theory. The idea is to reproduce the Bardeen potentials $\Psi$ and $\Upsilon$ or the Mukhanov-Sasaki variable $v$ with this formalism at first order in the perturbations. Moreover the observable power-series formula (\ref{eq:obsGR2}) can be used to directly calculate higher order gauge invariant perturbations. Lastly, finding appropriate clock fields leading to the commonly used gauge invariant quantities in linear cosmological perturbation theory might possibly get us an inside in the geometrical meaning of these clocks.

\section{Conclusions and outlook}
The aim of this article and our companion article \cite{Giesel:2018opa} 
is to reformulate linear canonical cosmological perturbation theory in the framework of the relational formalism, that is to derive a relation between Dirac observables, constructed in this framework, and the common gauge invariant quantities conventionally used in cosmological perturbation theory. As a first step towards this aim, we reviewed canonical cosmological perturbation theory at the linearized level. In the already existing literature one usually considers linearized perturbations around the reduced ADM phase space in which lapse and shift are not treated as dynamical variables, but as Lagrange multipliers, see for instance \cite{Langlois}.

However, in order to also rederive the canonical counterpart of for instance the Bardeen potentials it was necessary to extend the canonical analysis to the full or also called extended ADM phase space. Here we could build on techniques earlier developed in \cite{Pons1,Pons2} which have been reviewed in section \ref{se:obsGR} and allow to analyze in detail how diffeomorphisms act on the extended ADM phase space. Section \ref{se:cosmpert} includes a brief review on general relativistic perturbation theory, that is we did not fix a specific background yet and derived the full general relativistic linearized equations of motion for the reduced as well as extended phase space after a scalar-vector-tensor decomposition was performed. Afterwards these results were specialized to the case of a flat FLRW background solution yielding naturally to a canonical form of linearized cosmological perturbation theory that is consistent with the already existing results in the literature. In the new application of the extended ADM phase space formalism applied to cosmological perturbation theory we were able to compute the canonical form of the Bardeen potential as well as the Mukhanov-Sasaki variable analoge. Likewise to the Lagrangian picture the gauge invariant quantities are constructed from elementary configuration variables and specific combinations of the remaining variables that involve also momenta. The latter is expected since in the Lagrangian formulation also velocities are involved in these combinations. 
Furthermore, we could show that the resulting equations of motion for these linearized gauge invariant quantities agree with the ones in the literature derived in the Lagrangian framework.

Now if we go one step further towards our aim to reformulate canonical perturbation theory in terms of Dirac observables in the relational formalism, we certainly have to choose a set of clock fields (reference fields) that - when applying the observable map - yields the correct expression for the gauge invariant quantities such as the Bardeen potentials or the Mukhanov-Sasaki variable respectively. How perturbation theory can be formulated in terms of Dirac observables and how the choice of clock fields enters the construction has been discussed in section \ref{se:obsinpert}. 

The results obtained in this article provide the necessary basis to successfully implement canonical perturbation theory in terms of Dirac observables. 
The reason for this is that the construction of the canonical counterparts of the Bardeen potentials and the Mukhanov-Sasaki variable allows us to get an idea what are convenient choices for clocks fields associated to the Bardeen potentials and the Mukhanov-Sasaki variable respectively. Given this, in our companion paper we will use these clock fields together with the observable map and reconstruct the Bardeen potentials as well as the Mukhanov-Sasaki variable within the relational formalism as gauge invariant extensions of the corresponding phase space functions. As we will also present in our companion paper, each choice of clock fields can be related to the choice of a particular gauge. Hence, implementing cosmological perturbation theory within the relational formalism opens also the possibility to investigate the geometrical interpretation of gauges from a new perspective that is not necessarily available in the Lagrangian framework. 

Finally, let us briefly comment on the work in \cite{Dust1,Dust2,Han:2015jsa} where also the relational formalism was used in order to construct gauge invariant quantities and considered in the context of linearized cosmological perturbations theory. The difference to this work here is that on the one hand they consider the reduced ADM phase space only and on the other hand they worked with non-linear matter clocks. By this we mean the following: In \cite{Dust1,Dust2,Han:2015jsa} the authors consider gravity coupled to so called dust matter action at the non-linear level. Then they use the dust fields as clock fields and by means of them construct manifestly gauge invariant variables of the remaining geometric degrees of freedom at the full non-linear level and derive their associated equations of motion. Afterwards a perturbation of these equations of motion is considered and the resulting evolution equations for the perturbations are gauge invariant by construction. By restricting to linear perturbations it is shown that the results obtained are consistent with the results one usually gets in linearized cosmological perturbation theory. The fact that in principle one has non-linear clock fields available can be very useful as far as higher order cosmological perturbations theory is considered because once a set of clock fields has been chosen, it might be technically difficult but conceptually very clear how to construct also higher order gauge invariant quantities. 

However, if one uses geometrical clocks instead of matter clock fields as done in our work, the construction of a model for non-linear clock fields is much harder and has so far not been successfully formulated. The work done in this and our companion paper \cite{Giesel:2018opa} addresses this problem from a new angle, because given some candidates for linearized geometrical clocks, we might be able to get a better idea how their non-linear completion could look like and what kind of geometrical interpretation they have. 

\section*{Acknowledgements}
A.H. and K.G. thank Parampreet Singh for fruitful discussions, comments and suggestions for the manuscript. Furthermore, K.G. wants to thank Guillermo Mena Marugán for pointing out some references and the discussion during Loops'17 in Warsaw.
\appendix
\newpage

\section{Useful formulas for calculating Poisson brackets in perturbation theory}
\label{appendPoissonPert}

The Poisson brackets can be expanded perturbatively up to linear order as:
\begin{equation}
\label{PoissonPertLin}
\{ f(x),g(y) \} = \overline{\{ f(x),g(y) \}} + \delta \{ f(x),g(y) \} + \mathcal{O}(\delta^2).
\end{equation}
We use (\ref{eq:F(n)}) to express variations with respect to the full phase space variables in terms of the linearized phase space variables. To this end we rewrite the variation of a phase space function $F$ evaluated on the background $\Phi=\bar{\Phi}$ using (\ref{eq:F(n)}):
\begin{equation}
\frac{\delta F(x)}{\delta \Phi_I(y)}(\bar{\Phi}) = \frac{\delta F^{(1)}(x)}{\delta (\delta \Phi_I)(y)}(\bar{\Phi},\delta\Phi) = \frac{\delta (\delta F)(x)}{\delta (\delta \Phi_I)(y)}(\bar{\Phi},\delta\Phi).
\end{equation}
Therefore, we can formulate a relation between the Poisson bracket evaluated on the background and the Poisson bracket on the linearized phase space. For this purpose let us denote the configuration variables of the phase space by $\phi_I$ and the respective momenta by $\pi_\phi^I$. We find:
\begin{align}
\label{eq:backPoissonaslinearPoisson}
\overline{\{ f(x),g(y) \}} &:= \int_\Sigma\mathrm{d}^3z\overline{\left[ \frac{\delta f(x)}{\delta \phi_I(z)}\frac{\delta g(y)}{\delta \pi_\phi^I(z)} - \frac{\delta g(y)}{\delta \phi_I(z)}\frac{\delta f(x)}{\delta \pi_\phi^I(z)} \right]} \nonumber \\
~ &= \int_\Sigma\mathrm{d}^3z\left[ \frac{\delta (\delta f)(x)}{\delta (\delta\phi_I)(z)}\frac{\delta (\delta g)(y)}{\delta (\delta\pi_\phi^I)(z)} - \frac{\delta (\delta g)(y)}{\delta (\delta\phi_I)(z)}\frac{\delta (\delta f)(x)}{\delta (\delta\pi_\phi^I)(z)} \right]  =: \{ \delta f(x),\delta g(y) \}_\delta~,
\end{align}
where on the right-hand side the Poisson bracket $\{\cdot,\cdot\}_\delta$ is the Poisson bracket of the linearized phase space with variables $\delta\Phi_I$. As this is computed on the linearized phase space the final result of the Poisson bracket can depend on background quantities only and thus the equation above is well defined. Likewise we can use (\ref{eq:F(n)}) to compute the linear perturbations of the Poisson bracket involved as the second term on the righthand side in (\ref{PoissonPertLin}). This yields:
\begin{align}
\label{eq:delPoissonaslinearPoisson}
\delta \{ f(x),g(y) \} &= \{ \delta f(x), g^{(2)}(y) \}_\delta + \{ f^{(2)}(x), \delta g(y) \}_\delta \nonumber \\
~ &= \{ \delta f(x), \tfrac{1}{2}\delta^2g(y) \}_\delta + \{ \tfrac{1}{2}\delta^2f(x), \delta g(y) \}_\delta .
\end{align}

In particular this enables to express the perturbed equations of motion as a Poisson bracket on the linearized phase space:
\begin{equation}
\label{eq:deltadotPhi}
\delta \dot{\Phi}_I = \{ \delta \Phi_I,H^{(2)} \}_\delta ,
\end{equation}
where it was used that $\delta^2\Phi_I=0$. In this review and our subsequent paper \cite{Giesel:2018opa} we will drop the subscript $_\delta$ from the Poisson bracket of the linearized phase space for conciseness. 

\section{Calculation of the perturbed equations of motion}
\label{appendA}

We want to derive the perturbed equations of motion of canonical general relativity in (\ref{eq:dotq}), (\ref{eq:dotP}) using the Hamiltonian equations of motion given by:
\begin{align}
\delta \dot{q}_{ab} &= \{ \delta q_{ab},H^{(2)} \} & \delta \dot{P}^{ab} &= \{ \delta P^{ab},H^{(2)} \},
\end{align}
with
\begin{equation}
\label{eq:H(2)}
H^{(2)}=\frac{1}{\kappa} \int \mathrm{d}^3x~ \left[ \delta N \delta C + \bar{N}\tfrac{1}{2}\delta^2C +\delta N^a \delta C_a + \bar{N}^a \tfrac{1}{2}\delta^2C_a + \delta \lambda \delta \Pi + \delta \lambda^a \delta \Pi_a \right](x).
\end{equation}
Since the Hamiltonian $H^{(2)}$ involves the constraints, we have to perturb the constraints $C$  and $C_a$ in (\ref{eq:Cgeo}), (\ref{eq:Cphi}) (\ref{eq:Cageo}) and (\ref{eq:Caphi}) up to second order. Some straight forward but lengthy calculations yield the following expressions:
\begin{align}
\delta C_\text{geo} &= \left[-\frac{1}{2} \bar{C}_\text{geo}\bar{q}^{ab}+\sqrt{\det\bar{q}} (\bar{Q}^{-1})^{abcd} \bar{R}^{(3)}_{cd}+\frac{2}{\sqrt{\det\bar{q}}}(\bar{P}^{ac}\bar{P}^b_c-\frac{1}{2}\bar{P}\bar{P}^{ab}) \right]\delta q_{ab} \nonumber \\
~ &+ \frac{2}{\sqrt{\det\bar{q}}}\bar{Q}_{abcd}\bar{P}^{cd}\delta P^{ab} - \sqrt{\det\bar{q}} \bar{q}^{ab}\delta R^{(3)}_{ab}
\\
\delta^2 C_\text{geo} &= \left[\frac{1}{4}\bar{C}_\text{geo}(\bar{q}^{ab}\bar{q}^{cd}+2\bar{q}^{ac}\bar{q}^{bd})
+\sqrt{\det\bar{q}}(\bar{q}^{ab}\bar{q}^{ce}\bar{q}^{df}+\bar{q}^{ac}\bar{q}^{bd}\bar{q}^{ef}-2\bar{q}^{ac}\bar{q}^{be}\bar{q}^{df})\bar{R}^{(3)}_{ef} \right. \nonumber \\
~ &+ \left. \frac{2}{\sqrt{\det\bar{q}}}\left(\bar{P}^{ac}\bar{P}^{bd}-\frac{1}{2}\bar{P}^{ab}\bar{P}^{cd} - ( \bar{P}^{ae}\bar{P}^b_e - \tfrac{1}{2}\bar{P}\bar{P}^{ab} )\right) \right] \delta q_{ab} \delta q_{cd} \nonumber \\
~ &+ 2\sqrt{\det\bar{q}}(\bar{q}^{ac}\bar{q}^{bd}-\frac{1}{2}\bar{q}^{ab}\bar{q}^{cd}) \delta R^{(3)}_{ab} \delta q_{cd} - \sqrt{\det\bar{q}}\bar{q}^{ab}\delta^2R^{(3)}_{ab} \nonumber \\
~ &+ \left[ -\frac{2}{\sqrt{\det\bar{q}}}\bar{Q}_{abef}\bar{P}^{ef}\bar{q}^{cd} + \frac{2}{\sqrt{\det\bar{q}}}(4\delta^c_a\bar{P}^d_b-\delta^c_a\delta^d_b \bar{P}-\bar{q}_{ab}\bar{P}^{cd}) \right] \delta P^{ab} \delta q_{cd} \nonumber \\
~ &+ \frac{2}{\sqrt{\det\bar{q}}} \bar{Q}_{abcd}\delta P^{ab} \delta P^{cd}
\\
\delta C_\varphi &= \left[-\frac{1}{2} \bar{C}_\varphi\bar{q}^{ab}-\frac{\kappa}{2}\frac{\sqrt{\det\bar{q}}}{\lambda_\varphi}\left( (\bar{Q}^{-1})^{abcd}\bar{\varphi}_{,c}\bar{\varphi}_{,d}-V(\bar{\varphi})\bar{q}^{ab} \right) \right]\delta q_{ab} \nonumber \\
~ &+ \kappa \left[ \frac{\lambda_\varphi}{\sqrt{\det\bar{q}}}\bar{\pi}_\varphi \delta \pi_\varphi + \frac{\sqrt{\det\bar{q}}}{\lambda_\varphi} \left(\bar{q}^{ab}\bar{\varphi}_{,a}\delta \varphi_{,b}+\frac{1}{2}\frac{\mathrm{d}V}{\mathrm{d}\varphi}(\bar{\varphi})\delta \varphi \right) \right]
\\
\delta^2 C_\varphi &= \left[\frac{1}{4}\bar{C}_\varphi(\bar{q}^{ab}\bar{q}^{cd}+2\bar{q}^{ac}\bar{q}^{bd})
-\frac{\kappa}{2}\frac{\sqrt{\det\bar{q}}}{\lambda_\varphi} \left( (\bar{q}^{ab}\bar{q}^{ce}\bar{q}^{df}+\bar{q}^{ac}\bar{q}^{bd}\bar{q}^{ef}-2\bar{q}^{ac}\bar{q}^{be}\bar{q}^{df})\bar{\varphi}_{,e}\bar{\varphi}_{,f} \right. \right.  \nonumber \\
~ &+ \left. \left. \bar{q}^{ac}\bar{q}^{bd}V(\bar{\varphi}) \right) \right] \delta q_{ab} \delta q_{cd}  \nonumber \\
~ &- \kappa \left[ \frac{\lambda_\varphi}{\sqrt{\det\bar{q}}}\bar{q}^{ab}\bar{\pi}_\varphi \delta \pi_\varphi + \frac{\sqrt{\det\bar{q}}}{\lambda_\varphi} \left(2(\bar{q}^{ac}\bar{q}^{bd}-\frac{1}{2}\bar{q}^{ab}\bar{q}^{cd})\bar{\varphi}_{,c}\delta \varphi_{,d}-\frac{1}{2}\frac{\mathrm{d}V}{\mathrm{d}\varphi}(\bar{\varphi})\bar{q}^{ab}\delta \varphi \right) \right] \delta q_{ab} \nonumber \\
~ &+ \kappa \left[ \frac{\lambda_\varphi}{\sqrt{\det\bar{q}}} \delta \pi_\varphi \delta \pi_\varphi + \frac{\sqrt{\det\bar{q}}}{\lambda_\varphi} \left(\bar{q}^{ab}\delta \varphi_{,a}\delta \varphi_{,b}+\frac{1}{2}\frac{\mathrm{d}^2V}{\mathrm{d}\varphi^2}(\bar{\varphi}) \delta \varphi \delta \varphi \right) \right].
\end{align}

\begin{align}
\delta C_{a,\text{geo}} &= -2 (\bar{P}^{bc}_{|c}) \delta q_{ab} -2\bar{q}_{ab}(\delta P^{bc}_{|c} + \bar{P}^{cd}\delta \Gamma^b_{cd})
\\
\delta^2 C_{a,\text{geo}} &= -4\delta P^{bc}_{|c} \delta q_{ab} -4\delta \Gamma^b_{cd}\delta P^{cd}\bar{q}_{ab}
\\
\delta C_{a,\varphi} &=\kappa(\bar{\varphi}_{,a}\delta \pi_\varphi + \bar{\pi}_\varphi \delta \varphi_{,a})
\\
\delta^2 C_{a,\varphi} &=2\kappa \delta \pi_\varphi \delta \varphi_{,a}.
\end{align}
To obtain the results above we have frequently used:
\begin{align}
\delta \Gamma^c_{~ab}=\frac{1}{2}\bar{q}^{cd}(\delta q_{db|a}+\delta q_{ad|b}-\delta q_{ab|d})
\end{align}
as well as:
\begin{equation}
\delta^2 \Gamma^c_{~ab} = -2\bar{q}^{cd}\delta \Gamma^e_{~ab}\delta q_{de}.
\end{equation}
These results are already sufficient to calculate $\delta \dot{q}_{ab}$. We first calculate the Poisson brackets of $\delta q_{ab}$ with the different parts occurring in the formula for $H^{(2)}$ (see (\ref{eq:H(2)})):
\begin{align}
\{ \delta q_{ab},\tfrac{1}{\kappa}\delta C[\delta N] \} &= \frac{2}{\sqrt{\det\bar{q}}}\bar{Q}_{abcd}\bar{P}^{cd} \delta N
\nonumber \\
\{ \delta q_{ab},\tfrac{1}{2\kappa}\delta^2 C[\bar{N}] \} &= \bar{N}\left[ -\frac{1}{\sqrt{\det\bar{q}}}\bar{Q}_{abef} \bar{P}^{ef}\bar{q}^{cd}+\frac{1}{\sqrt{\det\bar{q}}}(4\delta^c_a\bar{P}^d_b-\delta^c_a\delta^d_b\bar{P}-\bar{q}_{ab}\bar{P}^{cd}) \right] \delta q_{cd} \nonumber \\
~ & \hspace{1.5em} + \bar{N} \frac{2}{\sqrt{\det\bar{q}}}\bar{Q}_{abcd}\delta P^{cd}
\nonumber \\
\{ \delta q_{ab},\tfrac{1}{\kappa}\delta C_c[\delta N^c] \} &= 2\delta N_{(a|b)} = (\mathcal{L}_{\delta\vec{N}}\bar{q})_{ab}
\nonumber \\
\{ \delta q_{ab},\tfrac{1}{2\kappa}\delta^2 C_c[\bar{N}^c] \} &= (\mathcal{L}_{\vec{\bar{N}}}\delta q)_{ab},
\end{align}
where integration by parts has been used for the last two equations. Adding up all individual terms yields the equation for $\delta \dot{q}_{ab}$ in (\ref{eq:deltadotq}). \\

For $\delta \dot{P}^{ab}$ we need to express the curvature perturbations $\delta R^{(3)}_{ab}$ and $\bar{q}^{ab}\delta^2 R^{(3)}_{ab}$ in terms of $\delta q_{ab}$ and its derivatives. Using the explicit expression of $\delta R^{(3)}_{ab}$ in (\ref{eq:delR}) and considering perturbations up to second order yields $\delta^2R^{(3)}_{ab}$. For the specific combination of the curvature perturbation involved in $\delta^2C_\text{geo}$ we can derive:
\begin{equation}
2(\bar{q}^{ac}\bar{q}^{bd}-\frac{1}{2}\bar{q}^{ab}\bar{q}^{cd}) \delta R^{(3)}_{ab} \delta q_{cd} - \bar{q}^{ab}\delta^2R^{(3)}_{ab} = A^{abcdef}\delta q_{cd|fe}\delta q_{ab} + B^{abcdef} \delta q_{ab|e}\delta q_{cd|f},
\end{equation}
with
\begin{equation}
A^{abcdef} = 2\bar{q}^{ae}\bar{q}^{bc}\bar{q}^{df} - \bar{q}^{ab}\bar{q}^{ce}\bar{q}^{df} + 2\bar{q}^{ac}\bar{q}^{bf}\bar{q}^{de} - 2\bar{q}^{ae}\bar{q}^{bf}\bar{q}^{cd} - 2\bar{q}^{ac}\bar{q}^{bd}\bar{q}^{ef} + \bar{q}^{ab}\bar{q}^{cd}\bar{q}^{ef}
\end{equation}
and
\begin{equation}
B^{abcdef} = 2\bar{q}^{ae}\bar{q}^{bc}\bar{q}^{df} - 2\bar{q}^{ab}\bar{q}^{ce}\bar{q}^{df} + \bar{q}^{ac}\bar{q}^{bf}\bar{q}^{de} - \frac{3}{2}\bar{q}^{ac}\bar{q}^{bd}\bar{q}^{ef} + \frac{1}{2}\bar{q}^{ab}\bar{q}^{cd}\bar{q}^{ef}.
\end{equation}

These preliminary results allow us to calculate the Poisson bracket of $\delta P^{ab}$ with the above combination of curvature perturbations appearing in $\delta^2C_\text{geo}$. We obtain:
\begin{align}
~ & \{ \delta P^{ab}(x), \frac{1}{2\kappa} \int \mathrm{d}^3y \left[ \sqrt{\det \bar{q}}\left(2(\bar{q}^{ce}\bar{q}^{df}-\frac{1}{2}\bar{q}^{cd}\bar{q}^{ef}) \delta R^{(3)}_{cd} \delta q_{ef} - \bar{q}^{cd}\delta^2R^{(3)}_{cd}\right)\bar{N}\right](y)  \} \nonumber \\
~ &= \frac{1}{2\kappa} \int \mathrm{d}^3y \sqrt{\det \bar{q}} \bar{N} \left\{ A^{ghcdef}\left[ \delta q_{cd|fe} \{ \delta P^{ab}(x),\delta q_{gh} \} + \delta q_{gh}\{ \delta P^{ab}(x),\delta q_{cd|fe} \} \right] \right. \nonumber \\
~ & \hspace{3em}+ \left. B^{ghcdef}\left[ \delta q_{gh|e}\{ \delta P^{ab}(x), \delta q_{cd|f} \} + \{ \delta P^{ab}(x),\delta q_{gh|e} \}\delta q_{cd|f} \right] \right\}(y).
\end{align}

Integration by parts and using  $\{ \delta P^{ab}(x),\delta q_{cd}(y) \} = -\kappa \delta^{(a}_c\delta^{b)}_d\delta(x,y)$ leads to:

\begin{align}
... &= -\tfrac{1}{2}\sqrt{\det \bar{q}} \left[ \left( A^{(ab)cdef}+A^{cd(ab)fe} - B^{(ab)cdef} - B^{cd(ab)fe} \right)\delta q_{cd|fe}\bar{N} \right. \nonumber \\
~~~ &+ \left. \left( 2A^{cd(ab)(ef)} - B^{(ab)cdef} - B^{cd(ab)fe} \right)\delta q_{cd|f}\bar{N}_{|e} + A^{cd(ab)fe}\delta q_{cd}\bar{N}_{|fe} \right].
\end{align}
Carefully using the definitions of $A$ and $B$, we arrive at the following expression:
\begin{align}
~ & \{ \delta P^{ab}(x), \frac{1}{2\kappa} \int \mathrm{d}^3y \left[ \sqrt{\det \bar{q}}\left(2(\bar{q}^{ce}\bar{q}^{df}-\frac{1}{2}\bar{q}^{cd}\bar{q}^{ef}) \delta R^{(3)}_{cd} \delta q_{ef} - \bar{q}^{cd}\delta^2R^{(3)}_{cd}\right)\bar{N}\right](y)  \} \nonumber \\
~ &= - \sqrt{\det \bar{q}} \left( \bar{q}^{ac}\bar{q}^{bd} -\frac{1}{2}\bar{q}^{ab}\bar{q}^{cd} \right) \delta R^{(3)}_{cd}\bar{N} \nonumber \\
~ &+ \sqrt{\det \bar{q}} \left( \bar{q}^{ab}\bar{q}^{ce}\bar{q}^{df} + \bar{q}^{ac}\bar{q}^{bd}\bar{q}^{ef} - 2\bar{q}^{c(a}\bar{q}^{b)e}\bar{q}^{df} +\frac{1}{2}\bar{q}^{cd}(\bar{Q}^{-1})^{abef} \right) \delta q_{cd} \bar{N}_{|ef} \nonumber \\
~ &-\sqrt{\det \bar{q}} (\bar{Q}^{-1} )^{abcd}\delta \Gamma^e_{~cd}\bar{N}_{|e}.
\end{align}
The other terms in $\{ \delta P^{ab}, \tfrac{1}{2\kappa}\delta^2C[\bar{N}] \}$ are straight forward to compute. For $\{ \delta P^{ab}, \tfrac{1}{\kappa}(\delta C[\delta N] + \delta C_a[\delta N^a]) \}$ one uses the Poisson bracket relation $\{ \delta f, \delta g \} = \overline{\{ f,g \}}$ and the form of $\dot{P}^{ab}$ in (\ref{eq:dotP}). Finally, a quick calculation using integration by parts yields:
\begin{equation}
\{ \delta P^{ab}(x), \tfrac{1}{2\kappa}\delta^2C_c[\bar{N}^c] \} = (\bar{N}^c \delta P^{ab})_{|c} - 2 \bar{N}^{(a}_{|c}\delta P^{b)c} = (\mathcal{L}_{\vec{\bar{N}}}\delta P)^{ab}.
\end{equation}
Summing up all individual contributions to $\{ \delta P^{ab},H^{(2)} \}$, we arrive at the lengthy expression for $\delta \dot{P}^{ab}$ in (\ref{eq:deltadotPgeo}) and (\ref{eq:deltadotPphi}).

Note that we compared the resulting Hamiltonian equations of $\delta \dot{q}_{ab}$ and $\delta \dot{P}^{ab}$  to the result that we obtain by directly perturbing the equations of motion for $q_{ab}$ and $P^{ab}$ in (\ref{eq:deltadotq}), (\ref{eq:deltadotPgeo}) and (\ref{eq:deltadotPphi}) for consistency and we obtain an exact agreement.

\section{Gauge invariant equations: Comparison to the Lagrangian approach}
\label{append:GIEOM}

In this appendix we will derive the equations of motion for $\Psi$ and $\Upsilon$ as well as the linearized constraint equations and compare the results to the corresponding equations of motion in the Lagrangian picture which are derived and discussed e.g. in the work of Mukhanov, Feldman and Brandenberger \cite{Mukhanov}. We start from the definition of $\Psi$ in (\ref{eq:Psi}) and use the equations of motion for $\tilde{P}$ in (\ref{eq:dotAdotPcosm}), as well as the scalar perturbations in (\ref{eq:dotpsiE}) and (\ref{eq:dotGammaSigma}) to derive the following equation of motion for the Bardeen potential:
\begin{equation}
\label{eq:dotPsi}
\dot{\Psi} = -2\tilde{\mathcal{H}}(\Psi-\Upsilon),
\end{equation}
with the gauge invariant perturbation $\Upsilon$ defined in (\ref{eq:Upsilon}). The calculation of $\dot{\Upsilon}$ is slightly more involved. Using the expressions for $\dot{\bar{\varphi}}$, $\dot{\bar{\pi}}_\varphi$ in (\ref{eq:dotbarvarphidotbarpivarphi}) and $\dot{p}$ in (\ref{eq:dotrhodotp}) we find:
\begin{align}
\label{eq:dotupsilonintermediate}
\dot{\Upsilon} &= -\frac{1}{2}\left( \tilde{\mathcal{H}}-\frac{\kappa}{2}\frac{\bar{N}^2}{\tilde{\mathcal{H}}}p \right) \Upsilon
+ \frac{1}{2}\left( \tilde{\mathcal{H}}+\frac{\kappa}{4}\frac{\bar{N}^2}{\tilde{\mathcal{H}}}3\left( \rho+p \right) \right) \Psi \nonumber \\
~ &+\frac{\kappa}{8}\frac{\bar{N}^2}{A^{3/2}\tilde{\mathcal{H}}}\left( \frac{\lambda_\varphi}{A^{3/2}}\bar{\pi}_\varphi^2 \Delta E -\bar{\pi}_\varphi \delta \pi_\varphi +\frac{1}{2}\frac{A^{3/2}}{\lambda_\varphi}\frac{\mathrm{d}V}{\mathrm{d}\varphi}(\bar{\varphi})\delta \varphi +4\frac{\tilde{\mathcal{H}}A}{\bar{N}}\bar{\pi}_\varphi (E+p_E) \right).
\end{align}
The first line of (\ref{eq:dotupsilonintermediate}) is already manifestly gauge invariant. Hence we have to take a closer look at the second line. We use the definitions of $\delta \varphi^{(gi)}$ and $\delta \pi_\varphi^{(gi)}$ from (\ref{eq:deltavarphigi}) and insert this back into the above equation leading to:
\begin{align}
\label{eq:dotupsilonintermediate1.5}
\dot{\Upsilon} &= -\frac{1}{2}\left( \tilde{\mathcal{H}}-\frac{\kappa}{2}\frac{\bar{N}^2}{\tilde{\mathcal{H}}}p \right) \Upsilon
+ \frac{1}{2}\left( \tilde{\mathcal{H}}+\frac{\kappa}{4}\frac{\bar{N}^2}{\tilde{\mathcal{H}}}3\left( \rho+p \right) \right) \Psi \nonumber \\
~ &+\frac{\kappa}{8}\frac{\bar{N}^2}{A^{3/2}\tilde{\mathcal{H}}}\left( \frac{1}{2}\frac{A^{3/2}}{\lambda_\varphi}\frac{\mathrm{d}V}{\mathrm{d}\varphi}(\bar{\varphi})\delta \varphi^{(gi)}-\frac{\lambda_\varphi}{A^{3/2}}\bar{\pi}_\varphi \delta \pi_\varphi^{(gi)} \right).
\end{align}
Now we can see that the second line is indeed gauge invariant. In order to compare this result with the literature \cite{Mukhanov} we have to rewrite this equation slightly in the following form:
\begin{align}
\label{eq:dotupsilonintermediate2}
\dot{\Upsilon} &= -\frac{1}{2}\left( \tilde{\mathcal{H}}-\frac{\kappa}{2}\frac{\bar{N}^2}{\tilde{\mathcal{H}}}p \right) \Upsilon
+ \frac{1}{2}\tilde{\mathcal{H}} \Psi \nonumber \\
~ &+\frac{\kappa}{8}\frac{\bar{N}^2}{A^{3/2}\tilde{\mathcal{H}}}\left( 3\frac{\lambda_\varphi}{A^{3/2}}\bar{\pi}_\varphi^2\Psi -\frac{\lambda_\varphi}{A^{3/2}}\bar{\pi}_\varphi \delta \pi_\varphi^{(gi)} + \frac{1}{2}\frac{A^{3/2}}{\lambda_\varphi}\frac{\mathrm{d}V}{\mathrm{d}\varphi}(\bar{\varphi})\delta \varphi^{(gi)} \right)~,
\end{align}
where the identity $\rho+p= \frac{\lambda_\varphi}{A^3}\bar{\pi}_\varphi^2$ has been used. The second line can be related to the spatial energy-momentum perturbation $\delta T^i_j$ by pulling $\delta T^i_j$ back to phase space. We can calculate $\delta T^i_j$ by perturbing the energy-momentum tensor of a scalar field. This results in the following expression:
\begin{equation}
\label{eq:perturbedT}
\delta T^i_j = \frac{1}{\lambda_\varphi} \left[ -\frac{1}{2} \delta^i_j\left( -\frac{1}{\bar{N}^4}\dot{\bar{\varphi}}^2\delta g_{tt} - 2 \frac{1}{\bar{N}^2}\dot{\bar{\varphi}}\delta\dot{\varphi}+\frac{\mathrm{d}V}{\mathrm{d}\varphi}(\bar{\varphi})\delta \varphi \right) \right].
\end{equation}
Let us define
\begin{equation}
\delta \tilde{T} := \mathcal{LM}\left( \tfrac{1}{3}\delta T^i_j\delta_i^j \right)
\end{equation}
A calculation using the form of $\delta T^i_j$ in (\ref{eq:perturbedT}) yields:
\begin{equation}
\label{eq:tildeTab}
\delta \tilde{T} = -\frac{1}{A^{3/2}} \left[ 3\frac{\lambda_\varphi}{A^{3/2}}\bar{\pi}_\varphi^2\psi -   \frac{\lambda_\varphi}{A^{3/2}} \bar{\pi}_\varphi \delta \pi_\varphi +\frac{1}{2}\frac{A^{3/2}}{\lambda_\varphi}\frac{\mathrm{d}V}{\mathrm{d}\varphi}(\bar{\varphi})\delta \varphi \right].
\end{equation}
The spatial energy-momentum perturbation is not gauge invariant yet, but one can substitute $\psi$, $\delta \varphi$ and $\delta \pi_\varphi$ by their corresponding gauge invariant expressions $\Psi$, $\delta \varphi^{(gi)}$ and $\delta \pi_\varphi^{(gi)}$. This motivates to define the gauge invariant extension of the spatial energy-momentum perturbation as:
\begin{equation}
\delta \tilde{T}^{(gi)} := -\frac{1}{A^{3/2}} \left[ 3\frac{\lambda_\varphi}{A^{3/2}}\bar{\pi}_\varphi^2\Psi -   \frac{\lambda_\varphi}{A^{3/2}} \bar{\pi}_\varphi \delta \pi_\varphi^{(gi)} +\frac{1}{2}\frac{A^{3/2}}{\lambda_\varphi}\frac{\mathrm{d}V}{\mathrm{d}\varphi}(\bar{\varphi})\delta \varphi^{(gi)} \right],
\end{equation}
which is up to an overall factor precisely the expression appearing in the second line of (\ref{eq:dotupsilonintermediate2}). Therefore, one can rewrite the equation of motion for $\Upsilon$ in a more concise form as follows:
\begin{equation}
\label{eq:dotupsilon}
\dot{\Upsilon} = -\frac{1}{2}\left( \tilde{\mathcal{H}}-\frac{\kappa}{2}\frac{\bar{N}^2}{\tilde{\mathcal{H}}}p \right) \Upsilon
+ \frac{1}{2}\tilde{\mathcal{H}} \Psi -\frac{\kappa}{8}\frac{\bar{N}^2}{\tilde{\mathcal{H}}}\delta \tilde{T}^{(gi)}.
\end{equation}
In order to compare our results with the literature (e.g. \cite{Mukhanov}) we have to derive Lagrangian equation of motion for $\Psi$ that involves second order time derivatives. One can use equation (\ref{eq:dotPsi}) to express $\Upsilon$ in terms of $\Psi$ and its velocity to get:
\begin{equation}
\mathcal{LM}^*\Upsilon = \frac{\dot{\Psi}}{2\mathcal{H}}+\Psi .
\end{equation}
Now we can derive the second order equation of motion by calculating $\ddot{\Psi}$ from (\ref{eq:dotPsi}) and pulling it back to the tangent bundle. We obtain
\begin{flalign}
\ddot{\Psi} &= -2\dot{\mathcal{H}}(\Psi-\Upsilon) - 2\mathcal{H}(\dot{\Psi}-\dot{\Upsilon}) \nonumber \\
~ &= -2\left( \mathcal{H} \frac{\dot{\bar{N}}}{\bar{N}}-\frac{3}{2}\mathcal{H}^2-\frac{\kappa}{4}\bar{N}^2p \right)(\Psi-\Upsilon) \nonumber \\
~ &~~~ +4\mathcal{H}^2(\Psi-\Upsilon)-\left( \mathcal{H}^2-\frac{\kappa}{2}\bar{N}^2p \right)\Upsilon +\mathcal{H}^2\Psi - \frac{\kappa}{4}\bar{N}^2\delta T^{(gi)} \nonumber \\
~ &= \left( \frac{\dot{\bar{N}}}{\bar{N}}-4\mathcal{H}\right)\dot{\Psi} +\frac{\kappa}{2}\bar{N}^2p\Psi - \frac{\kappa}{4}\bar{N}^2\delta T^{(gi)}.
\end{flalign}
with $\delta T^{(gi)} = \mathcal{LM}^*\delta \tilde{T}^{(gi)}$.
This yields the following second order equation of motion:
\begin{equation}
\label{eq:ddotPsi}
\ddot{\Psi} +\left(4\mathcal{H} - \frac{\dot{\bar{N}}}{\bar{N}}\right)\dot{\Psi} -\frac{\kappa}{2}\bar{N}^2p\Psi = - \frac{\kappa}{4}\bar{N}^2\delta T^{(gi)}.
\end{equation}

To compare this equation with the perturbed spatial Einstein-equations $\delta G^{i(gi)}_j = \frac{\kappa}{2}\delta T^{i(gi)}_j$ in \cite{Mukhanov}  first we have to ensure that our $\delta T^{i(gi)}_j$ is the same one that is used in \cite{Mukhanov} for our FLRW $k=0$ + scalar field model. First we calculate:
\begin{equation}
\delta \tilde{T}^{(gi)} = \delta \tilde{T} - \frac{A}{\bar{N}^2}\frac{1}{A^{3/2}}\left[ 3\tilde{\mathcal{H}}\frac{\lambda_\varphi}{A^{3/2}}\bar{\pi}_\varphi^2 + \bar{N}\bar{\pi}_\varphi\frac{\mathrm{d}V}{\mathrm{d}\varphi}(\bar{\varphi}) \right]4\tilde{\mathcal{H}}(E+p_E).
\end{equation}
The gauge invariant extension in the above equation can be expressed in terms of $\dot{\tilde{\bar{T}}}$. To show this we calculate $\tilde{\bar{T}}$ from the spatial background energy-momentum tensor of the scalar field:
\begin{equation}
\label{eq:barT}
\bar{T}^i_j=\frac{1}{\lambda_\varphi} \left[ -\frac{1}{2}\delta^i_j \left( -\frac{1}{\bar{N}^2}\dot{\bar{\varphi}}^2+V(\bar{\varphi}) \right)\right].
\end{equation}
Pulling back these quantities to phase space yields:
\begin{equation}
\tilde{\bar{T}} := \mathcal{LM}\left( \tfrac{1}{3} \delta^i_j \bar{T}^j_i \right) = - \frac{1}{2\lambda_\varphi}\left[ -\frac{\lambda_\varphi^2}{A^3}\bar{\pi}_\varphi^2 + V(\bar{\varphi}) \right].
\end{equation}
Its time derivative can be derived from the equations of motion for $\dot{\bar{\varphi}}$ and $\dot{\bar{\pi}}_\varphi$ in (\ref{eq:dotphidotpiphi}):
\begin{equation}
\dot{\tilde{\bar{T}}} = - \frac{1}{A^{3/2}}\left[ 3\tilde{\mathcal{H}}\frac{\lambda_\varphi}{A^{3/2}}\bar{\pi}_\varphi^2 + \bar{N}\bar{\pi}_\varphi\frac{\mathrm{d}V}{\mathrm{d}\varphi}(\bar{\varphi}) \right].
\end{equation}
Consequently, the gauge invariant spatial energy-momentum perturbation in phase space can be written as follows:
\begin{equation}
\delta \tilde{T}^{(gi)} = \delta \tilde{T} + \frac{A}{\bar{N}^2}\dot{\tilde{\bar{T}}}4\tilde{\mathcal{H}}(E+p_E).
\end{equation}
To compare this expression with the gauge invariant energy-momentum tensor in \cite{Mukhanov} we have to pull back $\delta \tilde{T}^{(gi)}$ to the tangent bundle, leading to:
\begin{equation}
\delta T^{(gi)} = \delta T + \frac{A}{\bar{N}^2} \dot{\bar{T}}(B-\dot{E}).
\end{equation}
This is precisely the expression appearing in \cite{Mukhanov} in equation (4.13) for the case of conformal time $\bar{N}=\sqrt{A}$ and $\bar{T}^i_j = \bar{T}\delta^i_j$. Next, we have to substitute $\bar{N}=\sqrt{A}$ into our equation for $\ddot{\Psi}$ in (\ref{eq:ddotPsi}). This leads us to the following relations:
\begin{align}
\frac{\dot{\bar{N}}}{\bar{N}} &= \mathcal{H} \nonumber \\
-\frac{\kappa}{2}\bar{N}^2p &= 2\dot{\mathcal{H}}+\mathcal{H}^2.
\end{align}
Further we use $\kappa = 16\pi G$ and get:
\begin{equation}
\ddot{\Psi} + 3\mathcal{H}\dot{\Psi} + (2\dot{\mathcal{H}}+\mathcal{H}^2)\Psi = -4\pi G A \delta T^{(gi)}.
\end{equation}
We compare this to the third equation in (4.15) in \cite{Mukhanov} for $D=0$, $k=0$ and find that the equations coincide.
~\\
~\\
\paragraph*{\centerline{C.a Perturbation of the Constraint equations}}
~\\

The perturbed constraints $\delta C$ and $\delta C_a$, given in (\ref{eq:deltaCdeltaCa}), are the secondary constraints on the linearized phase space. We will see that their pull-back to the tangent-bundle corresponds to the temporal-temporal and temporal-spatial linearized Einstein-equations respectively. We use the formula for the perturbed energy-momentum tensor and apply the Legendre map to its temporal-temporal and temporal-spatial components to find the corresponding expressions in phase space. Similarly to $\delta \tilde{T}$ we define:
\begin{align}
\delta \tilde{T}^0_0 &= \mathcal{LM}(\delta T^0_0) & \delta \tilde{T}^0_a &= \mathcal{LM}(\delta T^0_a).
\end{align}
A few steps of calculation yield:
\begin{equation}
\delta \tilde{T}_0^0 = \frac{1}{A^{3/2}}\left[ 3\frac{\lambda_\varphi}{A^{3/2}}\bar{\pi}_\varphi^2\psi - \frac{\lambda_\varphi}{A^{3/2}}\bar{\pi}_\varphi\delta \pi_\varphi -\frac{1}{2}\frac{A^{3/2}}{\lambda_\varphi}\frac{\mathrm{d}V}{\mathrm{d}\varphi}(\bar{\varphi})\delta \varphi \right]
\end{equation}
and
\begin{equation}
\label{eq:delTa0}
\delta \tilde{T}_a^0 = -\frac{1}{\bar{N}A^{3/2}}\bar{\pi}_\varphi\delta \varphi_{,a}.
\end{equation}
We use the explicit expression of the linearized constraints in (\ref{eq:deltaCdeltaCa}) and $\bar{C}_\varphi=\kappa A^{3/2}\rho$, where $\rho$ is defined in (\ref{eq:rhop}), to rewrite the above equations:
\begin{align}
\label{eq:TasC}
\delta \tilde{T}_0^0 &= \frac{1}{\kappa A^{3/2}} \left( \bar{C}_\varphi 3\psi-\delta C_\varphi \right) \nonumber \\
\delta \tilde{T}_a^0 &= -\frac{1}{\kappa \bar{N} A^{3/2}} \delta C_{a,\varphi}~,
\end{align}
where we also used the matter part of the linearized spatial diffeomorphism constraint shown in (\ref{eq:deltaCdeltaCa}). We use these results to express the linearized Hamiltonian constraint $\delta C$ in terms of $\delta T^0_0$:
\begin{equation}
\frac{\bar{N}^2}{4A^{3/2}}\delta C = 3\tilde{\mathcal{H}}^2(\psi-2p_\psi)+\frac{3\bar{N}^2}{4A^{3/2}}\bar{C}\psi + \frac{\bar{N}^2}{A}\Delta \left( \psi-\frac{1}{3}\Delta E \right) -\frac{\kappa}{4}\bar{N}^2\delta \tilde{T}_0^0.
\end{equation}
Using the definitions of the gauge invariant perturbations $\Psi$ and $\Upsilon$ this can be rewritten in the following way:
\begin{equation}
\frac{\bar{N}^2}{4A^{3/2}}\delta C = 3\tilde{\mathcal{H}}^2(\Psi-2\Upsilon)+\frac{\bar{N}^2}{4A^{3/2}}\bar{C}(3\Psi+\Delta E) + \frac{\bar{N}^2}{A}\Delta \Psi -\frac{\kappa}{4}\bar{N}^2\delta \tilde{T}_0^{0(gi)}
\end{equation}
with
\begin{equation}
\delta \tilde{T}_0^{0(gi)} := \delta \tilde{T}_0^0+3\frac{\tilde{\mathcal{H}}}{\bar{N}^2}\frac{\lambda_\varphi}{A^2}\bar{\pi}_\varphi^2 4\tilde{\mathcal{H}}(E+p_E).
\end{equation}
We also used the explicit form of $\bar{C}$ in (\ref{eq:barCbarCa}) to express the above equation mostly in terms of the Bardeen potentials. Pulling this equation back to the tangent-bundle ($\bar{C} \to 0$, $\delta C \to 0$, $\tilde{\mathcal{H}}\to \mathcal{H}$, $\Upsilon \to \frac{\dot{\Psi}}{2\mathcal{H}}+\Psi$) we get:
\begin{equation}
\label{eq:EOMconstraintC}
-3\mathcal{H}(\mathcal{H}\Psi+\dot{\Psi})+\frac{\bar{N}^2}{A}\Delta \Psi = \frac{\kappa}{4}\bar{N}^2\delta T_0^{0(gi)}.
\end{equation}
with $\delta T_0^{0(gi)} = \mathcal{LM}^*\delta \tilde{T}_0^{0(gi)}$.
It can be checked, that our definition of $\delta T_0^{0(gi)}$ coincides with the one in \cite{Mukhanov}.
Furthermore, equation (\ref{eq:EOMconstraintC}) becomes (for conformal time and $D=0$) exactly the first equation in (4.15) of \cite{Mukhanov}.

Now the equation corresponding to $\mathcal{LM}^*\delta C_a=0$ is still left. Using equation (\ref{eq:TasC}) we can rewrite the expression of $\delta C_a$ (\ref{eq:deltaCdeltaCa}) as follows:
\begin{equation}
\frac{\bar{N}}{4A^{3/2}}\delta C_a = 2\tilde{\mathcal{H}}\left[ p_\psi-\frac{1}{2}\psi+\frac{2}{3}\Delta\left( E+p_E \right) \right]_{,a} + \tilde{\mathcal{H}}\Delta(F_a+p_F^b\delta_{ab}) -\frac{\kappa}{4}\bar{N}^2\delta \tilde{T}_a^0.
\end{equation}
In terms of the gauge invariant perturbations this reads:
\begin{equation}
\frac{\bar{N}}{4A^{3/2}}\delta C_a = 2\tilde{\mathcal{H}}\left( \Upsilon-\frac{1}{2}\Psi \right)_{,a} + \tilde{\mathcal{H}}\Delta \nu^b\delta_{ab} -\frac{\kappa}{4}\bar{N}^2\delta \tilde{T}_a^{0(gi)} - \frac{\tilde{\mathcal{H}}}{\sqrt{A}}\bar{C}(E+p_E)_{,a},
\end{equation}
with $\nu^a = \delta^{ab}F_b+p_F^a$ and
\begin{equation}
\delta \tilde{T}_a^{0(gi)} := \delta \tilde{T}_a^0 -\frac{\lambda_\varphi}{\bar{N}^2A^2}\bar{\pi}_\varphi^24\tilde{\mathcal{H}}(E+p_E)_{,a}.
\end{equation}
On the tangent bundle this becomes:
\begin{equation}
\label{eq:dCagi}
\left( \mathcal{H}\Psi+\dot{\Psi} \right)_{,a} + \mathcal{H}\Delta\nu^b\delta_{ab} = \frac{\kappa}{4}\bar{N}^2\delta T_a^{0(gi)}.
\end{equation}
As $\hat{P}_{LT}\delta T_a^{0(gi)} = 0$ we get $\Delta\nu^a=0$.
After realizing that $\delta T_a^{0(gi)}$ is exactly the corresponding quantity in \cite{Mukhanov} we again find that this equation is precisely the second equation in (4.15) of \cite{Mukhanov}.
~\\
~\\
\paragraph*{\centerline{C.b Summary}}
~\\

We have shown that our derived Hamiltonian equations of motion, when pulled back to the tangent-bundle, become exactly the equations of motion for the Bardeen potentials presented in equation (4.15) of \cite{Mukhanov} for the special case $\bar{N}=\sqrt{A}$, $k=0$ and $D=\Psi+\Phi=0$. Thus, we can conclude that, in the case of linear perturbations around a flat ($k=0$) FLRW model minimally coupled to a scalar field, our formalism is equivalent to the conventional formalism of cosmological perturbation theory in the Lagrangian picture. 
Finally, for the benefit of the reader let us list our resulting Hamiltonian equation of motion here again:
\begin{align}
\dot{\Psi} &= -2\tilde{\mathcal{H}}(\Psi-\Upsilon) \nonumber \\
\dot{\Upsilon} &= -\frac{1}{2}\left( \tilde{\mathcal{H}}-\frac{\kappa}{2}\frac{\bar{N}^2}{\tilde{\mathcal{H}}}p \right) \Upsilon
+ \frac{1}{2}\tilde{\mathcal{H}} \Psi -\frac{\kappa}{8}\frac{\bar{N}^2}{\tilde{\mathcal{H}}}\delta \tilde{T}^{(gi)}.
\end{align}
\begin{align}
3\tilde{\mathcal{H}}^2(\Psi-2\Upsilon) + \frac{\bar{N}^2}{A}\Delta \Psi &\approx \frac{\kappa}{4}\bar{N}^2\delta \tilde{T}_0^{0(gi)} \nonumber \\
2\tilde{\mathcal{H}}\left( \Upsilon-\frac{1}{2}\Psi \right)_{,a} &\approx \frac{\kappa}{4}\bar{N}^2\delta \tilde{T}_a^{0(gi)}.
\end{align}

\newpage
%\nocite{*}
\bibliography{PaperReview}
\bibliographystyle{unsrt}

\end{document}